\documentclass[11pt,a4paper,openright,twoside]{report}
\usepackage{fancyhdr,a4wide}
\usepackage[latin1]{inputenc}
\usepackage[dvips]{graphicx,psfrag} %Zum Einbinden von Grafiken
\usepackage{amstext} %F�r Text-Indizes
\usepackage{amsmath}
\usepackage{dsfont}%F�r Einheitsmatritzen
\usepackage{picins}
\usepackage{amssymb} %F�r Mengensymbole wie IR
\usepackage{bbm} %F�r Einheitsmatritzen
\usepackage{geometry}
\usepackage{floatflt}
\usepackage{hhline} %Tabellenlinien
\usepackage{subfig}
\usepackage{wrapfig}
\usepackage{slashbox}
\usepackage{psfrag}
\usepackage{slashed} %Um Feynman-Slashes zu machen
\usepackage{tabls} 
\usepackage{caption}
\makeatletter
\def\cleardoublepage{\clearpage\if@twoside \ifodd\c@page\else
		\hbox{}
		\thispagestyle{empty}
		\newpage%
		\if@twocolumn\hbox{}\newpage\fi\fi\fi}
\makeatother

\begin{document}
\thispagestyle{empty}
\begin{titlepage}
\phantom{\Huge{ t}}
   \begin{center}
       \Huge{\textbf{Pion Configurations in the Presence of Baryons}} \\[5ex]
	 \Large \textbf{Diplomarbeit} \\[5ex]
       \large der Philosophisch-naturwissenschaftlichen Fakult\"at  \\[1ex]
       der Universit\"at Bern \\[3ex]
       vorgelegt von \\[5ex]
       \Large \textbf{Daniel Arnold} \\[5ex]
       \large 2008 \\[25ex]

     	Leiter der Arbeit: \\[2ex]
       \textbf{ Prof. Gilberto Colangelo}\\[1ex]
       Institut f\"ur theoretische Physik, Universit\"at Bern 
   \end{center}
\end{titlepage}

\newpage\thispagestyle{empty}~
\newpage
\thispagestyle{empty}

\begin{abstract}
In the framework of chiral perturbation theory we investigate whether, in the presence of nucleons, a many-particle system may lower its energy density by changing its mesonic vacuum from a homogeneous to a spiral configuration. Numerical values for the critical particle number densities for the spiral configuration to occur are given. They lie in the region of nuclear matter density where the applicability of chiral perturbation theory should be checked carefully.
\end{abstract}

\newpage\thispagestyle{empty}~
\newpage
\thispagestyle{empty}

\tableofcontents
\newpage
\chapter*{Introduction}
\addcontentsline{toc}{chapter}{Introduction}
Quantum chromodynamics, the field theory of the strong interaction, is an asymptotically free theory, i.e. the coupling constant decreases as the energy increases. On the other hand, at low energies the coupling constant becomes too large to make an expansion in it; QCD becomes non-perturbative. Besides lattice calculations, the effective field theory is the only systematic approach to deal with the low-energy dynamics of QCD. The symmetry of the QCD Lagrangian in the chiral limit (i.e. at vanishing quark masses) is supposed to be broken spontaneously leading to massless Goldstone bosons by virtue of the Goldstone theorem. Depending on whether the s-quark is considered as light or not, these Goldstone bosons may be identified either as the pseudoscalar octet or as the three pions. Since chiral symmetry is explicitly broken by non-zero quark masses the Goldstone bosons are not exactly massless but anyhow relatively light compared to $1\,\text{GeV}$, the typical scale of QCD. At low energies they are the fundamental degrees of freedom, whose dynamics is dictated by symmetry constraints. This leads to the effective field theory of QCD, chiral perturbation theory, formulated in its modern form by Weinberg, Gasser and Leutwyler.

Here, we consider the s-quark as heavy and deal with pions only. In the pure mesonic sector the ground state is a constant pionic field configuration. As we will discuss, the pions may be represented by a four-dimensional unit vector, such that the ground state just stands for a constant vector field. However, chiral perturbation theory may be extended to include also baryons (nucleons in the case where the s-quark is considered as heavy). This makes it possible to effectively describe the interactions between one nucleon and pions. It turns out that the fermion may lower its energy if the vector field representing the pions is not constant but varies throughout space. This behavior increases the pionic energy and for low nucleon densities this rise of the pionic energy outweighs the decrease of the fermionic energy. However, if we do not consider one nucleon only but a many-particle system (neglecting interactions between the fermions) and if the particle number density of nucleons exceeds some critical value the dropping of the fermionic energy becomes so large that the many-particle system may lower its total energy density by changing the pionic configuration from the homogeneous to an inhomogeneous phase.

In this thesis we investigate whether it is possible that the vector field representing the pions could show a spiral phase in the ground state if we take nucleons into account. The physical parameter we can control is the density of nucleons. In the framework of chiral perturbation theory we calculate the energy of a spiral configuration (to be introduced in section \ref{sec:aspiralconfiguration}) and try to find out what critical value the fermionic particle number density must exceed for the spiral phase instead of the homogeneous phase to be favored. The spiral configuration we consider is such that the nucleons see a constant background\footnote{Technically speaking they couple to a constant connection and vielbein, see \ref{sec:lowestordereffectivebaryoniclagrangian}.}, which simplifies the solution of the Dirac equation.

The idea of this thesis actually originates from the theoretical description of quantum antiferromagnets. Antiferromagnets doped with either electrons or holes are nowadays the only known approach to high temperature superconductors and have therefore been studied also from the theoretical point of view intensely. There are different models for the description of quantum antiferromagnets on the microscopic level. The most important are the so called Hubbard and $t-J$ model. However, currently these models can be simulated numerically only if they contain at most one single hole or electron. Additionally, due to severe fermion sign problems and other complications it is not possible to find the exact ground states of the two doped models, neither analytically nor numerically.

For these reasons one constructs a low energy (low temperature) effective theory for antiferromagnets, i.e. as in the case of QCD one considers only the low energy regime and the degrees of freedom appearing there. At low energies, the spin symmetry of the microscopic model is spontaneously broken giving rise to two massless Goldstone bosons (magnons) that can be represented by a three dimensional unit vector $\vec{e}(x)$, the so called local staggered magnetization. An effective theory in terms of $\vec{e}(x)$ for the pure antiferromagnet, i.e. for the magnon sector without holes or electrons, was developed by Hasenfratz and Niedermayer in \cite{Hasenfratz:1990jw}. Considering the effective action therein it becomes clear that the ground state of the antiferromagnet at low energies corresponds to a constant staggered magnetization.

Like chiral perturbation theory may be extended to include baryons, doped holes (\cite{Brugger:2006dz}) and doped electrons (\cite{Brugger:2006ej}) in the antiferromagnet may also be described effectively. In \cite{Brugger:2006fi} it was shown that through an inhomogeneous staggered magnetization, namely a spiral configuration, (which costs energy in the magnon sector) doped holes can lower their energy in such a way that the total energy of the system may get lower than in the case of a constant staggered magnetization. Depending on the values of the low energy constants and the density of holes, spiral configurations are hence a certain category of configurations that may then describe the ground state.

Due to the similarities of the two effective field theories, one for QCD and one for antiferromagnets, it was natural to ask whether the corresponding phenomenon to spiral phases observed in antiferromagnets could occur also in QCD.

The answer to this question is positive. However, as we will show, the critical particle number density for the spiral phase to occur will turn out to lie in the region of nuclear matter density. It is anything but obvious whether chiral perturbation theory remains valid at such high densities. In order to check the reliability of our results we investigate higher orders (in different ways) and calculate the corrections coming from these. If they are not too large we might get a reason to trust the results and to believe that the qualitative picture may be correct. Due to these uncertainties we will not calculate errors for the critical number densities originating from errors of the low energy constants. The calculations are performed using the low energy constants obtained in the chiral limit, but in order to analyze the sensitivity of the results with respect to higher orders we also do the calculations with the physical low energy constants.

The presence of a spiral configuration could be viewed as the phenomenon of pion condensation. This is supposed to occur in superdense matter like e.g. in nuclear matter or inside a neutron star. Spiral configurations in the mesonic vacuum could therefore be an alternative approach towards such phenomena.

\chapter{Some Remarks on Chiral Perturbation Theory}\label{ch:chpt}
This chapter is intended to give a brief summary of chiral perturbation theory as the effective theory of QCD. More detailed introductions can be found in \cite{ecker,scherer,leutwyler,Colangelo:2000zw,kubis,borasoy}.
\section{The Need for an Effective Field Theory}
The Lagrangian of quantum chromodynamics (QCD) reads
\begin{equation}
	\mathcal{L}_\text{QCD}=\sum_{f=1}^6\bar{q}_f(i\slashed{D}-m_f)q_f-\frac{1}{4}\mathcal{G}_{\mu\nu,a}\mathcal{G}^{\mu\nu}_a\label{eq:qcdlagr}.
\end{equation}
Here $f$ is the flavor index, $q_f$ a triplet of spinors (red, green and blue quarks),
\begin{equation}
	D_\mu=\partial_\mu-ig\sum_{a=1}^8\frac{\lambda_a}{2}\mathcal{A}_{\mu,a}
\end{equation}
is the covariant derivative with the eight gluon fields $\mathcal{A}_{\mu,a}$, the strong coupling constant $g$ and the Gell-Mann matrices $\lambda_a$, and
\begin{equation}
	\mathcal{G}_{\mu\nu,a}=\partial_\mu\mathcal{A}_{\nu,a}-\partial_\nu\mathcal{A}_{\mu,a}+gf_{abc}\mathcal{A}_{\mu,b}\mathcal{A}_{\nu,c}
\end{equation}
is the field tensor for the vector potentials, i.e. the kinetic term for the gluons with the $SU(3)$ structure constants $f_{abc}$. We have adopted Feynman's notation $\slashed{D}\equiv\gamma^\mu D_\mu$. %For a derivation of this Lagrangian from gauge principles see appendix \ref{app:gauge}.

Since QCD is a non-Abelian theory ($f_{abc}\neq 0$) the gluons carry the same color charge as the quarks and thus may interact with each other. The structure of QCD is therefore much more complicated than that of QED and, what is more important, due to the non-Abelian nature QCD is an asymptotically free theory, i.e. the $\beta$ function $\beta=\mu\partial g/\partial\mu$ (where $\mu$ is the renormalization scale, i.e. the typical energy) is negative.\footnote{One can show (see e.g. \cite{peskin} or \cite{donoghue}) that the $\beta$ function of QCD would be positive only if the number of quark flavors exceeded sixteen.} This means that, in contrast to QED, the coupling constant decreases with increasing energy. Since at low energies the strong coupling constant becomes quite large, perturbation theory, i.e. an expansion in the coupling constant, is possible only for high energies but will fail in the low energy regime. Only two rigorous approaches are known to treat the strong interaction at low energies: Lattice QCD and the effective field theory called chiral perturbation theory (ChPT).
\subsection{Effective Field Theories}\label{sec:eft}
Effective field theories (EFTs) are a very important tool in many areas of physics. The motivation for an EFT is that one does not have to know a theory at all energy scales in order to precisely describe a physical system. In general, EFTs are low energy approximations of a fundamental theory and allow an expansion of physical quantities in terms of $p/\Lambda$ where $p$ denotes momenta or masses being smaller than some scale $\Lambda$ (therefore an expansion is possible). The basis of EFTs is a statement given by Weinberg and proved by Leutwyler (\cite{leutwylerannals}) as well as d'Hoker and Weinberg (\cite{D'Hoker:1994ti}), which says that a quantum field theory has no content beyond unitarity, analyticity, cluster decomposition and symmetries.

Unitarity assures that $\sum_f|\langle f|S|i\rangle|^2=1$ ($f$: final state, $i$: initial state) and analyticity guarantees causality. The cluster decomposition theorem guarantees locality and states, in few words, that two experiments done in sufficiently separated regions of space-time do not correlate. Finally, by symmetry we mean Poincar\'e invariance, $C$, $P$, $T$ symmetries and internal symmetries like e.g. isospin symmetry.

Weinberg's statement means that a perturbative description in terms of the most general effective Lagrangian containing all possible terms compatible with assumed symmetry principles yields the most general $S$ matrix consistent with the fundamental principles mentioned above. Hence, an effective Lagrangian is constructed by writing down all possible terms allowed by the symmetry properties of the underlying theory. Of course, this will be an infinite number of terms and we need some method of assessing the importance of diagrams generated by the interaction terms of this Lagrangian. In ChPT this will be Weinberg's power counting scheme (see section \ref{sec:weinbergpowercounting}).

There are different types of EFTs (see e.g. \cite{ecker}) depending on the structure of the transition from the fundamental (energies $>\Lambda$) to the effective (energies $<\Lambda$) level:
\begin{itemize}
\item Decoupling: If we have a theory in which there are light and heavy degrees of freedom which are well separated by a scale $\Lambda$ and if we only consider energies well below $\Lambda$, then the heavy degrees of freedom will not matter. We can integrate them out of the generating functional and get a Lagrangian with the light degrees of freedom only (Appelquist-Carazzone theorem, see \cite{Appelquist:1974tg}). This effective Lagrangian will contain a renormalizable part and non-renormalizable couplings which are suppressed by inverse powers of $\Lambda$. An example for such an EFT might be the Standard Model (as an EFT of a yet unknown fundamental theory).
\item Non-decoupling: Due to some phase transition the degrees of freedom at low energy differ from the ones of the underlying theory. This is the case in QCD where the phase transition is spontaneous symmetry breaking of chiral symmetry. In the low energy regime we only deal with color-neutral hadronic states which will be the Goldstone bosons in the mesonic sector (originating from spontaneous symmetry breaking) and nucleons in the baryonic sector. An effective Lagrangian of that sort will not be renormalizable in the classical sense. The reason is that it contains infinitely many terms and due to the nonlinear transformation properties of the Goldstone bosons even the lowest order effective Lagrangian contains infinitely many fields. To get a dimensionless action  we thus need coupling constants of negative mass dimensionality. But such interactions are not renormalizable. However, this is not regarded as a serious problem since the infinities arising from loops in a given order of momentum expansion are absorbed in a renormalization of (a finite number of) coefficients in the Lagrangian. So the effective Lagrangian is renormalizable order by order.
\end{itemize}
Going from the fundamental to an effective level we lose track of the correct high energy behavior of the theory. An EFT will yield wrong results if the energy is too high and e.g. heavier degrees of freedom are produced. Thus the domain of utility of an EFT is necessarily bounded from above in the energy scale.

The effective coupling constants, i.e. the coefficients appearing in the effective Lagrangian, cannot be determined by symmetry considerations alone but should in principle be calculated from the fundamental theory. In the case of ChPT we cannot (yet) solve the fundamental theory (QCD) and the effective couplings stay as free parameters which have to be determined by some other means, e.g. matched by experiments.

\section{Symmetries of the QCD Lagrangian}
Since symmetry and symmetry breaking are the basic concepts in the construction of an effective Lagrangian we shall study the symmetries of the QCD Lagrangian in some detail.
\subsection{The Chiral Limit}
Due to the confinement of quarks into hadronic states the notion of quark masses is rather complicated and needs some extended discussion. However, without going into details, e.g. the Particle Physics Booklet \cite{pdg} tells us that there are three light (u,d and s) and three heavy (c,b and t) quarks. The masses of the light quarks are quite below $1\,\text{GeV}$ and the heavy quarks exceed this scale. $1\,\text{GeV}$ is the typical scale of QCD and is associated with the masses of the lightest hadrons containing light quarks that are not Goldstone bosons (e.g. $m_\rho=770\,$MeV). For spontaneous symmetry breaking the typical scale is $4\pi F_\pi\approx 1.2\,$GeV (see e.g. \cite{georgi}), which is of the same order. If we constrain ourselves to energies below $1\,\text{GeV}$ there will never appear any states containing heavy quarks and we can ignore them in the Lagrangian, i.e. we set their masses to infinity.

If we compare e.g. the mass of a proton ($m_p=938\,\text{MeV}$) with the sum of two up and one down quark we see that the quark masses contribute almost nothing and that the proton gets its mass through another complicated mechanism. This is the motivation to send the light quark masses to zero as a first approximation. This is the so-called chiral limit and the Lagrangian (\ref{eq:qcdlagr}) then reads
\begin{equation}
	\mathcal{L}_\text{QCD}^0=\sum_{f=u,d,s}\bar{q}_fi\slashed{D}q_f-\frac{1}{4}\mathcal{G}_{\mu\nu,a}\mathcal{G}^{\mu\nu}_a\label{eq:qcdlagr0}.
\end{equation}
Spinors of massless or ultrarelativistic fermions are eigenstates of the chirality operator $\gamma_5=i\gamma^0\gamma^1\gamma^2\gamma^3$, where particles with eigenvalue $+1$ are called right-handed and those with eigenvalue $-1$ are called left-handed. In this extreme relativistic case chirality equals to helicity. Defining the projection operators $P_R=\frac{1}{2}(1+\gamma_5)$ and $P_L=\frac{1}{2}(1-\gamma_5)$ we can decompose an arbitrary spinor into its right- and left-handed parts which then are eigenstates of $\gamma_5$. Thus, with $q_{R,L}=P_{R,L}q$ and $\bar{q}_{R,L}=\bar{q}P_{L,R}$ we can then write (\ref{eq:qcdlagr0}) as
\begin{equation}	\mathcal{L}_\text{QCD}^0=\sum_{f=u,d,s}(\bar{q}_{R,f}i\slashed{D}q_{R,f}+\bar{q}_{L,f}i\slashed{D}q_{L,f})-\frac{1}{4}\mathcal{G}_{\mu\nu,a}\mathcal{G}^{\mu\nu}_a\label{eq:qcdlagr0rl}.
\end{equation}
We see that the right- and left-handed quark fields have completely decoupled. This is not the case if we have a mass term which mixes right- and left-handed fields. Besides Lorentz invariance, $SU(3)_c$ gauge invariance and $P$, $T$ and $C$ symmetries, the Lagrangian (\ref{eq:qcdlagr0rl}) shows also a chiral symmetry: It is invariant under global $U(3)_L\times U(3)_R$ flavor transformations $(q_L,q_R)\rightarrow (Lq_L,Rq_R)$ and $(\bar{q}_L,\bar{q}_R)\rightarrow (q_LL^\dagger,q_RR^\dagger)$ where $L,R\in U(3)_{L,R}$ and $q_{L,R}$ are flavor triplets. Since we can make the decomposition $U(N)=SU(N)\times U(1)$, we see that $\mathcal{L}_\text{QCD}^0$ shows a global $SU(3)_L\times SU(3)_R\times U(1)_L\times U(1)_R$ symmetry.
\subsection{Symmetry Currents}\label{sec:symmetrycurrents}
From these symmetries we get several conserved currents due to Noether's theorem (see e.g. \cite{scherer}). The $U(1)_{R,L}$ symmetries will lead to the currents $\bar{q}_{R,L}\gamma^\mu q_{R,L}$. By adding and subtracting respectively these currents we get a singlet vector and a singlet axialvector current (according to to their behavior under parity transformation):
\begin{align}
V^\mu & = \bar{q}_{R}\gamma^\mu q_{R}+\bar{q}_{L}\gamma^\mu q_{L}=\bar{q}\gamma^\mu q\\
A^\mu & = \bar{q}_{R}\gamma^\mu q_{R}-\bar{q}_{L}\gamma^\mu q_{L}=\bar{q}\gamma^\mu\gamma_5 q
\end{align}
The current $V^\mu$ can be interpreted as total current when doing the $U(1)_L$ and $U(1)_R$ transformation with the same phase, while $A^\mu$ comes from transformations with opposite phases (if a $U(1)$ transformation with phase $\theta$ yields the current $J^\mu$, then a transformation with phase $k\theta$ ($k\in\mathbb{R}$) leads to $kJ^\mu$).

From $V^\mu$ the conserved charge is given by $Q_V=\int d^3xV^0=\int d^3x q^\dagger q$ which counts the number of quarks minus antiquarks (see e.g. \cite{scherereft}). Thus $Q_V/3$ is the number $B$ of baryons and the $U(1)_V$ symmetry simply expresses baryon number conservation and leads to a classification of hadrons into mesons ($B=0$) and baryons ($B=1$).

The current $A^\mu$ is conserved only at the classical level. This symmetry is broken by quantization due to anomalies (see e.g. \cite{georgi}).

The interesting symmetry is $SU(3)_L\times SU(3)_R$ which leads to eight left-handed and eight right-handed conserved currents:
\begin{align}
L^{\mu,a} & = \bar{q}_L\gamma^\mu\frac{\lambda^a}{2}q_L\\
R^{\mu,a} & = \bar{q}_R\gamma^\mu\frac{\lambda^a}{2}q_R.
\end{align}
The charge operators read
\begin{align}
Q_L^a &= \int d^3x L^{0,a}=\int d^3x q_L^\dagger\frac{\lambda^a}{2}q_L\\
Q_R^a &= \int d^3x R^{0,a}=\int d^3x q_R^\dagger\frac{\lambda^a}{2}q_R
\end{align}
and are the generators of $SU(3)_L\times SU(3)_R$ since they satisfy the commutation relations corresponding to the Lie algebra of this group.
Again one considers the vector and axialvector linear combinations
\begin{align}
V^{\mu,a} & = R^{\mu,a}+L^{\mu,a}=\bar{q}\gamma^\mu\frac{\lambda^a}{2}q\\
A^{\mu,a} & = R^{\mu,a}-L^{\mu,a}=\bar{q}\gamma^\mu\gamma_5\frac{\lambda^a}{2}q,
\end{align}
as well as
\begin{align}
Q_V^a&=Q_R^a+Q_L^a\\
Q_A^a&=Q_R^a-Q_L^a.
\end{align}
Note that the charge operators $Q_V^a$ satisfy the commutation relations for $SU(3)_V$, whereas $Q_A^a$ do not form a closed algebra, i.e. the commutator of two axial charge operators is not again an axial charge operator and there is nothing like $SU(3)_A$.

\section{Spontaneous Symmetry Breaking}
Spontaneous symmetry breaking takes place if the Lagrangian of a system has a given symmetry but the ground state is not invariant under that symmetry; rather the system has chosen one of (possibly many) ground states, all related by symmetry transformations. A classical example is a ferromagnet described by spin-spin interaction, whose Hamiltonian is invariant under rotations. However, below the critical temperature the ground state is one in which all the spins point in the same direction, which is surely not rotationally invariant. The direction of the spins is random (in the absence of an external magnetic field), hence the system has infinitely many degenerated ground states.

As we saw that $\mathcal{L}_\text{QCD}^0$ is invariant under $SU(3)_L\times SU(3)_R\times U(1)_V$, it is interesting to ask whether the ground state of QCD is invariant under the same symmetry group. To study this, let $|\psi\rangle$ be an eigenstate of $H_\text{QCD}^0$ with some energy $E$ and with positive parity, i.e. $H_\text{QCD}^0|\psi\rangle=E|\psi\rangle$ and $P|\psi\rangle=+|\psi\rangle$. Then the state $Q_A^a|\psi\rangle$ is also an eigenstate of $H_\text{QCD}^0$ with the same eigenvalue $E$ but with opposite parity:
\begin{align*}
H_\text{QCD}^0Q_A^a|\psi\rangle & = Q_A^aH_\text{QCD}^0|\psi\rangle=EQ_A^a|\psi\rangle\\
PQ_A^a|\psi\rangle & = PQ_A^aP^{-1}P|\psi\rangle=-Q_A^a|\psi\rangle,
\end{align*}
where we have used $\left[H_\text{QCD}^0,Q_A^a\right]=0$ since $Q_A^a$ is time independent and $PQ_A^aP^{-1}=-Q_A^a$ since $Q_A^a$ is an axialvector.

Consequently, for any state of positive parity one would expect the existence of a degenerate state of negative parity. However, this is not observed in the low-energy spectrum of hadrons. Though, for these arguments we (not obviously) assumed that the axial charges $Q_A^a$ annihilate the ground state, i.e. $Q_A^a|0\rangle=0$ (see \cite{scherer}). But since there is no experimental evidence for parity doubling one concludes that the $Q_A^a$ do not annihilate the vacuum, i.e. the ground state of QCD is not invariant under $SU(3)_L\times SU(3)_R\times U(1)_V$. In \cite{Vafa:1983tf} it was shown that in the chiral limit the ground state must be invariant under $SU(3)_V\times U(1)_V$, i.e. the vector charges $Q_V^a$ and $Q_V$ annihilate the ground state; $Q_V^a|0\rangle=Q_V|0\rangle=0$. Thus, while $\mathcal{L}_\text{QCD}^0$ is invariant under $SU(3)_L\times SU(3)_R\times U(1)_V$, the ground state is invariant under $SU(3)_V\times U(1)_V$ only, which is spontaneous symmetry breaking.

\subsection{Goldstone's Theorem}\label{sec:goldstone}
Goldstone theorem states that for every spontaneously broken continuous symmetry, i.e. for every generator that does not annihilate the ground state, there is one particle of zero mass and spin in the spectrum. These are called Goldstone bosons.

Let us consider a theory involving several fields $\phi_i(x)$, described by a Lagrangian of the form $\mathcal{L}=(\text{derivatives})-V(\vec{\phi})$. Suppose that the potential $V(\vec{\phi})$ has a set of degenerate ground states and that the system has chosen spontaneously one of them, let us call it $\vec{\phi_0}$. Expanding $V$ about this minimum yields
\begin{equation}
	V(\vec{\phi})=V(\vec{\phi_0})+\frac{1}{2}(\vec{\phi}-\vec{\phi_0})_i(\vec{\phi}-\vec{\phi_0})_j\frac{\partial^2V}{\partial\phi_i\partial\phi_j}\Bigr|_{\vec{\phi}=\vec{\phi_0}}+...,
\end{equation}
where the linear terms disappear since $\vec{\phi_0}$ is a minimum. The coefficients
\begin{equation}
	\frac{\partial^2V}{\partial\phi_i\partial\phi_j}\Bigr|_{\vec{\phi}=\vec{\phi_0}}=:m^2_{ij}
\end{equation}
define a symmetric matrix whose eigenvalues give the squared masses of the particle excitations above the ground state (vacuum) $\vec{\phi_0}$. This can be seen by performing an orthogonal transformation of the fields such that the mass matrix gets diagonal. Since $m^2_{ij}$ is a positive semidefinite matrix (as $\vec{\phi_0}$ is a minimum) these eigenvalues cannot be negative. We show now, that for every continuous symmetry of $\mathcal{L}$ which is not a symmetry of $\vec{\phi_0}$ the matrix $m^2_{ij}$ yields a zero eigenvalue and we therefore have a massless particle. Suppose that $\mathcal{L}$ is invariant under a symmetry group $G$ of order $n_G$ (thus $G$ having $n_G$ generators) while $\vec{\phi_0}$ is invariant under a subgroup $H\subset G$ of order $n_H$ only. Under $G$ an infinitesimal symmetry transformation of the fields takes the form
\begin{equation}
	\phi_i\rightarrow \phi_i'=\phi_i+i\epsilon_aT^a_{ij}\phi_j,
\end{equation}
where $T^a$ are the representation matrices of the generators of $G$. Since the potential $V(\vec{\phi})$ is invariant under $G$, we have
	\[V(\vec{\phi})=V(\vec{\phi}')=V(\vec{\phi}+i\epsilon_aT^a\vec{\phi})=V(\phi_1+i\epsilon_aT^a_{1j}\phi_j,\phi_2+i\epsilon_aT^a_{2j}\phi_j,...)=V(\vec{\phi})+\frac{\partial V}{\partial\phi_i}i\epsilon_aT^a_{ij}\phi_j
\]
which yields
\begin{equation}
	\frac{\partial V}{\partial\phi_i}i\epsilon_aT^a_{ij}\phi_j=0.
\end{equation}
Differentiating with respect to $\phi_k$ and evaluating the resulting expression at $\vec{\phi}=\vec{\phi_0}$ we get
\begin{equation}
	\underbrace{\frac{\partial^2V}{\partial\phi_i\partial\phi_k}\Bigr|_{\vec{\phi}=\vec{\phi_0}}}_{m^2_{ki}}i\epsilon_aT^a_{ij}\phi_{0,j}+\underbrace{\frac{\partial V}{\partial\phi_i}\Bigr|_{\vec{\phi}=\vec{\phi_0}}}_{0}i\epsilon_aT^a_{ik}=0\quad\Rightarrow\quad m^2_{ki}T^a_{ij}\phi_{0,j}=0,
\end{equation}
since the equation holds for every $\epsilon_a$. This can be written more compactly as $m^2_{ki}(T^a\vec{\phi_0})_i=0$. This equation is trivially fulfilled if $a\in\{1,2,...,n_H\}$, i.e. if $T^a$ is a generator of the subgroup $H$, because $\vec{\phi_0}$ is invariant under $H$ and hence $T^a\vec{\phi_0}=0$. But for $a\in\{n_H+1,...,n_G\}$, i.e. for the generators belonging to $G$ but not to $H$, we have $T^a\vec{\phi_0}\neq 0$, since $\vec{\phi_0}$ is not invariant under the full group $G$. Then the above equation tells us that $T^a\vec{\phi_0}$ is an eigenvector of $m^2_{ki}$ with eigenvalue zero. Thus the matrix has $n_G-n_H$ zero eigenvalues and therefore there are $n_G-n_H$ massless particles (Goldstone bosons) in the spectrum. These Goldstone bosons carry the same quantum numbers as the generators that do not annihilate the vacuum. Furthermore, it follows (see \cite{scherer}) that the matrix element of the symmetry currents $J_\mu^a(x)$ that lead to the generators of $G$ not annihilating the ground state between the vacuum and the massless one-particle states $|\pi^a\rangle$ is non-zero,
\begin{equation}
	\langle 0|J_\mu^a(0)|\pi^a\rangle\neq 0.
\end{equation}
Using this, it can be shown that, in the chiral limit, the Goldstone bosons do not interact with each other at zero momentum (see. e.g. \cite{borasoy} or \cite{leutwyler}). This feature is essential for the consistency of ChPT; otherwise the power counting of ChPT (see \ref{sec:weinbergpowercounting}) would break down.

\subsection{The Scalar Quark Condensate}
Remembering the example of the ferromagnet below the critical temperature, we see that, as soon as the symmetry of the system has broken down spontaneously, the magnetization $\langle \vec{M}\rangle$ becomes non-zero. $\langle \vec{M}\rangle$ can therefore be identified as an order parameter of the system. In a similar manner we can detect a spontaneous breakdown of chiral symmetry by investigating the scalar quark condensate $\langle \bar{q}q\rangle=\langle 0|\bar{q}q|0\rangle$ which is the order parameter of QCD (see e.g. \cite{leutwyler}). A non-zero value of this order parameter in the chiral limit is a sufficient (but not necessary) condition for spontaneous symmetry breaking in QCD (see \cite{scherer} or \cite{gasser}). If the axial charges $Q^a_A$ do not annihilate the vacuum it is consistent to assume that $\langle\bar{q}q\rangle\neq 0$ (see e.g. \cite{gasser}).

As we have 8 generators $Q^a_A$ which do not annihilate the ground state, Goldstone's theorem now allows us to conclude that there must be 8 massless pseudoscalar particles $|\pi^a\rangle$ whose coupling to the currents $A_\mu^a$ is non-vanishing:
\begin{equation}
	\langle 0|A_\mu^a(x)|\pi^b(p)\rangle = ip_\mu F_0\delta^{ab}e^{-ipx}\label{eq:matrixelement}.
\end{equation}
This form of the matrix element comes from symmetry considerations: Due to translational invariance we have $\langle 0|A_\mu^a(x)|\pi^b\rangle=e^{-ipx}\langle 0|A_\mu^a(0)|\pi^b\rangle$ and the only quantity in question bearing a Lorenz index is the momentum $p_\mu$ of the pseudoscalar state. The constant of proportionality $F_0$ is called the pion-decay constant in the chiral limit. It measures the strength with which the Goldstone boson $|\pi^b\rangle$ decays via the axial vector current into the hadronic vacuum.

Acting in (\ref{eq:matrixelement}) on both sides with $\partial^\mu$ we conclude that, in the chiral limit (where $\partial^\mu A_\mu^a=0$), the Goldstone bosons are indeed massless:
\begin{equation}
	0=\langle 0|\partial^\mu A_\mu^a(x)|\pi^b(p)\rangle=p^2F_0\delta^{ab}e^{-ipx}\quad\Rightarrow\quad p^2=m^2=0.
\end{equation}
\subsection{Goldstone Bosons of QCD}\label{sec:goldstoneqcd}
Since in the chiral limit of QCD we suppose a spontaneous symmetry breakdown from $SU(3)_L\times SU(3)_R\times U(1)_V$ to $SU(3)_V\times U(1)_V$ and since $SU(3)$ has eight generators, we expect 8 massless Goldstone bosons. Their symmetry properties are tightly connected to the generators which are responsible for them, i.e. to those which do not annihilate the vacuum. In QCD this are, as seen above, the axial generators $Q_A^a$. Thus the Goldstone bosons are, as already mentioned, expected to transform with a negative sign under parity and hence to be pseudoscalars. Indeed, the spectrum of QCD shows eight candidates for the Goldstone bosons, namely the pseudoscalar octet $(\pi,K,\eta)=(\pi^0,\pi^{\pm},K^0,\bar{K}^0,K^{\pm},\eta)$. These particles are of course not massless but this is interpreted as a consequence of the explicit breaking of chiral symmetry due to finite u-, d- and s-quark masses in the Lagrangian (\ref{eq:qcdlagr}). E.g. in \cite{scherer} the resulting masses of the Goldstone bosons are calculated (the results are quoted in section \ref{sec:effectivelagrangian}). Nonetheless, the Goldstone bosons of QCD are much lighter than all other hadrons. This statement becomes even more accurate if we restrict ourselves only to u- and d-quarks and neglect the s-quark (which is considerably heavier than the u- and d-quarks) in the Lagrangian, i.e. we only consider energies where states including s-quarks never appear. Then $\mathcal{L}_\text{QCD}^0$ shows an $SU(2)_L\times SU(2)_R\times U(1)_V$ symmetry which breaks down to $SU(2)_V\times U(1)_V$, resulting in three Goldstone bosons, as $SU(2)$ has three generators. These are then the pions which, since the u- and d-quarks are indeed very light and thus the chiral limit is a better approximation than in the case of $SU(3)$, show strikingly small masses (about $139\,$MeV for $\pi^{\pm}$ and about $135\,$MeV for $\pi^0$).
\section{Construction of the Effective Lagrangian}
Since in QCD at low energies the Goldstone bosons are the only degrees of freedom, this are the objects to be contained in an effective Lagrangian.
\subsection{Representation of the Goldstone Bosons}\label{sec:representationofgoldstonebosons}
Let us denote the symmetry group of the Lagrangian by $G$ and the smaller symmetry group of the ground state by $H$. Due to spontaneous symmetry breaking we have $n=n_G-n_H$ Goldstone bosons $\phi_i$ which we collect in the vector $\vec{\Phi}=(\phi_1,...,\phi_n)$. The set of all these vectors is a vector space, let us call it $M$. We can define the action of the symmetry group $G$ on $M$ by a mapping $\vec{f}:G\times M\rightarrow M$ with
\begin{equation}
	\vec{\Phi}\rightarrow \vec{\Phi}'=\vec{f}(g,\vec{\Phi}),\quad g\in G.
\end{equation}
This defines an operation of $G$ on $M$ if the mapping $\vec{f}$ satisfies
\begin{align}
\vec{f}(e,\vec{\Phi})&=\vec{\Phi}\quad\forall\vec{\Phi}\\
\vec{f}(g_1,\vec{f}(g_2,\vec{\Phi}))&=\vec{f}(g_1g_2,\vec{\Phi})\quad\forall g_1,g_2\in G,\forall \Phi,
\end{align}
where $e$ is the identity of $G$. Since the subgroup $H$ leaves the ground state $\vec{\Phi}=0$ invariant we have $\vec{f}(h,0)=0$ for all $h\in H$ and hence
\begin{equation}
	\vec{f}(gh,0)=\vec{f}(g,\vec{f}(h,0))=\vec{f}(g,0)\quad\forall g\in G,h\in H.
\end{equation}
This shows that for all elements of a given coset $gH=\{gh|h\in H\}$ of $g$, $\vec{f}$ maps the origin onto the same vector in $M$. Furthermore, this mapping is injective; one simply shows that for two elements $g$ and $g'$ of $G$ with $g'\notin gH$ it follows $\vec{f}(g,0)\neq\vec{f}(g',0)$ (see e.g. \cite{scherer}). So the mapping is bijective on the image of $\vec{f}(g,0)$ (as it is automatically surjective on the image) and can there be inverted. Hence, to each $\vec{\Phi}$ corresponds a coset $\tilde{g}H$ with appropriate $\tilde{g}$ such that $\vec{\Phi}=\vec{f}(\tilde{g}h,0)$. We have therefore found an isomorphic mapping between the set $\{gH|g\in G\}$ of all cosets and the Goldstone boson fields. The set of all cosets is called quotient $G/H$ and therefore the Goldstone boson fields are said to live on the coset space $G/H$.

The transformation behavior of a $\vec{\Phi}=\vec{f}(\tilde{g}h,0)$ under a group element $g\in G$ is easily found by applying $\vec{f}(g,\cdot)$:
\begin{equation}
	\vec{\Phi}'=\vec{f}(g,\vec{\Phi})=\vec{f}(g,\vec{f}(\tilde{g}h,0))=\vec{f}(g\tilde{g}h,0)
\end{equation}
The coset corresponding to $\vec{\Phi}'$ is thus obtained by multiplication of the coset corresponding to $\vec{\Phi}$ with $g$.

Now, in the case of QCD we have $G=SU(N)\times SU(N)=\{(L,R)|L,R\in SU(N)\}$ and $H=\{(V,V)|V\in SU(N)\}$ which is isomorphic to $SU(N)_V$, with $N=3$ or $N=2$ depending on whether we include the s-quark or not. The coset of $\tilde{g}=(\tilde{L},\tilde{R})\in G$ is $\tilde{g}H=\{(\tilde{L}V,\tilde{R}V)|V\in SU(N)\}$ and since we have
	\[(\tilde{L}V,\tilde{R}V)=(\tilde{L}V,\tilde{R}\tilde{L}^\dagger\tilde{L}V)=(1,\tilde{R}\tilde{L}^\dagger)\underbrace{(\tilde{L}V,\tilde{L}V)}_{\in H},
\]
it can be written as $\tilde{g}H=(1,\tilde{R}\tilde{L}^\dagger)H$. Hence the coset and therefore the Goldstone bosons may be uniquely characterized through the $SU(N)$ matrix $U=\tilde{R}\tilde{L}^\dagger$. Its transformation behavior under an element $g=(L,R)\in G$ is, as seen above, obtained by multiplication of the coset with $g$:
	\[g\underbrace{\tilde{g}H}_{\widehat{=}\vec{\Phi}}=(L,R\tilde{R}\tilde{L}^\dagger)H=(1,R\tilde{R}\tilde{L}^\dagger L^\dagger)\underbrace{(L,L)}_{\in H}H=(1,R\underbrace{(\tilde{R}\tilde{L}^\dagger)}_{U}L^\dagger)H,
\]
which yields
\begin{equation}
	U\stackrel{G}{\longrightarrow} RUL^\dagger\label{eq:transform}.
\end{equation}
This mapping defines a nonlinear realization of $SU(N)_L\times SU(N)_R$.
One possible and convenient representation of the matrix $U$ is the exponential representation:
\begin{equation}
	U(x)=\exp\left(i\frac{\phi(x)}{F_0}\right)=\exp\left(i\frac{t_a\phi_a}{F_0}\right)\label{eq:parametrisierung},
\end{equation}
where $t_a$ are the generators of $SU(N)$ and $F_0$ will turn out to be the pion decay constant in the chiral limit (see section \ref{sec:effectivelagrangian}). In terms of the physical fields we have for $SU(3)$
\begin{equation}
	\phi(x)=\phi_a\lambda_a=\begin{pmatrix} \pi^0+\frac{1}{\sqrt{3}}\eta & \sqrt{2}\pi^+ & \sqrt{2}K^+ \\ \sqrt{2}\pi^- & -\pi^0+\frac{1}{\sqrt{3}}\eta & \sqrt{2}K^0 \\ \sqrt{2}K^- & \sqrt{2}\bar{K}^0 & -\frac{2}{\sqrt{3}}\eta \end{pmatrix}
\end{equation}
and for $SU(2)$
\begin{equation}
	\phi(x)=\phi_a\tau_a=\begin{pmatrix} \pi^0 & \sqrt{2}\pi^+ \\ \sqrt{2}\pi^- & -\pi^0 \end{pmatrix}.
\end{equation}
\subsection{The Effective Lagrangian}\label{sec:effectivelagrangian}
As mentioned in section \ref{sec:eft}, the effective Lagrangian contains all possible terms allowed by the symmetry properties of the underlying theory. For our case this means that the Lagrangian should be invariant under $SU(N)_L\times SU(N)_R\times U(1)_V$ with $N=2,3$, while the ground state should only be invariant under $SU(N)_V\times U(1)_V$. Incorporating the $SU(N)$ matrix $U$ it contains eight (three) pseudoscalar degrees of freedom, which, under $SU(N)_V$ transform as an octet (triplet). This corresponds to the observations in nature as the hadrons organize themselves into approximately degenerate multiplets transforming under irreducible representations of $SU(N)$.

It is easily checked that the ground state $\phi(x)=0$ and thus $U_0=\mathds{1}$ is invariant under $SU(N)_V$, since in this case we have $R=L=V$ and (\ref{eq:transform}) yields
	\[U_0'=VU_0V^\dagger=VV^\dagger=\mathds{1}=U_0.
\]
On the other hand $U_0$ is not invariant under axial transformations, i.e. for $L=A$ and $R=A^\dagger$ (corresponding to the case of opposite phases, see section \ref{sec:symmetrycurrents}):
	\[U_0'=A^\dagger U_0 A^\dagger = A^\dagger A^\dagger \neq \mathds{1} = U_0,
\]
just as we expect it due to spontaneous symmetry breaking. The transformation behavior of the Goldstone boson fields $\phi_a$ may be verified to be that of a multiplet under $SU(N)_V$. Thus, we can start writing down all possible allowed terms containing $U$, $U^\dagger$ and derivatives thereof. As mentioned in section \ref{sec:eft} ChPT is an expansion in momenta (and quark masses). Momenta are generated by derivatives in the Lagrangian, thus we have a chain of terms with an increasing number of derivatives. In order to get Lorentz invariant terms there must be an even number of derivatives, thus we have
\begin{equation}
	\mathcal{L}_\text{eff}=\mathcal{L}_0+\mathcal{L}_2+\mathcal{L}_4+...\label{eq:string},
\end{equation}
where the subscript denotes the number of derivatives. The only possible building block in $\mathcal{L}_0$ is $UU^\dagger=\mathds{1}$ which contributes only a constant and can therefore be dropped. The lowest order is therefore $\mathcal{L}_2$ and reads
\begin{equation}
	\mathcal{L}_2=\frac{F_0^2}{4}\langle \partial_\mu U\partial^\mu U^\dagger \rangle\label{eq:effectivelagrangian},
\end{equation}
where $\langle...\rangle$ denotes the trace which is necessary in order to get an invariant. The prefactor is chosen such that we recover the standard form of the kinetic term $\frac{1}{2}\partial_\mu\phi_a\partial^\mu\phi_a$ when expanding $U$ in powers of $\phi$. The invariance of $\mathcal{L}_2$ under global $SU(N)_L\times SU(N)_R$ transformations is easily shown using (\ref{eq:transform}) and the cyclicity of traces.

A term proportional to $\langle(\partial_\mu\partial^\mu U)U^\dagger\rangle$ contains also two derivatives and satisfies all conditions of symmetry but is equivalent to that already written as can be shown by partial integration.

Using a parametrization of the $SU(N)$ matrices $R$ and $L$ we find via Noether's theorem the conserved vector and axial-vector currents. Expanding the axial-vector currents $A^{\mu,a}$ in the Goldstone fields we find
\begin{equation}
	A^{\mu,a}(x)=-F_0\partial^\mu\phi_a(x)+\mathcal{O}(\phi^3),
\end{equation}
which leads to a non-vanishing matrix element when evaluated between the vacuum and a one-Goldstone boson state:
\begin{equation}
	\langle 0|A^{\mu,a}(x)|\phi^b(p)\rangle=ip^\mu F_0e^{-ipx}\delta^{ab}.
\end{equation}
Comparing this with (\ref{eq:matrixelement}) we conclude that $F_0$ (the constant introduced in (\ref{eq:parametrisierung})) is the pion decay constant in the chiral limit. It can be measured in the pion decay $\pi^+\rightarrow \mu^+\nu_\mu$ (see \cite{scherer}) and is found to be $92.4\,$MeV (actually what is measured there is $F_\pi=F_0(1+\mathcal{O}(m_q))$ which slightly differs from $F_0$ since we are not living in a chiral limit world).

Up to now we have constructed the effective Lagrangian under the assumption of perfect chiral symmetry. But as mentioned in section \ref{sec:goldstoneqcd} chiral symmetry is explicitly broken by finite quark masses which mix left- and right-handed quark fields. The mass term in (\ref{eq:qcdlagr}) can be written as
\begin{equation}
	\mathcal{L}_M=-\bar{q}_R\mathcal{M} q_L-\bar{q}_L\mathcal{M}^\dagger q_R,
\end{equation}
with $\mathcal{M}=\text{diag}(m_u,m_d,m_s)$. Although $\mathcal{M}$ is just a constant matrix, $\mathcal{L}_M$ would be chirally invariant if $\mathcal{M}$ transformed as
\begin{equation}
	\mathcal{M}\rightarrow R\mathcal{M}L^\dagger\label{eq:spurion}
\end{equation}
(we formally replace the mass matrix with a so called spurion). One now constructs terms of the effective Lagrangian being invariant under (\ref{eq:transform}) and (\ref{eq:spurion}). The simplest non-constant term reads
\begin{equation}
	\mathcal{L}_{2}^M=\frac{F_0^2B_0}{2}\langle \mathcal{M}U^\dagger+U\mathcal{M}^\dagger\rangle\label{eq:effmass},
\end{equation}
where $B_0$ is an new parameter. Setting $\mathcal{M}$ now back to the constant mass matrix, the chiral symmetry of the effective Lagrangian is explicitly broken in the same manner as in QCD. The parameter $B_0$ appearing in (\ref{eq:effmass}) is related to the scalar quark condensate $\langle\bar{q}q\rangle$ by\footnote{In the case of $SU(2)$ one writes $F$ and $B$ instead of $F_0$ and $B_0$.}
\begin{align}
F_0^2B_0 & = -\langle\bar{u}u\rangle_3\quad\text{for } SU(3)\\
F^2B & = -\langle\bar{u}u\rangle_2\quad\text{for } SU(2).
\end{align}

Using (\ref{eq:effmass}) together with (\ref{eq:effectivelagrangian}), the masses (at lowest order) of the Goldstone bosons are found to be
\begin{eqnarray}
M_\pi^2 & = & 2B_0m\label{eq:masse1}\\
M_K^2 & = & B_0(m+m_s)\label{eq:masse2}\\
M_\eta^2 & = & \frac{2}{3}B_0(m+2m_s)\label{eq:masse3},
\end{eqnarray}
where $m_u=m_d=m$ (isospin limit for simplicity). These relations and the on-shell condition $p^2=M^2$ are the reason why terms in the effective Lagrangian containing one quark mass contribute to the same order as terms with two derivatives and why therefore the effective mass term in (\ref{eq:effmass}) got the subscript 2.
\section{Higher Orders and Loops}\label{sec:higherorderandloops}
Using the same principles as in the previous section, effective Lagrangian terms of higher orders are constructed. As indicated in (\ref{eq:string}) the next order is $\mathcal{L}_4$ which contains four derivatives, two derivatives and one power of the quark mass matrix or two powers of the quark mass matrix. It was written down by Gasser and Leutwyler, see \cite{Gasser:1983yg}, and is found in section $\ref{sec:chirallagrangianorderp4}$. 

Calculating now a physical process to a given order of momentum, we have to know which diagrams from which orders of the Lagrangian must be taken into account. This is what Weinberg's power counting scheme tells us.
\subsection{Weinberg's Power Counting Scheme}\label{sec:weinbergpowercounting}
Given an arbitrary loop diagram with $L$ loops, $I$ internal lines and $V_d$ vertices of order $d$ (i.e. originating from $\mathcal{L}_d$), the amplitude will be of the form $\int (d^4p)^L/(p^2)^I\cdot\prod_d(p^d)^{V_d}$. The so called chiral order $D$ (or chiral dimension) of the diagram counts the dimension of momenta and is therefore given by $D=4L-2I+\sum_ddV_d$. Using the identity $L=I-\sum_dV_d+1$ one can eliminate $I$ and finds
\begin{equation}
	D=\sum_dV_d(d-2)+2L+2\label{eq:chiraldimension}
\end{equation}
This is equivalent to the following procedure: Consider a given diagram and rescale all external momenta as $p_i\rightarrow tp_i$ and the quark masses as $m_q\rightarrow t^2m_q$ (which corresponds to a rescaling of the Goldstone boson masses as $M\rightarrow tM$). The chiral dimension $D$ of the diagram with amplitude $\mathcal{M}(p_i,m_q)$ is then defined by $\mathcal{M}(tp_i,t^2m_q)=t^D\mathcal{M}(p_i,m_q)$, see \cite{scherer}.

For small momenta only diagrams of low chiral order $D$ will dominate. The most simple diagram just contains a vertex from $\mathcal{L}_2$ and is therefore of chiral order $D=2$, i.e. $\mathcal{O}(p^2)$, which is the leading order. Going on to $D=4$ we will have to consider one-loop graphs composed only of $\mathcal{L}_2$ vertices and tree graphs with one $\mathcal{L}_4$ vertex. Thus, for fixed $D$ there is only a finite number of diagrams which have to be taken into account.
\subsection{Renormalization Scheme}
When doing calculations at chiral order $D=4$ we get loop integrals with $\mathcal{L}_2$ vertices which diverge. Since these infinities are of $\mathcal{O}(p^4)$, they cannot be absorbed by a renormalization of the low energy constants $F_0$ and $B_0$ of $\mathcal{L}_2$. However, it is possible to absorb them in the coefficients of $\mathcal{L}_4$ by renormalizing just these.

Hence, in order to get finite results at $\mathcal{O}(p^4)$ we must not only consider one-loop graphs with $\mathcal{L}_2$ vertices but also include tree graphs obtained from $\mathcal{L}_4$, i.e. all possible diagrams of that order. There is always only a finite number of necessary `counterterms' to cancel the divergences. This is meant by the statement that ChPT is renormalizable order by order.
\section{Local Invariance and External Fields}\label{sec:localinvarianceandexternalfields}
So far we have considered effective Lagrangians being invariant under global chiral transformations (\ref{eq:transform}) only. However, in \cite{Gasser:1983yg} it was shown that global symmetry does not suffice to determine the full low energy structure and that one needs to consider off-shell Green functions (i.e. involving momenta with no correspondence to the Goldstone boson masses), where a study of Ward-Takahashi identities (relations between Green functions originating from symmetries) becomes necessary. All possible Green functions may be obtained via a generating functional depending on external fields and the Ward identities are equivalent to the invariance of the generating functional under \textit{local} transformations.

Following Gasser and Leutwyler, the QCD Lagrangian is extended by coupling the quarks to external Hermitian matrix fields ($3\times 3$ matrices in flavor space) $v_\mu(x)$, $a_\mu(x)$, $s(x)$ and $p(x)$:
\begin{equation}
	\mathcal{L}=\mathcal{L}^0_\text{QCD}+\mathcal{L}_\text{ext}=\mathcal{L}^0_\text{QCD}+\bar{q}\gamma^\mu(v_\mu+\gamma_5 a_\mu)q-\bar{q}(s-i\gamma_5 p)q\label{eq:lagrangianexternalfields}.
\end{equation}
The ordinary three flavor QCD Lagrangian is recovered by setting $v_\mu=a_\mu=p=0$ and $s=\text{diag}(m_u,m_d,m_s)$. The generating functional $Z[v,a,s,p]$ mentioned above is defined as
\begin{equation}
	\exp(iZ[v,a,s,p])=\langle 0|T\exp\left[i\int d^4x\mathcal{L}_\text{ext}(x)\right]|0\rangle\label{eq:generatingfunctional}
\end{equation}
and is related to the vacuum-to-vacuum transition amplitude in the presence of external fields. Green functions are calculated by performing functional derivatives of the expression (\ref{eq:generatingfunctional}) with respect to the external fields. Since the generating functional may be represented by the path integral
\begin{equation}
	\exp(iZ[v,a,s,p])=\int [DU]\exp\left[i\int d^4x\mathcal{L}_\text{eff}\right]
\end{equation}
and since we request $Z[v,a,s,p]$ to be locally invariant, we will have to promote the global $SU(N)_L\times SU(N)_R$ symmetry of $\mathcal{L}_\text{eff}$ to a local one and thus include the same external fields as in QCD (see \cite{eckerchpt}).

There is another reason why it is very convenient to introduce external fields: In this way the coupling of an external photon field $A_\mu$ to the quarks (i.e. electromagnetic interaction) as well as the coupling of the massive charged weak boson $W_\mu$ to the quarks (i.e. weak interaction) may be easily incorporated. To do so, the Lagrangian (\ref{eq:lagrangianexternalfields}) is split up in left- and right-handed parts, yielding
\begin{equation}
	\mathcal{L}=\mathcal{L}^0_\text{QCD}+\bar{q}_L\gamma^\mu l_\mu q_L+\bar{q}_R\gamma^\mu r_\mu q_R-\bar{q}_R(s+ip)q_L-\bar{q}_L(s-ip)q_R,
\end{equation}
where $r_\mu=v_\mu+a_\mu$ and $l_\mu=v_\mu-a_\mu$. This Lagrangian remains invariant under local transformation of the $q_{L,R}$ if the external fields are subject to transform in a certain manner, see e.g. \cite{scherer}.

Then, e.g. for the electromagnetic interaction we set
\begin{equation}
	r_\mu=l_\mu=-eQA_\mu\label{eq:externalphotons},
\end{equation}
where $Q=\text{diag}(2/3, -1/3, -1/3)$ is the quark charge matrix (three flavors). The weak interaction is included in a similar easy way.
\subsection{Locally Invariant Effective Lagrangian}
As in the case of gauge theories, the external fields are contained in a covariant derivative
\begin{equation}
	D_\mu U=\partial_\mu U-ir_\mu U+iUl_\mu,
\end{equation}
which, under a local transformation $U\rightarrow R(x)UL^\dagger(x)$, transforms in the same way as $U$. Beside this new building block there are further objects to be used to construct a locally invariant effective Lagrangian: The fields strength tensors
\begin{align}
f^R_{\mu\nu} &= \partial_\mu r_\nu-\partial_\nu r_\mu-i[r_\mu,r_\nu]\\
f^L_{\mu\nu} &= \partial_\mu l_\nu-\partial_\nu l_\mu-i[l_\mu,l_\nu],
\end{align}
as well as the linear combination
\begin{equation}
	\chi=2B_0(s+ip)\label{eq:defchi}.
\end{equation}
As before $D_\mu$ is of $\mathcal{O}(p)$. The external fields $r_\mu$ and $l_\mu$ are of $\mathcal{O}(p)$ to match $\partial_\mu U$; therefore the fields strength tensors are of $\mathcal{O}(p^2)$ and will not contribute to the lowest order effective Lagrangian $\mathcal{L}_2$ since the Lorentz indices have to be contracted. Finally, $\chi$ is of $\mathcal{O}(p^2)$ because of (\ref{eq:masse1})-(\ref{eq:masse3}).

The most general, locally invariant, effective Lagrangian at lowest order now reads
\begin{equation}
	\mathcal{L}_2=\frac{F_0^2}{4}\langle D_\mu U (D^\mu U)^\dagger\rangle+\frac{F_0^2}{4}\langle \chi U^\dagger+U\chi^\dagger\rangle.\label{eq:lowestorderlocallyinvlagr}
\end{equation}
It still has two free parameters; $B_0$ is hidden in $\chi$. (\ref{eq:lowestorderlocallyinvlagr}) reduces to (\ref{eq:effectivelagrangian}) and (\ref{eq:effmass}) if all the external fields except for $s$ are switched off and $s$ is set to $s=\mathcal{M}=\text{diag}(m_u,m_d,m_s)$.
\section{The Chiral Lagrangian at Order $\mathcal{O}(p^4)$}\label{sec:chirallagrangianorderp4}
As already mentioned in section \ref{sec:higherorderandloops}, $\mathcal{L}_4$ can be constructed applying the same ideas that where used to find $\mathcal{L}_2$. It does not only contain two low-energy constants but 12. Using the same covariant derivative and field strength tensors as above, $\mathcal{L}_4$ reads
\begin{align}
	\mathcal{L}_4&=L_1\langle D_\mu U\left(D^\mu U\right)^\dagger\rangle^2+L_2\langle D_\mu U\left(D_\nu U\right)^\dagger\rangle\langle D^\mu U\left(D^\nu U\right)^\dagger\rangle\nonumber\\
	&+L_3\langle D_\mu U\left(D^\mu U\right)^\dagger D_\nu U\left(D^\nu U\right)^\dagger\rangle+L_4\langle D_\mu U\left(D^\mu U\right)^\dagger\rangle\langle\chi U^\dagger+U\chi^\dagger\rangle\nonumber\\
	&+L_5\langle D_\mu U\left(D^\mu U\right)^\dagger\left(\chi U^\dagger+U\chi^\dagger\right)\rangle+L_6\langle\chi U^\dagger+U\chi^\dagger\rangle^2\nonumber\\
	&+L_7\langle\chi U^\dagger-U\chi^\dagger\rangle^2+L_8\langle U\chi^\dagger U\chi^\dagger+\chi U^\dagger\chi U^\dagger\rangle\nonumber\\
	&-iL_9\langle f^R_{\mu\nu}D^\mu U\left(D^\nu U\right)^\dagger+f^L_{\mu\nu}\left(D^\mu U\right)^\dagger D^\nu U\rangle+L_{10}\langle Uf^L_{\mu\nu}U^\dagger f_R^{\mu\nu}\rangle\nonumber\\	
	&+H_1\langle f^R_{\mu\nu}f_R^{\mu\nu}+f^L_{\mu\nu}f_L^{\mu\nu}\rangle+H_2\langle\chi\chi^\dagger\rangle
\end{align}
In the case of $SU(2)$ the Cayley-Hamilton theorem (see e.g. \cite{Bijnens:1999sh}) can be applied to reduce the number of terms.

\section{Chiral Perturbation Theory with Baryons}\label{sec:chptwithbaryons}
So far we have considered an effective description for Goldstone bosons only. However, it is possible to extend ChPT to include also nucleons ($SU(2)$) or the baryon octet ($SU(3)$). There is one important difficulty: The nucleon mass is a heavy mass scale that does not vanish in the chiral limit. The 3-momenta of the baryons need to be small in order to keep ChPT valid.

The task is to find an effective Lagrangian $\mathcal{L}_{\pi N}$ that describes the interaction between baryons and Goldstone bosons. We only consider matrix elements with  a single baryon in the initial and final states.
\subsection{Representation of the Goldstone Bosons and the Baryons}\label{sec:reprofgoldstonebosandbaryons}
As in the case with Goldstone bosons we first need a suitable representation of the particles of interest. It turns out to be advantageous not to represent the Goldstone bosons by the familiar matrix $U$ but rather by its square root $u$, thus $u^2=U$. Under $G=SU(N)_L\times SU(N)_R$ this field transforms as
\begin{equation}
	u\rightarrow\sqrt{LUR^\dagger}=LuK^\dagger(L,R,U)=K(L,U,R)uR^\dagger,
\end{equation}
where we introduce the compensator field $K(L,U,R)$ which is a $SU(N)$ matrix that nontrivially depends on $L$, $R$ and $U$, see \cite{bernardkaisermeissner}. Only for $SU(N)_V$ transformations where $L=R$ we have the simple relation $K=L=R$.

We will only consider two quark flavors, thus $SU(2)$, and therefore only the proton and the neutron. These can be represented by an isospinor
\begin{equation}
	\Psi=\begin{pmatrix} p\\ n\end{pmatrix}\label{eq:fermionicpsi}.
\end{equation}
The transformation behavior of the pair (U,$\Psi$) under $SU(2)_L\times SU(2)_R\times U(1)_V$ reads
\begin{equation}
	\begin{pmatrix} U\\ \Psi\end{pmatrix}\rightarrow \begin{pmatrix} RUL^\dagger\\ \exp(i\Theta)K[L,R,U]\Psi\end{pmatrix},
\end{equation}
where $\Theta$ parametrizes $U(1)_V$ transformations. For more details see \cite{eckerchpt,bernardkaisermeissner}.
\subsection{Lowest-Order Effective Baryonic Lagrangian}\label{sec:lowestordereffectivebaryoniclagrangian}
As in the mesonic sector, in order to incorporate Ward identities we have to construct the most general effective Lagrangian coupled to external fields with local $SU(2)_L\times SU(2)_R\times U(1)_V$ symmetry (see \cite{scherer}). Local transformations imply a covariant derivative $D_\mu\Psi$ with the usual property to transform in the same way as $\Psi$; $D_\mu\Psi(x)\rightarrow \exp(i\Theta(x))K[L(x),R(x),U(x)]\Psi(x)$.

The covariant derivative is given by
\begin{equation}
	D_\mu\Psi=(\partial_\mu+\Gamma_\mu)\Psi\label{eq:covariatderivativebaryons},
\end{equation}
where the vector
\begin{equation}
	\Gamma_\mu=\frac{1}{2}\left[u^\dagger(\partial_\mu-ir_\mu)u+u(\partial_\mu-il_\mu)u^\dagger\right]\label{eq:defconnection}
\end{equation}
is the so called connection which contains the same external fields as introduced in section \ref{sec:localinvarianceandexternalfields}. Another building block is the chiral vielbein (which is an axial vector)
\begin{equation}
	u_\mu=i\left[u^\dagger(\partial_\mu-ir_\mu)u-u(\partial_\mu-il_\mu)u^\dagger\right].
\end{equation}
Knowing the transformation behavior of $\Psi$ under $G$ it follows that the most general effective $\pi N$ Lagrangian with a single nucleon in the initial and final states needs to be of the type $\bar{\Psi}\hat{O}\Psi$, where $\hat{O}$ is an operator transforming as $\hat{O}\rightarrow K\hat{O}K^\dagger$ under $G$.

Due to different Lorentz structure of meson and baryon fields, the chiral expansion of $\mathcal{L}_{\pi N}$ contains terms of all orders of $p$ and not only of even ones, hence we have $\mathcal{L}_{\pi N}=\mathcal{L}_{\pi N}^{(1)}+\mathcal{L}_{\pi N}^{(2)}+...$.

The most general such Lagrangian with the smallest number of derivatives being additionally a Hermitian Lorentz scalar and even under $C$, $P$ and $T$ reads
\begin{equation}
	\mathcal{L}^{(1)}_{\pi N}=\bar{\Psi}\left(i\slashed{D}-\mathring{m}_N+\frac{\mathring{g}_A}{2}\gamma^\mu\gamma_5 u_\mu\right)\Psi\label{eq:effbaryonlagrangian}.
\end{equation}
The nucleon mass $\mathring{m}_N$ and the axial vector coupling constant $\mathring{g}_A$ appear as two free parameters. They are both meant to be in the chiral limit (denoted by $\circ$); $m=\mathring{m}[1+\mathcal{O}(m_q)]$, $g_A=\mathring{g}_A[1+\mathcal{O}(m_q)]$ with $m=939\,\text{MeV}$ and $g_A\simeq 1.26$ (known from neutron beta decay). The Lagrangian (\ref{eq:effbaryonlagrangian}) reduces to that of a free nucleon of mass $\mathring{m}_N$ in the case of no external fields and no pion fields.

The power counting rules for the new quantities are
\begin{equation}
	\bar{\Psi},\Psi=\mathcal{O}(1),\quad D_\mu\Psi=\mathcal{O}(1),\quad (i\slashed{D}-\mathring{m}_N)\Psi=\mathcal{O}(p)
\end{equation}
(to be explained in \cite{scherer}). Especially, the fact that the covariant derivative is not counted as $\mathcal{O}(p)$ (i.e. as a small quantity) anymore is due to the fact that the nucleon mass does not vanish in the chiral limit and therefore the zeroth component of the partial derivative acting on the nucleon field does not produce a small quantity. This fact leads to an inconvenience: We lose the correspondence between the loop and the chiral expansion as we had it in the mesonic sector. The contribution from loops is not automatically suppressed and an amplitude with given chiral dimension $D$ may get contributions from diagrams with arbitrary many loops.

\subsection{The next Order of the Baryonic Lagrangian}\label{sec:nextorderbaryoniclagrangian}
Without any further comments on how to derive it, $\mathcal{L}_{\pi N}^{(2)}$ is quoted here (see \cite{Becher:2001hv}):
\begin{multline}
	\mathcal{L}_{\pi N}^{(2)}=c_1\langle u^\dagger\chi u^\dagger+u\chi^\dagger u\rangle\bar{\Psi}\Psi-\frac{c_2}{4m_N^2}\langle u_\mu u_\nu\rangle\left(\bar{\Psi}D^\mu D^\nu\Psi+\text{h.c.}\right)\\
	+\frac{c_3}{2}\langle u_\mu u^\mu\rangle\bar{\Psi}\Psi-\frac{c_4}{4}\bar{\Psi}\gamma^\mu\gamma^\nu\left[u_\mu,u_\nu\right]\Psi.\label{eq:secondorderfermioniclagrangian}
\end{multline}
Since at $\mathcal{O}(p^2)$ only tree graphs from $\mathcal{L}_{\pi N}^{(2)}$ must be considered, there are no divergent loop integrals and therefore the low energy constants $c_i$ cannot contain a divergent part. They can be determined by comparison to some $\pi N$ threshold parameters which have been measured. The following two sets of the coupling constants are taken from \cite{Becher:2001hv}:
\begin{align*}
c_1&=-0.6m_N^{-1}\;,\; c_2=1.6m_N^{-1}\;,\; c_3=-3.4m_N^{-1}\;,\; c_4=2.0m_N^{-1}\\
c_1&=-0.9m_N^{-1}\;,\; c_2=2.5m_N^{-1}\;,\; c_3=-4.2m_N^{-1}\;,\; c_4=2.3m_N^{-1}
\end{align*}

\chapter{Chiral Spirals - A First Approach}\label{ch:firstapproach}
In this chapter we construct a single-particle Hamiltonian from the fermionic part of the effective Lagrangian. This will be diagonalized in order to find the energy of one nucleon in the background of pions, leading to two eigenstates with different energies. We introduce the spiral configuration for pions and fix the gauge such that both the connection and the vielbein become constant fields. Next, in the momentum space we fill the two fermionic eigenstates up to their corresponding Fermi energy. To do so, the sophisticated relativistic expression for the energy of one eigenstate is expanded in inverse powers of the nucleon mass up to $\mathcal{O}(m_N^{-1})$. We express the Fermi momenta and hence the fermionic energy density in terms of the particle number densities of the two eigenstates. The total energy density may then be written in terms of these and some spiral parameters. First in the chiral limit and then away from it, we find critical values for the total particle number density that need to be exceeded in order to allow the many-particle system to lower its energy density by changing the pionic configuration from a homogeneous configuration to a spiral one.
\section{Pion-Nucleon Effective Theory}
The leading terms of the effective Lagrangian describing a pion-nucleon system with two quark flavors and including explicit chiral symmetry breaking due to quark masses is given by (\ref{eq:effectivelagrangian}), (\ref{eq:effmass}) and (\ref{eq:effbaryonlagrangian}):
\begin{equation}
	\mathcal{L}=\frac{F^2}{4}\langle \partial_\mu U\partial^\mu U^\dagger \rangle+\frac{F^2B}{2}\langle \mathcal{M}U^\dagger+U\mathcal{M}^\dagger\rangle+\bar{\Psi}\left(i\slashed{D}-\mathring{m}_N+\frac{\mathring{g}_A}{2}\gamma^\mu\gamma_5 u_\mu\right)\Psi\label{eq:vollstefflagr}.
\end{equation}
We will restrict ourselves to pion field configurations $U(x)$ which lead to a constant connection and vielbein, such that the Dirac equation may be solved more easily;
\begin{equation}
	u_\mu(x)=c_\mu,\quad \Gamma_\mu(x)=d_\mu\label{eq:konstconvielb}.
\end{equation}
Furthermore we consider static pion fields which implies $\Gamma_0=u_0=0$.
\subsection{The Single-Particle Hamiltonian}\label{sec:singleparticlehamiltonian}
Under these conditions the fermionic part of (\ref{eq:vollstefflagr}) is given by (dropping the symbol $\circ$ on $\mathring{m}_N$ and $\mathring{g}_A$)
\begin{equation}
	\mathcal{L}_f=i\bar{\Psi}\gamma^\mu\partial_\mu\Psi+i\bar{\Psi}\gamma^id_i\Psi-m_N\bar{\Psi}\Psi+\frac{g_A}{2}\bar{\Psi}\gamma^i\gamma_5c_i\Psi.
\end{equation}
The equations of motion for $\Psi$ are given by the Euler-Lagrange equation
\begin{equation}
	\partial_\mu\frac{\partial\mathcal{L}_f}{\partial\partial_\mu\bar{\Psi}}-\frac{\partial\mathcal{L}_f}{\partial\bar{\Psi}}=0,
\end{equation}
which yields
\begin{equation}
	i\gamma^\mu\partial_\mu\Psi+i\gamma^id_i\Psi-m_N\Psi+\frac{g_A}{2}\gamma^i\gamma_5c_i\Psi=0\label{eq:eom}.
\end{equation}
The Dirac equation, on the other hand, reads
\begin{equation}
	i\partial_0\Psi=\hat{H}\Psi.
\end{equation}
Thus, separating (\ref{eq:eom}) such that on one side of the equation we have $i\partial_0\Psi$ only, we can read off the Hamiltonian $\hat{H}$ on the other side:
\begin{equation}
	\hat{H}=m_N\gamma^0+\gamma^0\gamma^ip_i-i\gamma^0\gamma^id_i-\frac{g_A}{2}\gamma^0\gamma^i\gamma_5c_i\label{eq:fermionhamiltonian}.
\end{equation}
We have used $-i\partial_i\Psi=p_i\Psi$ (momentum of the nucleon). For $\hat{H}$ to be Hermitian we find that the $c_i$ need to be Hermitian while the $d_i$ are antihermitian,
\begin{equation}
	c_i=c_{i,a}\tau^a,\quad id_i=d_{i,a}\tau^a\quad \text{with}\quad c_{i,a},d_{i,a}\in\mathbb{R}\label{eq:ciunddi}.
\end{equation}
Since $\Psi$ is a doublet containing two Dirac spinors, $\hat{H}$ is an $8\times 8$ matrix, i.e. by the term $m_N\gamma^0$ we actually mean $m_N\text{diag}(\gamma^0,\gamma^0)$ etc.
\section{A Spiral Configuration}\label{sec:aspiralconfiguration}
\subsection{The Spiral}
As explained in section \ref{sec:representationofgoldstonebosons} the pions live in the coset space $SU(2)_L\times SU(2)_R/SU(2)_V=SU(2)$ which is isomorphic (as a manifold) to $S^3$. Each $SU(2)$ matrix $U$ can be written in the form
\begin{equation}
	U=\xi_0\mathds{1}+i\vec{\xi}\cdot\vec{\tau},\quad \xi_\mu\in\mathbb{R},
\end{equation}
where we require the condition
\begin{equation}
	\xi_0^2+\vec{\xi}^2=1\label{eq:su2condition}
\end{equation}
in order to have $\det U=1$. The condition (\ref{eq:su2condition}) explicitly shows the relation between $SU(2)$ and $S^3$: $U$ can be seen as a four-dimensional unit vector. Thus, the pions (described by $U$) can be thought to live on the three-dimensional surface of the four-dimensional unit sphere.

We now choose the following particular form for the matrix $U$:
\begin{equation}
	U(\vec{x})=\cos\alpha_0\mathds{1}+i\sin\alpha_0\left(\cos\varphi(\vec{x})\tau^1+\sin\varphi(\vec{x})\tau^2\right),\quad\varphi(\vec{x})=\vec{a}\cdot\vec{x}+\varphi_0,\label{eq:spiralconfiguration}
\end{equation}
which satisfies the condition (\ref{eq:su2condition}). $\vec{a}$ is a constant three-dimensional vector. In appendix \ref{app:spiral} it is shown that the condition of having a constant connection as well as a constant mass term in the Lagrangian naturally leads us to the configuration (\ref{eq:spiralconfiguration}).

Let us interpret this configuration: Since we have $\xi_3=0$ we restrict ourselves to the two-dimensional hypersurface $\xi_0^2+\xi_1^2+\xi_2^1=1$, i.e. to $S^2$. Hence, the matrix describing the pion fields in each point of the three-dimensional space can be respresented by a three-dimensional vector lying on the unit sphere at that point. The ground state (no pions) corresponds to $U=\mathds{1}$ (see section \ref{sec:effectivelagrangian}) which is equivalent to $\alpha_0=0$. So, $\alpha_0$ may be interpreted as the polar angle of the unit sphere, if we choose the ground state to point in the third direction of a cartesian coordinate system. Then, $\varphi(\vec{x})$ can be thought of as the azimuthal angle and we can indeed describe the vector representing $U$ by the ordinary spherical coordinates.

Now let us walk around in space, namely in a plane perpendicular to the vector $\vec{a}$. This plane may be parametrized as $\vec{x}=\vec{x}_0+\lambda\vec{v}+\mu\vec{w}$ where $\vec{x}_0$, $\vec{v}$ and $\vec{w}$ are constant three vectors ($\vec{v}$ and $\vec{w}$ linearly independent) and $\lambda$, $\mu\in\mathbb{R}$. Since $\vec{a}$ is perpendicular to both $\vec{v}$ and $\vec{w}$ we get $\vec{x}\cdot\vec{a}=\vec{x}_0\cdot\vec{a}=\text{const}$. Thus, in the plane perpendicular to $\vec{a}$ the azimuth angle $\varphi$ (see (\ref{eq:spiralconfiguration})) does not depend on $\vec{x}$ and the unit vector respresenting $U$ will not change throughout space. However, if we walk in any other direction the azimuth angle $\varphi(\vec{x})$ becomes position-dependent and the vector starts to rotate about the ``north pole'' of the unit sphere. For a given speed of our walk the scalar product $\vec{a}\cdot\vec{x}$ and so the azimuth angle increases most quickly in the direction of $\vec{a}$ which we therefore may identify as the direction of the spiral.

Having such a configuration, we can choose an arbitrary vector to represent the ground state and hence to define our third direction. In all neighboring space points (except those lying on the plane perpendicular to $\vec{a}$) the vector will have rotated.

For our calculations we need the field $u$ rather than $U$. We make the ansatz
\begin{equation}
	u(\vec{x})=\cos\alpha\mathds{1}+i\sin\alpha\,\vec{e}(\vec{x})\cdot\vec{\tau},
\end{equation}
where $\vec{e}(\vec{x})$ is a unit vector. Using $(\vec{e}\cdot\vec{\tau})^2=e_ie_j\tau^i\tau^j=e_ie_j(i\epsilon^{ijk}\tau^k+\delta^{ij}\mathds{1})=e_ie_i\mathds{1}=\mathds{1}$ (where we used the symmetry of $e_ie_j$ and the antisymmetry of $\epsilon^{ijk}$ in the indices $i$ and $j$ and the fact that $\vec{e}$ is a unit vector) we find
	\[u^2(\vec{x})=\left(\cos^2\alpha-\sin^2\alpha\right)\mathds{1}+2i\sin\alpha\cos\alpha\,\vec{e}(\vec{x})\cdot\vec{\tau}=\cos 2\alpha\mathds{1}+i\sin 2\alpha\,\vec{e}(\vec{x})\cdot\vec{\tau}
\]
and the requirement $u^2=U$ yields
\begin{equation}
	\alpha=\frac{\alpha_0}{2},\quad \vec{e}(\vec{x})=(\cos\varphi(\vec{x}),\sin\varphi(\vec{x}),0)\label{eq:winkeleinheitsvektor}.
\end{equation}
\subsection{Connection and Vielbein}\label{sec:connectionandvielbein}
From this we may calculate the connection and the vielbein (we set the external fields $r_\mu$ and $l_\mu$ to zero). For convenience we write $e_i=\delta_{1i}\cos\varphi+\delta_{2i}\sin\varphi$ and, using the particular form of $\varphi$ of (\ref{eq:spiralconfiguration}), $\partial_ie_j=a_i(\delta_{2j}\cos\varphi-\delta_{1j}\sin\varphi)$. We then have
\begin{align*}
\Gamma_i & = \frac{1}{2}\left[u^\dagger\partial_i u+u\partial_iu^\dagger\right]\\
& = \frac{1}{2}\left[(\cos\alpha\mathds{1}-i\sin\alpha\,e_i\tau^i)i\sin\alpha\partial_ie_j\tau^j-(\cos\alpha\mathds{1}+i\sin\alpha\,e_i\tau^i)i\sin\alpha\partial_ie_j\tau^j\right]\\
& = \frac{1}{2}\cdot 2\sin^2\alpha e_k\tau^k\partial_i e_j\tau^j=a_i\sin^2\alpha(\delta_{1k}\cos\varphi+\delta_{2k}\sin\varphi)(\delta_{2j}\cos\varphi-\delta_{1j}\sin\varphi)\tau^k\tau^j\\
& = a_i\sin^2\alpha\big[\delta_{1k}\delta_{2j}\cos^2\varphi-\delta_{1k}\delta_{1j}\cos\varphi\sin\varphi+\delta_{2k}\delta_{2j}\sin\varphi\cos\varphi\\
&\hspace{8cm}-\delta_{2k}\delta_{1j}\sin^2\varphi\big](i\epsilon^{kjl}\tau^l+\delta^{kj}\mathds{1}).
\end{align*}
The second and third term in the squared bracket vanish because of (anti-)symmetry of the indices $k$ and $j$ when multiplied with $\epsilon^{kjl}$ and because they cancel when multiplied with $\delta^{kj}$. The first and the fourth term disappear when multiplied with $\delta^{kj}$ because the indices can not coincide in both Kronecker deltas at the same time. Thus, what is left is
\begin{equation}
	\Gamma_i=ia_i\sin^2\alpha(\cos^2\varphi\epsilon^{12l}\tau^l-\sin^2\varphi\epsilon^{21l}\tau^l)=ia_i\sin^2\alpha\tau^3.
\end{equation}
The vielbein:
\begin{align}
u_i & = i\left[u^\dagger\partial_iu-u\partial_i u^\dagger\right]\nonumber\\
&=i\left[(\cos\alpha\mathds{1}-i\sin\alpha\,e_i\tau^i)i\sin\alpha\partial_ie_j\tau^j+(\cos\alpha\mathds{1}+i\sin\alpha\,e_i\tau^i)i\sin\alpha\partial_ie_j\tau^j\right]\nonumber\\
& = -2\cos\alpha\sin\alpha\partial_ie_j\tau^j=-2a_i\cos\alpha\sin\alpha(-\sin\varphi\tau^1+\cos\varphi\tau^2).
\end{align}
Obviously $u_i$ is not a constant field in contrast to what we demanded in (\ref{eq:konstconvielb}). It can however be made constant by an appropriate $SU(2)_V$ gauge transformation. As can be shown, the transformation behavior of $u_\mu$ under local $SU(2)_L\times SU(2)_R$ is given by
\begin{equation}
	u_\mu\rightarrow Ku_\mu K^\dagger\label{eq:vielbeintrafo},
\end{equation}
where $K$ is the compensator field introduced in section \ref{sec:reprofgoldstonebosandbaryons}. Starting from the equation
\begin{equation}
	\partial_\mu K=K\Gamma_\mu-\Gamma'_\mu K
\end{equation}
(derived in \cite{scherer}) and using the relation $\partial_\mu KK^\dagger=-K\partial_\mu K^\dagger$, we find the transformation behavior of the connection to be
\begin{equation}
	\Gamma_\mu\rightarrow K(\Gamma_\mu+\partial_\mu)K^\dagger\label{eq:connectiontrafo}.
\end{equation}
In the case of the isospin transformation $L=R=V$ we have $K=V$ (see \cite{scherer}). Writing out $u_i$ in matrix form,
\begin{align*}
u_i(x) & = -2a_i\cos\alpha\sin\alpha\begin{pmatrix} 0 & -\sin\varphi(x)-i\cos\varphi(x) \\ -\sin\varphi(x)+i\cos\varphi(x) & 0 \end{pmatrix}\\
& = -2a_i\cos\alpha\sin\alpha\begin{pmatrix} 0 & -i\exp(-i\varphi(x)) \\ i\exp(i\varphi(x)) & 0 \end{pmatrix},
\end{align*}
and parametrizing the $SU(2)$ matrix $V$ in the following way,
\begin{equation}
	V(x)=\begin{pmatrix} \exp(i\chi(x)) & 0 \\ 0 & \exp(-i\chi(x)) \end{pmatrix},
\end{equation}
we find, using (\ref{eq:vielbeintrafo}), for the transformed vielbein
\begin{equation}
	u_i'(x)=-2a_i\cos\alpha\sin\alpha\begin{pmatrix} 0 & -i\exp(-i\varphi(x)+2i\chi(x)) \\ i\exp(i\varphi(x)-2i\chi(x)) & 0 \end{pmatrix}\label{eq:transformedvielbein}
\end{equation}
and, using (\ref{eq:connectiontrafo}), for the transformed connection
\begin{equation}
	\Gamma_i'(x)=ia_i\sin^2\alpha\tau^3-i\partial_i\chi(x)\tau^3.
\end{equation}
We have dropped the arrow on $\vec{x}$ and will go on like this for convenience (remember that we do not have time evolution). (\ref{eq:transformedvielbein}) leads to the choice $\chi(x)=\frac{1}{2}\varphi(x)$; then the vielbein gets constant and we obtain
\begin{align}
\Gamma_i' & = ia_i(\sin^2\alpha-\frac{1}{2})\tau^3=d_i=\text{const}\label{eq:finaltransformedconnection}\\
u_i' & = -2a_i\cos\alpha\sin\alpha\tau^2=c_i=\text{const}\label{eq:finaltransformedvielbein}.
\end{align}
For (\ref{eq:finaltransformedconnection}) we have used the form of $\varphi(x)$ introduced in (\ref{eq:spiralconfiguration}).
\subsection{Pion Contribution to the Energy Density}\label{sec:pioncontributiontotheenergydensity}
Making use of $U=u^2$, $u^\dagger u=\mathds{1}$, the identities\footnote{both following from $uu^\dagger=u^\dagger u=\mathds{1}$} $u\partial_\mu u^\dagger=-\partial_\mu uu^\dagger$ and $u^\dagger\partial_\mu u=-\partial_\mu u^\dagger u$ as well as the cyclicity of the trace, it is easy to show that
\begin{equation}
	\langle \partial_\mu U\partial^\mu U^\dagger\rangle=\langle u_\mu u^\mu\rangle.
\end{equation}
With (\ref{eq:finaltransformedvielbein}) we thus find (dropping the prime)
\begin{equation}
	\langle \partial_\mu U\partial^\mu U^\dagger\rangle=\langle u_iu^i\rangle=4\cos^2\alpha\sin^2\alpha a_ia^i\langle\mathds{1}\rangle=8\cos^2\alpha\sin^2\alpha a_ia^i\label{eq:spurvielbeinsumme}.
\end{equation}
We finally need
\begin{equation}
	\langle \mathcal{M}U^\dagger+U\mathcal{M}^\dagger\rangle=\langle\mathcal{M}(U^\dagger+U)\rangle=2\cos\alpha_0\langle \mathcal{M}\rangle=2\cos\alpha_0(m_u+m_d)\label{eq:massentermberechnen},
\end{equation}
where we have used $\mathcal{M}^\dagger=\mathcal{M}$. Thus, remembering (\ref{eq:winkeleinheitsvektor}), the mesonic part of the Lagrangian reads
\begin{align}
	\mathcal{L}_p & = 2F^2\cos^2\alpha\sin^2\alpha a_ia^i+F^2B\cos\alpha_0(m_u+m_d)\nonumber\\
	& = \frac{F^2}{2}\sin^2\alpha_0a_ia^i+F^2B\cos\alpha_0(m_u+m_d).
\end{align}
Since we have $\partial_0 U=0$ we obtain for the energy density due to the pions
\begin{align}
\epsilon_p & = \mathcal{H}_p=-\mathcal{L}_p=-\frac{F^2}{2}\sin^2\alpha_0a_ia^i-F^2B\cos\alpha_0(m_u+m_d)\nonumber\\
& = \frac{F^2}{2}a^2\sin^2\alpha_0-F^2B\cos\alpha_0(m_u+m_d),\label{eq:pionicenergydensity}
\end{align}
where in the last step we used $a_ia^i=-a_ia_i=-|\vec{a}|^2=\mathrel{\mathop:}-a^2$.
\section{Fermionic Contribution to the Energy}\label{sec:fermioniccontributiontotheenergy}
The fermionic energy is found by calculating the eigenvalues of the single-particle Hamiltonian (\ref{eq:fermionhamiltonian}). In principle this corresponds to the calculation done in section \ref{sec:pioncontributiontotheenergydensity} for pions but since these are scalar degrees of freedom, this calculation was much simpler.

Comparing (\ref{eq:ciunddi}) with (\ref{eq:finaltransformedconnection}) and (\ref{eq:finaltransformedvielbein}), we find that only $c_{i,2}\neq 0$ and $d_{i,3}\neq 0$ and we set
\begin{align}
\beta_i\mathrel{\mathop:}=c_{i,2} & = -2a_i\cos\alpha\sin\alpha=-a_i\sin\alpha_0\label{eq:defbetai}\\
\delta_i\mathrel{\mathop:}=d_{i,3} & = -a_i(\sin^2\alpha-\frac{1}{2})=\frac{a_i}{2}\cos\alpha_0\label{eq:defdeltai}.
\end{align}
The Hamiltonian can then be written as
\begin{equation}
	\hat{H}=m_N\gamma^0\mathds{1}+\gamma^0\gamma^ip_i\mathds{1}-\gamma^0\gamma^i\delta_i\tau^3-\frac{g_A}{2}\gamma^0\gamma^i\gamma_5\beta_i\tau^2,
\end{equation}
where $\mathds{1}$ means the $2\times 2$ unit matrix. The eigenvalues of the Hamiltonian are found by Mathematica and read

\begin{align}
E^+_{\pm}&=\sqrt{p^2+\delta^2+m_N^2+\frac{1}{4}g_A^2\beta^2\pm\sqrt{4(p_i\delta_i)^2+g_A^2\{m_N^2\beta^2+\delta^2\beta^2+(p_i\beta_i)^2-(\delta_i\beta_i)^2\}}}\\
E^-_{\pm} &= -\sqrt{p^2+\delta^2+m_N^2+\frac{1}{4}g_A^2\beta^2\pm\sqrt{4(p_i\delta_i)^2+g_A^2\{m_N^2\beta^2+\delta^2\beta^2+(p_i\beta_i)^2-(\delta_i\beta_i)^2\}}},
\end{align}
where $p^2=p_1^2+p_2^2+p_3^2$ etc. Each of these four eigenvalues is twofold degenerate, such that we have eight eigenvalues altogether. Although in the original basis $\Psi$ has the form (\ref{eq:fermionicpsi}), in the new basis where $\hat{H}$ is diagonal, the eigenvectors contain linear combinations of the different $p$- and $n$ spinor components. It is therefore not straightforward to decide which eigenvalue belongs to which state. However, negative energy states (which are the antistates to the positive energy states) should not be considered in the effective theory, since we work at an energy scale of $m_N$ and consider only small deviations from it. Considering the negative energies would mean a deviation of $2m_N$ which is way to large to do ChPT. Hence we neglect the $E^-_{\pm}$.

Due to (\ref{eq:defbetai}) and (\ref{eq:defdeltai}) the two vectors $\vec{\beta}$ and $\vec{\delta}$ are parallel and therefore the two terms $\delta^2\beta^2$ and $(\delta_i\beta_i)^2=(\vec{\delta}\cdot\vec{\beta})^2$ cancel, such that the fermionic energy reads
\begin{equation}
	E_\pm=\sqrt{p^2+\delta^2+m_N^2+\frac{1}{4}g_A^2\beta^2\pm\sqrt{4(\vec{p}\cdot\vec{\delta})^2+g_A^2\{m_N^2\beta^2+(\vec{p}\cdot\vec{\beta})^2\}}}.\label{eq:energieplus}
\end{equation}
Making an expansion of (\ref{eq:energieplus}) in powers of $1/m_N$ yields
\begin{equation}
	E_\pm=m_N+\frac{p^2}{2m_N}+\frac{\delta^2}{2m_N}\pm\frac{1}{2}g_A\beta+\mathcal{O}\left(\frac{1}{m_N^2}\right)\label{eq:energie},
\end{equation}
where $\beta=\sqrt{\beta_1^2+\beta_2^2+\beta_3^2}\geq 0$. $E_\pm$ is the energy of one linear combined state. We now want to calculate the energy due to a given amount of states. To do this, we fill, in a given volume $L^3$, the fermions up to some Fermi energy. Let us suppose the temperature $T=0$ or at least a sufficiently low temperature to assume that we have occupied states only up to the Fermi energy. Since the dispersion relation (\ref{eq:energie}) depends only on $p^2$, in momentum space the Fermi surface is a sphere and to the Fermi energy corresponds a Fermi momentum $p_F$.

The number of possible modes in a region $d^3p$ is
	\[\left(\frac{L}{2\pi}\right)^3 d^3p,
\]
which (neglecting of course the interaction between the fermions) remains the same even if we consider two different particles (since all the creation and annihilation operators of different particles anticommute with each other, the statistic of one particle is completely independent of the other). Hence, we can talk about two separate Fermi momenta $p_{F+}$ and $p_{F-}$. Given these, for the particle numbers $N_\pm$ we then find
\begin{equation}
	N_\pm=2\left(\frac{L}{2\pi}\right)^3\int\limits_{\hidewidth\substack{\text{Fermi}\\\text{sphere}}\hidewidth}\! d^3p=2\left(\frac{L}{2\pi}\right)^3\frac{4\pi}{3}p_{F\pm}^3\label{eq:particlenumberssphere}.
\end{equation}
The factor 2 appears because the states $E_\pm$ are each twofold degenerate. We calculate $N_\pm$ since we want to express the fermionic energy in terms of the particle number or particle number density respectively rather than in terms of Fermi momenta. The particle number density is
\begin{equation}
	n_\pm=\frac{N_\pm}{L^3}=\frac{p_{F\pm}^3}{3\pi^2}\label{eq:particlenumberdensity}.
\end{equation}
Now, using (\ref{eq:energie}) the total energy of all fermions of a given sort up to the Fermi momentum is
\begin{align}
E^\pm_\text{tot}&=2\left(\frac{L}{2\pi}\right)^3\int\limits_{\hidewidth\substack{\text{Fermi}\\\text{sphere}}\hidewidth}\!d^3p\, E_\pm(p)\nonumber\\
&=\left(m_N+\frac{\delta^2}{2m_N}\pm\frac{1}{2}g_A\beta\right)\cdot\underbrace{2\left(\frac{L}{2\pi}\right)^3\int\limits_{\hidewidth\substack{\text{Fermi}\\\text{sphere}}\hidewidth}\!d^3p}_{N_\pm}\,+\,2\left(\frac{L}{2\pi}\right)^3\int\limits_{\hidewidth\substack{\text{Fermi}\\\text{sphere}}\hidewidth}\!d^3p\,\frac{p^2}{2m_N}\nonumber\\
&=\left(m_N+\frac{\delta^2}{2m_N}\pm\frac{1}{2}g_A\beta\right)N_\pm+\frac{1}{m_N}\left(\frac{L}{2\pi}\right)^3 4\pi\int\limits_0^{p_{F\pm}} dp\,p^4\nonumber\\
&=\left(m_N+\frac{\delta^2}{2m_N}\pm\frac{1}{2}g_A\beta\right)N_\pm+\frac{L^3}{2\pi^2 m_N}\frac{p_{F\pm}^5}{5}.\label{eq:totaleenergiepm}
\end{align}
Using now (\ref{eq:particlenumberdensity}) to replace $p_{F\pm}$ in terms of $n_\pm$ we find for the energy densities
\begin{equation}
	\epsilon_\text{tot}^\pm=\frac{E_\text{tot}^\pm}{L^3}=\left(m_N+\frac{\delta^2}{2m_N}\pm\frac{1}{2}g_A\beta\right)n_\pm+\frac{(3\pi^2 n_\pm)^{5/3}}{10\pi^2 m_N}\label{eq:fermionischeenergiedichtenersterordnung}.
\end{equation}
The total fermionic contribution to the energy density $\epsilon_f=\epsilon_\text{tot}^++\epsilon_\text{tot}^-$ finally reads
\begin{align}
\epsilon_f &=\left(m_N+\frac{\delta^2}{2m_N}+\frac{1}{2}g_A\beta\right)n_++\frac{(3\pi^2 n_+)^{5/3}}{10\pi^2 m_N}+\left(m_N+\frac{\delta^2}{2m_N}-\frac{1}{2}g_A\beta\right)n_-+\frac{(3\pi^2 n_-)^{5/3}}{10\pi^2 m_N}\nonumber\\
&=\left(m_N+\frac{\delta^2}{2m_N}\right)n+\frac{g_A\beta}{2}\left(n_+-n_-\right)+\frac{(3\pi^2)^{5/3}}{10\pi^2m_N}\left(n_+^{5/3}+n_-^{5/3}\right)\label{eq:fermionicenergy},
\end{align}
where in the last step we set the total particle number density $n=n_++n_-$.
\section{The Total Energy Density}\label{sec:thetotalenergydensity}
The total energy density is given by this fermionic part and the pionic part of (\ref{eq:pionicenergydensity}). Assuming for simplicity the isospin limit $m_u=m_d$ we can use (\ref{eq:masse1}). Together with (\ref{eq:defbetai}) and (\ref{eq:defdeltai}) and remembering that $n_++n_-=n$ fixed and so $n_-=n-n_+$ we then get for the total energy density the following expression:
\begin{multline}
\epsilon_\text{tot}= \epsilon_f+\epsilon_p=\left(m_N+\frac{a^2\cos^2\alpha_0}{8m_N}\right)n-\frac{g_A}{2}a\sin\alpha_0(n-2n_+)\\
+\frac{(3\pi^2)^{5/3}}{10\pi^2m_N}\left(n_+^{5/3}+(n-n_+)^{5/3}\right)+\frac{F^2}{2}a^2\sin^2\alpha_0-F^2M_\pi^2\cos\alpha_0.\label{eq:etotvona}
\end{multline}
Let us find the value of $a$ that minimizes $\epsilon_\text{tot}$:
\begin{eqnarray}
\frac{\partial\epsilon_\text{tot}}{\partial a}=\frac{an\cos\alpha_0}{4m_N}-\frac{g_A}{2}\sin\alpha_0(n-2n_+)+F^2a\sin^2\alpha_0\stackrel{!}{=}0\nonumber\\
\Rightarrow\quad a_\text{min}=\frac{2m_Ng_A\sin\alpha_0(n-2n_+)}{n\cos^2\alpha_0+4m_NF^2\sin^2\alpha_0}.\label{eq:aminvonetot}
\end{eqnarray}
The second derivative is always positive for $0\leq\alpha_0\leq\pi/2$, showing that we have found a minimum. After plugging $a_\text{min}$ into $\epsilon_\text{tot}$ the first derivative with respect to $\alpha_0$ reads
\begin{equation}
	\frac{\partial\epsilon_\text{tot}}{\partial\alpha_0}=F^2M_\pi^2\sin\alpha_0-\frac{2g_A^2 m_Nn(n-2n_+)^2\sin 2\alpha_0}{(4F^2m_N+n+(n-4F^2m_N)\cos 2\alpha_0)^2}.\label{eq:ableitungnachalpha0}
\end{equation}
For a minimum this expression needs to vanish and we get
\begin{equation}
	F^2M_\pi^2\sin\alpha_0\left(4F^2m_N+n+(n-4F^2m_N)\cos 2\alpha_0\right)^2=2g_A^2m_Nn(n-2n_+)^2\sin 2\alpha_0.\label{eq:winkeleinstellen}
\end{equation}
This equation determines how the angle $\alpha_0$ will adjust itself for given densities $n$ and $n_+$.
\section{Chiral Limit}\label{sec:ersteordnungchiralerlimes}
\begin{figure}[h!]
\centering
\psfrag{eps}{{\scriptsize $10^4\cdot\epsilon_\text{tot}^*$}}
\psfrag{npdens}{{\scriptsize $n_+/n$}}
\hspace{1mm}
\subfloat[$n<\mathring{n}_c$]{\includegraphics[width=0.285\textwidth]{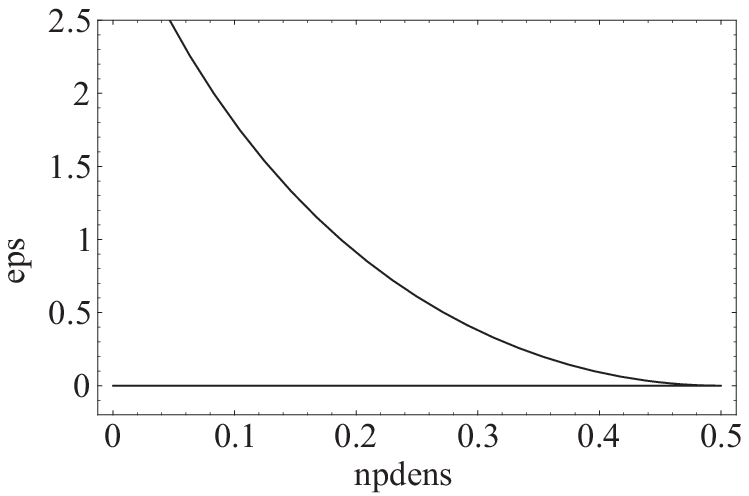}\label{fig:plot70000}}
\hspace{6mm}
\psfrag{eps}{{\scriptsize $10^5\cdot\epsilon_\text{tot}^*$}}
\subfloat[$\mathring{n}_c<n<\mathring{n}_c^u$]{\includegraphics[width=0.285\textwidth]{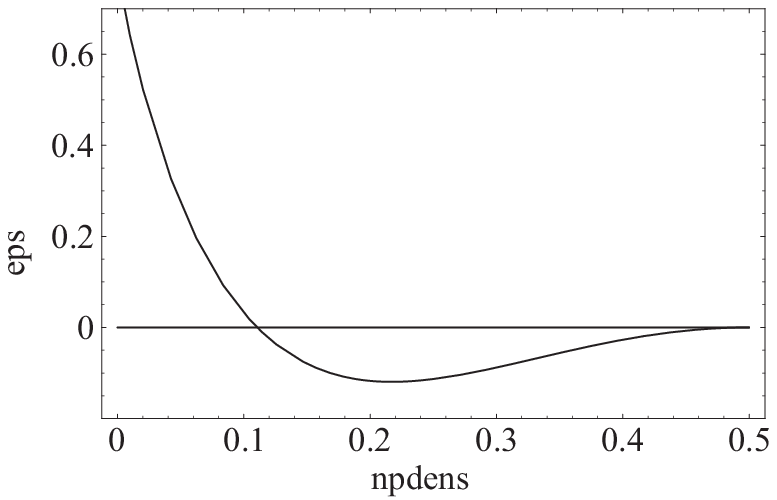}\label{fig:plot110000}}
\hspace{7.5mm}
\psfrag{eps}{{\scriptsize $10^4\cdot\epsilon_\text{tot}^*$}}
\subfloat[$n>\mathring{n}_c^u$]{\includegraphics[width=0.28\textwidth]{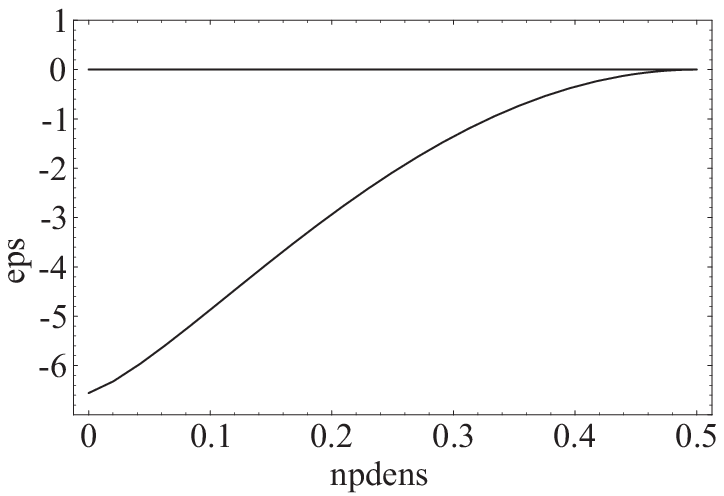}\label{fig:plot190000}}
\caption{In the upper line $\partial\epsilon_\text{tot}/\partial n_+$ is plotted against $n_+/n$. The lower line shows $\epsilon_\text{tot}^*(n_+)=\frac{\epsilon_\text{tot}(n_+)}{\epsilon_\text{tot}(n_+=n/2)}-1$, multiplied by an appropriate constant indicated in the plot. The horizontal line in the lower plots marks $\epsilon_\text{tot}^\text{ns}$ and shows that for $n>\mathring{n}_c$ a spiral configuration is favored energetically.}
\end{figure}

Let us now consider the chiral limit, i.e. set $M_\pi=0$. Then the left hand side of (\ref{eq:winkeleinstellen}) vanishes and for $n_+\neq n/2$ we conclude\footnote{For $n_+=n/2$ we have $a=0$ according to (\ref{eq:aminvonetot}). However, this implies a non-spiral configuration, see the remark preceding (\ref{eq:nonspiralenergydensity}).} $\sin 2\alpha_0=0$. In the interval $0\leq\alpha_0\leq\pi/2$ this has two solutions; $\alpha_0=0$ and $\alpha_0=\pi/2$. Let us examine the second derivative of $\epsilon_\text{tot}$ with respect to $\alpha_0$ at $\alpha_0=\pi/2$:
\begin{equation}
	\frac{\partial^2\epsilon_\text{tot}}{\partial\alpha_0^2}\Bigr|_{\alpha_0=\frac{\pi}{2}}=\frac{g_A^2n(n-2n_+)^2}{16F^4m_N}\label{eq:ersteordnungzweiteablnachalpha0beipizweitel}
\end{equation}
This is always positive (except for $n_+=n/2$ where it vanishes) and thus the minimum of $\epsilon_\text{tot}$ is at $\alpha_0=\pi/2$. So, in the chiral limit the vector representing the pions will move in the equatorial plane. There (\ref{eq:defbetai}) and (\ref{eq:defdeltai}) become $\delta_i=0$ and $\beta_i=-a_i$, hence $a^2=\beta^2$. We therefore get
\begin{equation}
	\epsilon_p=\frac{F^2}{2}a^2=\frac{F^2}{2}\beta^2
\end{equation}
and
\begin{equation}
	\epsilon_f=m_Nn-\frac{g_A\beta}{2}(n-2n_+)+\frac{(3\pi^2)^{5/3}}{10\pi^2 m_N}\left(n_+^{5/3}+(n-n_+)^{5/3}\right).
\end{equation}
Let us find the value for $\beta$ such that $\epsilon_\text{tot}=\epsilon_p+\epsilon_f$ gets minimized:
\begin{equation}
	\frac{\partial\epsilon_\text{tot}}{\partial\beta}=F^2\beta-\frac{g_A}{2}(n-2n_+)\stackrel{!}{=}0\quad\Rightarrow\quad\beta=\frac{g_A}{2F^2}(n-2n_+)\label{eq:betamin}
\end{equation}
Since $\beta\geq 0$ we must have $n_+\leq n/2$, which makes sense as the lower energy states $E_-$ are energetically favored and $n_+>n/2$ would only increase the energy density. Plugging the expression for $\beta$ into $\epsilon_\text{tot}$ yields
\begin{equation}	\epsilon_\text{tot}=m_Nn-\frac{gA^2}{8F^2}(n-2n_+)^2+\frac{(3\pi^2)^{5/3}}{10\pi^2 m_N}\left(n_+^{5/3}+(n-n_+)^{5/3}\right)\label{eq:totaleenergiedichterewritten}.
\end{equation}
At a fixed total density $n$ we now vary $n_+$ in order to find extrema of the total energy density:
\begin{equation}
	\frac{\partial\epsilon_\text{tot}}{\partial n_+}=\frac{(3\pi^2)^{5/3}}{6\pi^2m_N}\left(n_+^{2/3}-(n-n_+)^{2/3}\right)-\frac{g_A^2}{2F^2}(2n_+-n)\stackrel{!}{=}0\label{eq:nplusnminusbedingung}.
\end{equation}
From a plot of $\partial\epsilon_\text{tot}/\partial n_+$ (see fig. \ref{fig:plot70000}) one finds that for small $n$ this equation has the only solution $n_+=n_-=n/2$. The second derivative at $n_+=n/2$ reads
\begin{equation}
	\frac{\partial^2\epsilon_\text{tot}}{\partial n_+^2}\Bigr|_{n_+=\frac{n}{2}}=\frac{2(3\pi^2)^{5/3}}{9\pi^2m_N}\left(\frac{2}{n}\right)^{1/3}-\frac{g_A^2}{F^2},
\end{equation}
which is positive for
\begin{equation}
	n<\mathring{n}_c=\frac{16F^6\pi^4}{3g_A^6 m_N^3}\label{eq:krit1}.
\end{equation}
The symbol $\circ$ reminds us that we are in the chiral limit.

Thus, for densities $n$ smaller than this value $n_+=n_-=n/2$ is the only minimum of $\epsilon_\text{tot}$. In this case we have, using (\ref{eq:betamin}), $\beta=0$. But since $\beta=a=|\vec{a}|$ it follows that $\vec{a}=0$ and so $\varphi(\vec{x})=\varphi_0$ (see (\ref{eq:spiralconfiguration})). In this case we therefore do not have a spiral. The energy density of this non-spiral configuration is
\begin{equation}
	\epsilon_\text{tot}^\text{ns}=m_Nn+\frac{(3\pi^2n)^{5/3}}{5\cdot 2^{5/3}\pi^2m_N}=m_Nn+\frac{3}{10m_N}\left(\frac{9}{4}\pi^4n^5\right)^{1/3}\label{eq:nonspiralenergydensity}.
\end{equation}
\begin{figure}[h!]
\centering
\psfrag{abl}{{\scriptsize $\partial\epsilon_\text{tot}/\partial n_+$}}
\psfrag{npdens}{{\scriptsize $n_+/n$}}
\hspace{0.01mm}
\subfloat{\includegraphics[width=0.3\textwidth]{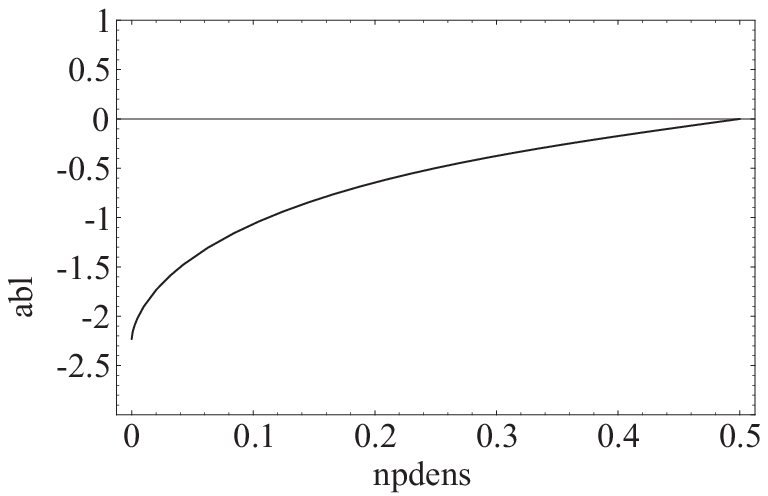}}
\hspace{4mm}
\subfloat{\includegraphics[width=0.3\textwidth]{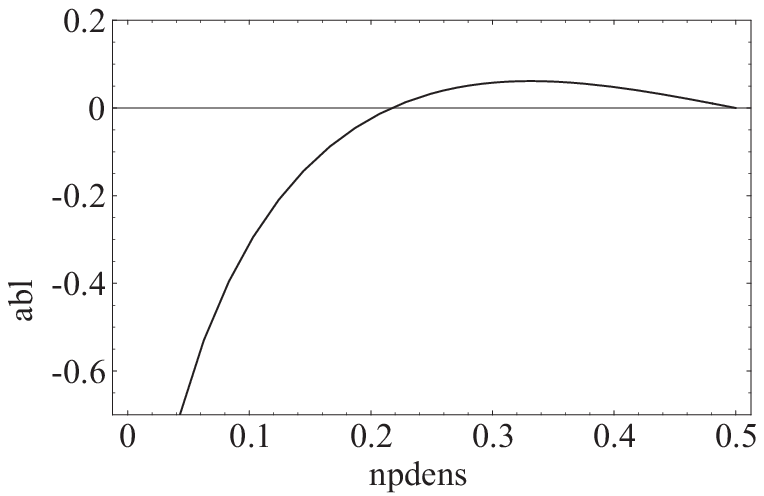}}
\hspace{5mm}
\subfloat{\includegraphics[width=0.3\textwidth]{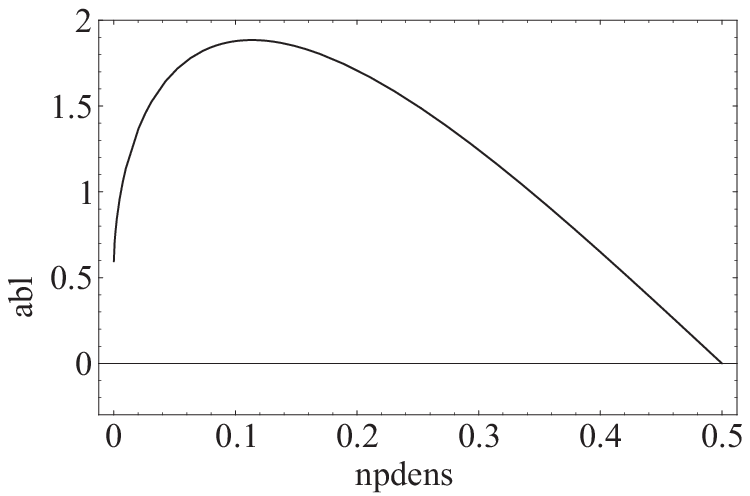}}
\end{figure}
\\
However, for densities $n>\mathring{n}_c$ an interesting thing happens: Since in this case the second derivative of $\epsilon_\text{tot}$ at $n_+=n/2$ becomes negative, $n_+=n/2$ becomes a maximum. Additionally, (\ref{eq:nplusnminusbedingung}) now has two roots (see fig. \ref{fig:plot110000}), where, since the one at $n_+=n/2$ is a maximum, the other is a minimum.

With increasing $n$ the additional root of (\ref{eq:nplusnminusbedingung}) moves to the left of the interval $0\leq n_+\leq n/2$ and eventually leaves it. The ``upper critical density'' $\mathring{n}_c^u$ where this happens is found by solving $\partial\epsilon_\text{tot}/\partial n_+(0)=0$ for $n$, which yields
\begin{equation}
	\mathring{n}_c^u=\frac{9F^6\pi^4}{g_A^6m_N^3}\label{eq:uppercritdens}.
\end{equation}
For densities $n>\mathring{n}_c^u$ $\partial\epsilon_\text{tot}/\partial n_+$ looks like in fig. \ref{fig:plot190000} and we do not have a local minimum in the interval $0\leq n_+\leq n/2$ anymore but instead a global one at $n_+=0$. Hence, $\mathring{n}_c^u$ is the density from where on $E_-$ is completely filled.

The relevant fact is that for $n>\mathring{n}_c$ the energy density gets lower than $\epsilon_\text{tot}^\text{ns}$ for some $n_+\neq n/2$ and hence for a spiral configuration.

For the calculation of the numerical value of $\mathring{n}_c$ we use, as mentioned in the introduction, the values of the low energy constants in the chiral limit, as well as the physical ones. The value of the pion decay constant in the chiral limit was found in \cite{Colangelo:2003hf}, we take $F=86\,\text{MeV}$, while the physical value is taken to be $F=93\,\text{MeV}$. The nucleon mass $\mathring{m}_N$ in the chiral limit was calculated in \cite{Procura:2006bj}; we set $\mathring{m}_N=882\,\text{MeV}$. For the physical nucleon mass we choose $m_N=939\,\text{MeV}$ (neutron mass). Finally, the axial-vector coupling constant $\mathring{g}_A$ in the chiral limit was extrapolated in \cite{Procura:2006gq} with a rather large uncertainty. We take $\mathring{g}_A=1.20$, while the physical value is more precisely known; we take $g_A=1.26$.

If $\mathring{n}_c$ denotes the critical particle number density in the chiral limit, (\ref{eq:krit1}) then yields
\begin{equation}
	\mathring{n}_c=102.6\,\text{GeV}^3\quad\text{or}\quad \mathring{n}_c^{1/3}=46.81\,\text{MeV}\label{eq:erstekritdichte}
\end{equation}
when using the values of the low energy constants in the chiral limit and
\begin{equation}
	\mathring{n}_c=101.5\,\text{GeV}^3\quad\text{or}\quad \mathring{n}_c^{1/3}=46.64\,\text{MeV}
\end{equation}
when calculating with the physical low energy constants. We see that the difference between these two results is quite small, i.e. the value of the critical particle density does not strongly depend on uncertainties in the low energy constants. This will also be true for all the further results and for this reason we will quote the result obtained using the low energy constants in the chiral limit only.

Let us compare the value in (\ref{eq:erstekritdichte}) to the nuclear matter density. The radius of a typical nucleus of an atom with $A$ nucleons is $R\approx A^{1/3}r_0$, where $r_0\approx 1.3\,\text{fm}$. With
	\[A=\frac{4\pi}{3}nR^3
\]
it follows for the nuclear matter density
\begin{equation}
	n_\text{nuclear}=\frac{3}{4\pi r_0^3}=0.11\,\text{fm}^{-3}\label{eq:nuclearmatterdensityinfm},
\end{equation}
or, using $1\,\text{fm}=1/197.3\,\text{MeV}^{-1}$,
\begin{equation}
	n_\text{nuclear}^{1/3}\approx 95\,\text{MeV}.
\end{equation}
Hence, at least in the chiral limit, we can expect a spiral configuration to occur already at densities smaller than ordinary nuclear matter.
\section{Away from the Chiral Limit}\label{sec:awayfromthechirallimit}
As soon as we allow the quarks to bear masses, $M_\pi$ will no longer be zero and the vector representing the pions will no longer move in the equatorial plane. Rather, $\alpha_0$ will adjust according to (\ref{eq:winkeleinstellen}).

Plugging (\ref{eq:aminvonetot}) into (\ref{eq:etotvona}) and varying the resulting expression with respect to $n_+$ leads to the equation
\begin{multline}
5(9\pi^4)^{1/3}n\cos^2\alpha_0\left(n_+^{2/3}-(n-n_+)^{2/3}\right)\\
+4m_N\sin^2\alpha_0\left(g_A^2m_N(n-2n_+)+(9\pi^4)^{1/3}F^2\left(n_+^{2/3}-(n-n_+)^{2/3}\right)\right)=0.\label{eq:nonchiralextr}
\end{multline}
Again, this equation is always solved by $n_+=n/2$. According to (\ref{eq:aminvonetot}) this means $a=0$ and hence a non-spiral configuration.

Since we have two parameters now, $n_+$ and $\alpha_0$, there does not seem to exist a straightforward way to analytically determine a critical fermionic density from where on a spiral configuration ($n_+\neq n/2$) is energetically favored. Hence, we approach this problem numerically.

\begin{figure}[h]
\centering
\psfrag{a}{\scriptsize $\alpha_0$}
\psfrag{np}{\scriptsize $n_+/n$}
\psfrag{d}{\scriptsize $d\cdot 10^{-3}\,[\text{MeV}]$}
\includegraphics[scale=0.8]{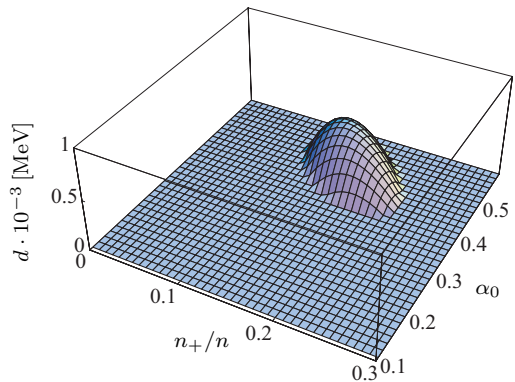}
\caption{$d(n_+,\alpha_0)$ for $M_\pi=2\,\text{MeV}$ and $n^{1/3}=49.7\,\text{MeV}$}\label{fig:3dplot}
\end{figure}

The energy density of reference we aim to undercut is the one of the non-spiral-configuration where $n_+=n/2$ and $a=0$. From (\ref{eq:etotvona}) we find
\begin{equation}
	\epsilon_\text{tot}\bigr|_{a=0,n_+=n/2}=m_Nn+\frac{3}{10m_N}\left(\frac{9}{4}\pi^4n^5\right)^{1/3}-F^2M_\pi^2\cos\alpha_0\label{eq:energydensitymitanullundnpgleichnzweitel},
\end{equation}
which is minimized by $\alpha_0=0$. The non-spiral energy density therefore reads
\begin{equation}
	\epsilon_\text{tot}^\text{ns}=m_Nn+\frac{3}{10m_N}\left(\frac{9}{4}\pi^4n^5\right)^{1/3}-F^2M_\pi^2\label{firstordernonspiralenergynonchiral}.
\end{equation}
If we take the total energy density as a function of the two parameters, $\epsilon_\text{tot}(n_+,\alpha_0)$, then the interesting quantity to look at is
\begin{equation}
	d(n_+,\alpha_0)\mathrel{\mathop:}=\epsilon_\text{tot}^\text{ns}-\epsilon_\text{tot}(n_+,\alpha_0)=\epsilon_\text{tot}(n/2,0)-\epsilon_\text{tot}(n_+,\alpha_0)\label{eq:differenzspiralnonspiral},
\end{equation}
which denotes the difference of energy densities between the non-spiral configuration and a particular configuration with parameters $n_+$ and $\alpha_0$. Whenever $d(n_+,\alpha_0)<0$, the non-spiral configuration will be energetically favored. But as soon as we can find a configuration such that $d(n_+,\alpha_0)>0$, this configuration will have $n_+\neq n/2$, since (\ref{eq:energydensitymitanullundnpgleichnzweitel}) tells us that for $n_+=n/2$ the parameter $\alpha_0$ will go to zero and hence $d=0\ngtr 0$. This configuration will therefore be a spiral configuration and will have a lower total energy density than the non-spiral one. The fermionic density where this is possible for exactly one configuration will be our critical density $n_c$. Let us refer to the corresponding parameters as the critical ones; $n_+^c$ and $\alpha_0^c$. Fig. \ref{fig:3dplot} shows the function $d(n_+,\alpha_0)$ for $M_\pi=2\,\text{MeV}$ and $n^{1/3}=49.7\,\text{MeV}$, using the low energy constants in the chiral limit.
\\ 
Obviously we are in this case already above the critical density since $d(n_+,\alpha_0)>0$ for a whole region in the parameter space.

\begin{figure}[h!]
\centering
\hspace{0.01mm}
\psfrag{a}{\scriptsize $\alpha_0$}
\psfrag{np}{\scriptsize $n_+/n$}
\subfloat[][$\begin{aligned} M_\pi &= 2\,\text{MeV}\\ n^{1/3} &= 49.7\,\text{MeV} \end{aligned}$]{\includegraphics[width=0.3\textwidth]{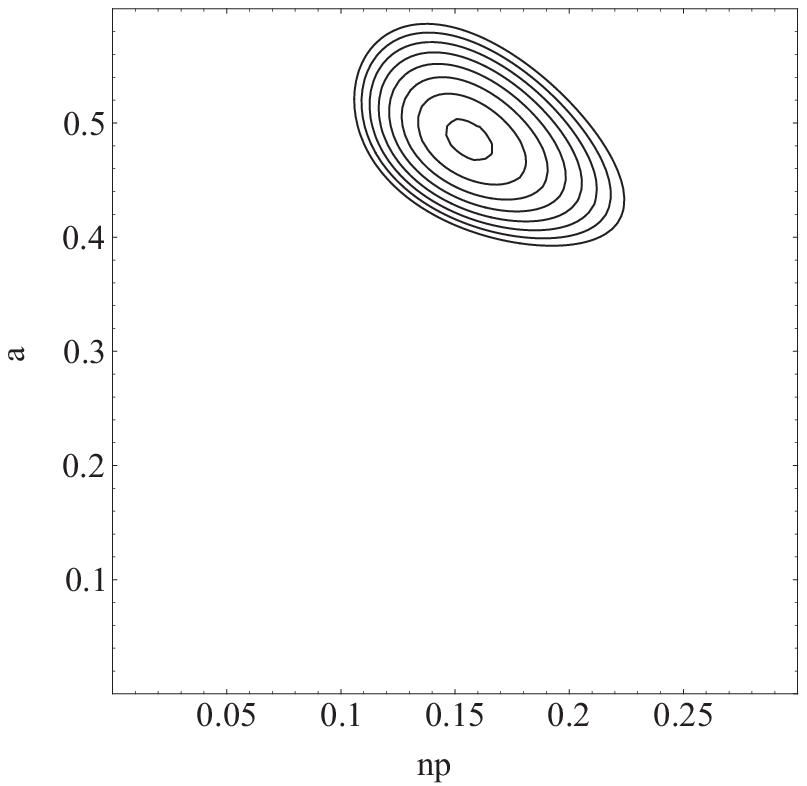}\label{fig:erstercontourplot}}
\hspace{4mm}
\subfloat[][$\begin{aligned} M_\pi &= 8\,\text{MeV}\\ n^{1/3} &= 54.25\,\text{MeV} \end{aligned}$]{\includegraphics[width=0.3\textwidth]{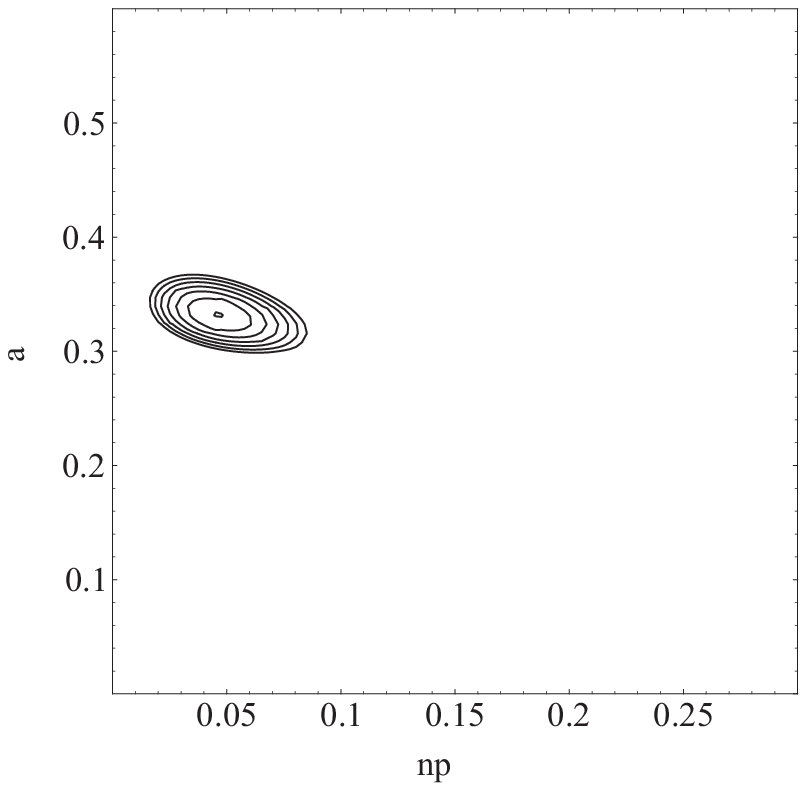}}
\hspace{5mm}
\subfloat[][$\begin{aligned} M_\pi &= 18\,\text{MeV}\\ n^{1/3} &= 60\,\text{MeV} \end{aligned}$]{\includegraphics[width=0.3\textwidth]{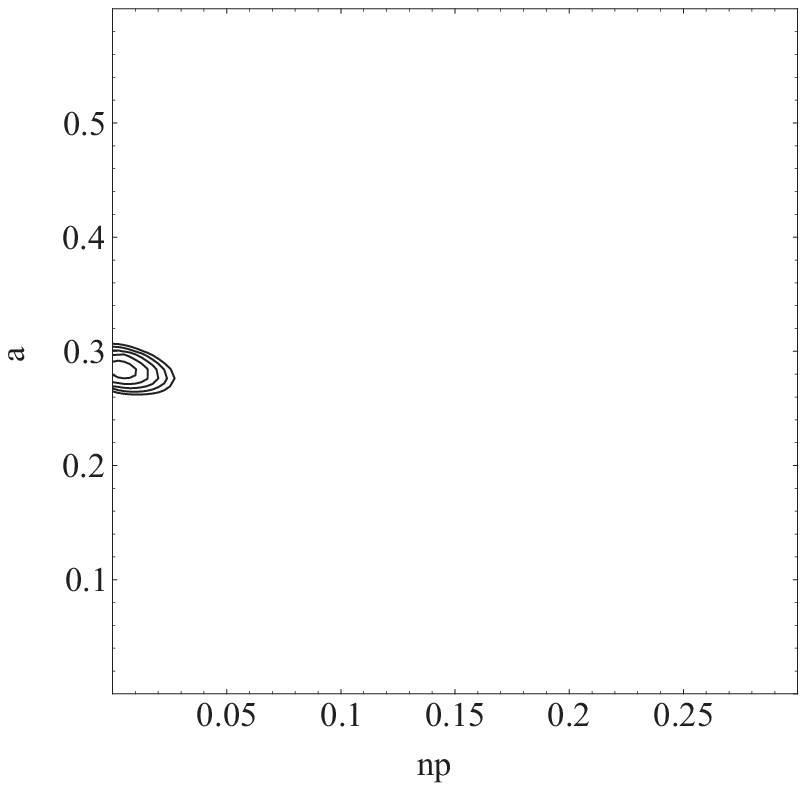}}
\caption{Contour plots of $d(n_+,\alpha_0)$ over the parameter space for different pion masses. The fermionic densities are chosen to be slightly above the critical ones.}\label{fig:contourplots}
\end{figure}

A first observation is the following: For small values of $M_\pi$ the critical parameter $n_+^c$ will lie somewhere in the interval $0<n_+^c<n/2$. For increasing values it will move towards lower values and eventually reach $n_+^c=0$ for a certain pion mass $M_\pi^c$. This is visualized in fig. \ref{fig:contourplots} where contour plots of $d(n_+,\alpha_0)$ for different values of $M_\pi$ just above the critical density $n_c$ are shown. For comparison: Fig. \ref{fig:erstercontourplot} shows the same	situation as fig. \ref{fig:3dplot}.
From this plots the above statement is obvious. For pion masses larger than $M_\pi^c\approx 20\,\text{MeV}$ the critical parameter $n_+^c$ drops down to zero. For an investigation including physical pion masses we can therefore clearly constrain our analysis to the $n_+=0$ parameter subspace.

Although we now have just one parameter $\alpha_0$ to consider, the problem of finding the critical fermionic density $n_c$ from where on $\epsilon_\text{tot}(0,\alpha_0)<\epsilon_\text{tot}(n/2,0)$ still seems to be nontrivial (since the condition $\partial\epsilon_\text{tot}(0,\alpha_0)/\partial\alpha_0=0$ is complicated to solve for $\alpha_0$). We therefore go on numerically in the following way: For each value of $n$ in a given interval we numerically calculate the maximum value of $d(0,\alpha_0)$, which we call $d_\text{max}(n)$. As long as $d_\text{max}<0$, the non-spiral configuration will have a lower energy density. For large enough values of $n$, we will however find $d_\text{max}>0$, which means, that the spiral configuration will appear. The critical density $n_c$ is therefore just the root of $d_\text{max}(n)$.

\begin{figure}[h]
\centering
\psfrag{nn}{$n^{1/3}\,[\text{MeV}]$}
\psfrag{d}{$d_\text{max}\cdot 10^{-5}\,[\text{MeV}]$}
\subfloat[][$M_\pi=135\,\text{MeV}$]{\includegraphics[width=0.4\textwidth]{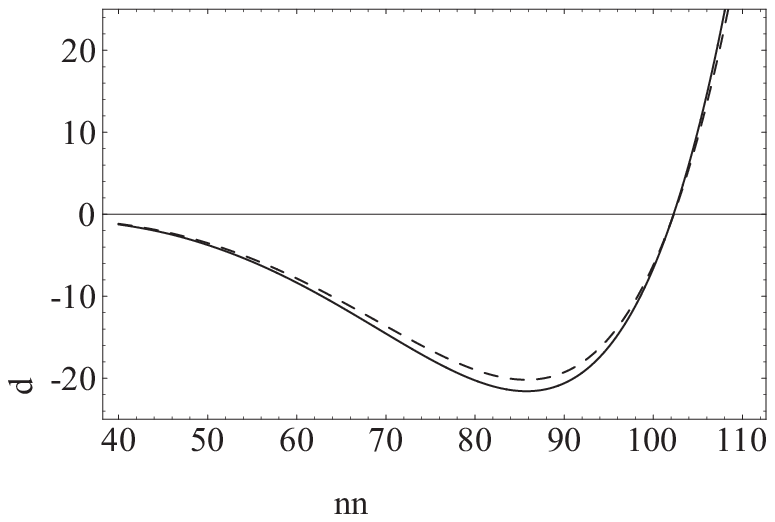}\label{fig:fmin}}
\hspace{8mm}
\subfloat[][$M_\pi=40\,\text{MeV}$]{\includegraphics[width=0.4\textwidth]{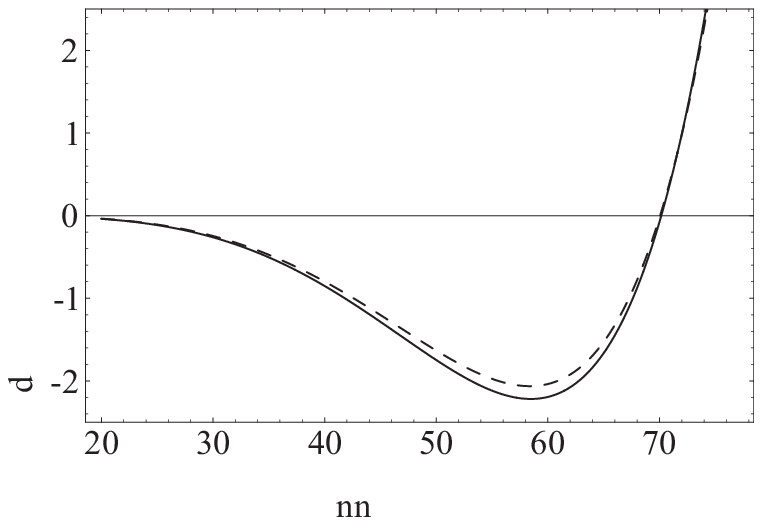}\label{fig:fmin_40}}
\caption{The maximum values of $d(0,\alpha_0)$ as a function of n for physical pion mass as well as for $M_\pi=40\,\text{MeV}$. Continuous graph: Low energy constants in the chiral limit; Dashed graph: Physical low energy constants.}
\end{figure}

The plot in fig. \ref{fig:fmin} shows $d_\text{max}(n)$, numerically found by Mathematica, using a physical pion mass of $M_\pi=135\,\text{MeV}$. The continuous graph was obtained by using the low energy constants in the chiral limit, while the dashed graph shows $d_\text{max}(n)$ using the physical values (again showing the weak dependence of the result on variations of the low energy constants). For small values of $n$ $d_\text{max}$ is small but negative, actually $d_\text{max}\to 0$ as $n\to 0$. This is consistent, since at small densities the system will increase its total energy density only slightly when choosing a spiral configuration instead of a homogeneous one. For increasing particle density this effect however becomes stronger, i.e. the difference between $\epsilon_\text{tot}$ and $\epsilon_\text{tot}^\text{ns}$ becomes larger. However, for densities larger than around $85^3\,\text{MeV}^3$ the difference starts to shrink again since the lowering of the fermionic energy due to the spiral configuration (that increases the energy in the pionic sector) starts to carry weight. For densities higher than around $100^3\,\text{MeV}^3$ the decrease of the fermionic energy becomes larger than the increase of the pionic energy, i.e. the total energy density is lowered for a spiral configuration and $d_\text{max}$ becomes positive. The root, i.e. the critical density is found to be located at
\begin{equation}
	n_c=1069.6\,\text{GeV}^3\quad\text{or}\quad n_c^{1/3}=102.27\,\text{MeV}.\label{eq:lowerorderresult}
\end{equation}

Beside vanishing and physical pion mass, let us find the critical density additionally for some intermediate pion mass. We take $M_\pi=40\,\text{MeV}$, which is considerably lower than the physical mass but still higher than the critical mass of about $20\,\text{MeV}$ from where on we can set $n_+=0$. Fig. \ref{fig:fmin_40} shows again the function $d_\text{max}(n)$. The critical density (let us denote it by $\bar{n}_c$) is
\begin{equation}
	\bar{n}_c=345.4\,\text{GeV}^3\quad\text{or}\quad \bar{n}_c^{1/3}=70.16\,\text{MeV}.
\end{equation}

\newpage

\chapter{Higher Orders}
The calculations in chapter \ref{ch:firstapproach} showed that the critical particle number density for the spiral configuration to occur lies in the region of nuclear matter density. It is not obvious that we can still trust ChPT at such high fermionic densities. In this chapter we therefore want to take higher orders into account and check their influence on the results. First, we continue the expansion of the relativistic expression for the fermionic one-particle energy up to $\mathcal{O}(m_N^{-2})$. This will point out the so far uncared problem that the expansion in the way we do it is not valid for all values of the spiral parameters. After having realized that the simple continuation of the energy expansion will yield no results, we perform the expansion explicitly for small values of the parameter $\alpha_0$.

Next, we take higher orders of the baryonic Lagrangian into account, breaking chiral symmetry explicitly also in this sector.
\section{Higher Orders in the Energy Expansion}
First, let us examine how the above results will be affected if we continue the non-relativistic expansion (\ref{eq:energie}) of $E_\pm$ up to order $1/m_N^2$. We then have
\begin{multline}
	E_\pm=m_N+\frac{p^2}{2m_N}+\frac{\delta^2}{2m_N}\pm\frac{1}{2}g_A\beta\\
	\pm\frac{1}{4g_A\beta m_N^2}\left\{4(\vec{p}\cdot\vec{\delta})^2+g_A^2\left[(\vec{p}\cdot\vec{\beta})^2-\beta^2(p^2+\delta^2)\right]\right\}+\mathcal{O}\left(\frac{1}{m_N^3}\right).\label{eq:energiezweiteordnung}
\end{multline}
The first term proportional to $m_N^{-2}$ shows a severe problem: The $\beta$ in the denominator may get small for small values of $\alpha_0$ and the expansion fails as the term explodes. This will be analyzed further, see below.

Unlike $E_\pm$ of (\ref{eq:energie}) the energies now not only depend on the absolute value $p$ of the fermion momentum, but also on its direction. The Euclidean scalar products $\vec{p}\cdot\vec{\delta}=p_i\delta_i$ and $\vec{p}\cdot\vec{\beta}=p_i\beta_i$ may be rewritten in terms of the absolute values of $\vec{p}$, $\vec{\delta}$ and $\vec{\beta}$ and the angles $\angle(\vec{p},\vec{\delta})$ and $\angle(\vec{p},\vec{\beta})$ respectively, which turn out to be the same, namely $\angle(\vec{p},\vec{a})$, because of (\ref{eq:defbetai}) and (\ref{eq:defdeltai}). Denoting this angle by $\theta$ and choosing the coordinate system such that the third axis points into the direction of $\vec{a}$, we have
\begin{align}
\vec{p}\cdot\vec{\delta} &= p\delta\cos\theta=p_3\delta\label{eq:skalarproduktpdelta}\\
\vec{p}\cdot\vec{\beta} &= p\beta\cos\theta=p_3\beta\label{eq:skalarproduktpbeta}.
\end{align}
Since $E_\pm$ is now not only a function of $p^2$ but also of $\theta$ or $p_3$ explicitly, the Fermi surface is not a sphere anymore. In order to see its shape in the momentum space, let us rewrite (\ref{eq:energiezweiteordnung}) as
\begin{equation}
	E_\pm=w_1^\pm+w_2^\pm p^2+w_3^\pm p_3^2\label{eq:energiezweiteordnungkurz},
\end{equation}
where
\begin{align}
w_1^\pm &= m_N+\frac{\delta^2}{2m_N}\pm\frac{1}{2}g_A\beta\mp\frac{g_A\beta\delta^2}{4m_N^2}\label{eq:defk1pm}\\
w_2^\pm &= \frac{1}{2m_N}\mp\frac{\beta g_A}{4 m_N^2}\label{eq:defk2pm}\\
w_3^\pm &= \pm\frac{4\delta^2+g_A^2\beta^2}{4g_A\beta m_N^2}\label{eq:defk3pm}
\end{align}
are constants with respect to $p_i$. Fixing now $E_\pm$ to certain values (e.g. the Fermi energies) $E_F^\pm$, we may write (\ref{eq:energiezweiteordnungkurz}) as
\begin{equation}
	\frac{w_2^\pm}{E_F^\pm-w_1^\pm}p_1^2+\frac{w_2^\pm}{E_F^\pm-w_1^\pm}p_2^2+\frac{w_2^\pm+w_3^\pm}{E_F^\pm-w_1^\pm}p_3^2=1\label{eq:fermiellipsoid},
\end{equation}
which is the equation of an ellipsoid in momentum space. A general ellipsoid
\begin{equation}
	\frac{p_1^2}{a^2}+\frac{p_2^2}{b^2}+\frac{p_3^2}{c^2}=1
\end{equation}
has the volume
\begin{equation}
	V_\text{ellipsoid}=\frac{4\pi}{3}abc,
\end{equation}
so the volume of our Fermi ellipsoid is
\begin{equation}
	V_\text{FE}=\frac{4\pi(E_F^\pm-w_1^\pm)^{3/2}}{3w_2^\pm\cdot\sqrt{w_2^\pm+w_3^\pm}}.
\end{equation}
When calculating the particle numbers $N_\pm$ (cf. (\ref{eq:particlenumberssphere})) and the total energies $E^\pm_\text{tot}$ (cf. (\ref{eq:totaleenergiepm})) we have to integrate over this Fermi ellipsoid. Let us first find the particle number:
\begin{equation}
	N_\pm=2\left(\frac{L}{2\pi}\right)^3\int\limits_{\hidewidth\substack{\text{Fermi}\\\text{ellipsoid}}\hidewidth}\! d^3p=2\left(\frac{L}{2\pi}\right)^3V_\text{FE}=\frac{L^3(E_F^\pm-w_1^\pm)^{3/2}}{3\pi^2w_2^\pm\cdot\sqrt{w_2^\pm+w_3^\pm}}\label{eq:particlenumberszweiteordnung}
\end{equation}
In order to calculate the total energy of all fermions of a given sort, we have to integrate $E_\pm$ of (\ref{eq:energiezweiteordnungkurz}) over the Fermi ellipsoid. From (\ref{eq:fermiellipsoid}) we see that the integration variable transformation
\begin{align}
p_1^2 &\rightarrow p_1^2\nonumber\\
p_2^2 &\rightarrow p_2^2\nonumber\\
p_3^2 &\rightarrow p_3'^2=\frac{w_2^\pm+w_3^\pm}{w_2^\pm}p_3^2
\end{align}
maps the ellipsoid to a sphere $S$ of radius
\begin{equation}
	r_{S\pm}=\sqrt{\frac{E_F^\pm-w_1^\pm}{w_2^\pm}}\label{eq:radiustransformiertesfermiellipsoid},
\end{equation}
which is an integration region we can easily deal with by changing to spherical coordinates; $p_1=p\sin\theta\cos\phi$, $p_2=p\sin\theta\sin\phi$, $p_3=p\cos\theta$. $r_{S\pm}$ is the analogon of $p_{F\pm}$ in section \ref{sec:fermioniccontributiontotheenergy}.
\begin{align}
E_\text{tot}^\pm &= 2\left(\frac{L}{2\pi}\right)^3\int\limits_{\hidewidth\substack{\text{Fermi}\\\text{ellipsoid}}\hidewidth}\!d^3p\, E_\pm(\vec{p})=2\left(\frac{L}{2\pi}\right)^3\int\limits_{\hidewidth\substack{\text{Fermi}\\\text{ellipsoid}}\hidewidth}\!d^3p\, \left(w_1^\pm+w_2^\pm p^2+w_3^\pm p_3^2\right)\nonumber\\
&= w_1^\pm\cdot\underbrace{2\left(\frac{L}{2\pi}\right)^3\int\limits_{\hidewidth\substack{\text{Fermi}\\\text{ellipsoid}}\hidewidth}\!d^3p}_{N_\pm}+2\left(\frac{L}{2\pi}\right)^3\int\limits_{\hidewidth\substack{\text{Fermi}\\\text{ellipsoid}}\hidewidth}\!d^3p\,\left(w_2^\pm p^2+w_3^\pm p_3^2\right)\nonumber\\
&= w_1^\pm N_\pm+2\left(\frac{L}{2\pi}\right)^3\sqrt{\frac{w_2^\pm}{w_2^\pm+w_3^\pm}}\int\limits_S\!d^3p'\left(w_2^\pm p_1'^2+w_2^\pm p_2'^2+\left(w_2^\pm+w_3^\pm\right)\frac{w_2^\pm}{w_2^\pm+w_3^\pm} p_3'^2\right)\nonumber\\
&= w_1^\pm N_\pm+2\left(\frac{L}{2\pi}\right)^3 w_2^\pm\sqrt{\frac{w_2^\pm}{w_2^\pm+w_3^\pm}}\int\limits_S\!d^3p'p'^2\nonumber\\
&= w_1^\pm N_\pm+\frac{L^3}{5\pi^2}r_{S\pm}^5 w_2^\pm\sqrt{\frac{w_2^\pm}{w_2^\pm+w_3^\pm}}.\label{eq:totenergiemitrs}
\end{align}
Now we replace $r_{S\pm}$ such that we have the total energy in terms of the particle number (density), in analogy to what we did in section \ref{sec:fermioniccontributiontotheenergy}. Dividing (\ref{eq:particlenumberszweiteordnung}) by $L^3$ we get the particle number densities $n_\pm$ which we may solve for $E_F^\pm-w_1^\pm$. Plugging this expression into (\ref{eq:radiustransformiertesfermiellipsoid}), we find
\begin{equation}
	r_{S\pm}=w_2^{\pm -1/6}\left(3\pi^2 n_\pm\cdot\sqrt{w_2^\pm+w_3^\pm}\right)^{1/3}.
\end{equation}
After using this to eliminate $r_{S\pm}$ in (\ref{eq:totenergiemitrs}) and dividing by $L^3$ we are led to
\begin{equation}
	\epsilon_\text{tot}^\pm=w_1^\pm n_\pm+\frac{3^{5/3}\pi^{4/3}}{5}n_\pm^{5/3}w_2^\pm\left(1+\frac{w_3^\pm}{w_2^\pm}\right)^{1/3}.
\end{equation}
Let us split up the terms $w_i^\pm$; $w_i^\pm = x_i\pm y_i$. We have (see (\ref{eq:defk1pm})-(\ref{eq:defk3pm}))
\begin{align}
x_1 &= m_N+\frac{\delta^2}{2m_N}\label{eq:defx1}\\
x_2 &= \frac{1}{2m_N}\\
x_3 &= 0\\
y_1 &= \frac{1}{2}g_A\beta\left(1-\frac{\delta^2}{2m_N^2}\right)\\
y_2 &= -\frac{g_A\beta}{4m_N^2}\\
y_3 &= \frac{\delta^2}{g_A\beta m_N^2}+\frac{g_A\beta}{4m_N^2}=\frac{\delta^2}{g_A\beta m_N^2}-y_2.\label{eq:defy3}
\end{align}
The total fermionic contribution may then be written as
\begin{align}
	\epsilon_f &= \epsilon_\text{tot}^++\epsilon_\text{tot}^-\nonumber\\
	&= w_1^+n_++w_1^-n_-+\frac{3^{5/3}\pi^{4/3}}{5}\left[n_+^{5/3}w_2^+\left(1+\frac{w_3^+}{w_2^+}\right)^{1/3}+n_-^{5/3}w_2^-\left(1+\frac{w_3^-}{w_2^-}\right)^{1/3}\right]\nonumber\\
	&= x_1n+y_1(n_+-n_-)+\frac{3^{5/3}\pi^{4/3}}{5}\left[(x_2+y_2)\left(1+\frac{y_3}{x_2+y_2}\right)^{1/3}n_+^{5/3}\right.\nonumber\\
	&\hspace{7cm}+ \left.(x_2-y_2)\left(1-\frac{y_3}{x_2-y_2}\right)^{1/3}n_-^{5/3}\right].\label{eq:zweiteordnungepsilonf}
\end{align}
Since we are only interested in terms up to order $m_N^{-2}$ we expand
\[(x_2\pm y_2)\left(1\pm\frac{y_3}{x_2\pm y_2}\right)^{1/3}\]
in powers of $m_N^{-1}$:
\begin{equation}
	(x_2\pm y_2)\left(1\pm\frac{y_3}{x_2\pm y_2}\right)^{1/3}= \underbrace{\frac{1}{2m_N}}_{x_2}-\underbrace{\frac{g_A\beta}{6m_N^2}}_{-\frac{2}{3}y_2}+\underbrace{\frac{\delta^2}{3g_A\beta m_N^2}}_{\frac{y_3}{3}+\frac{y_2}{3}}+\mathcal{O}\left(\frac{1}{m_N^3}\right)=x_2\pm y_2\pm\frac{y_3}{3}.
\end{equation}
Plugging this into (\ref{eq:zweiteordnungepsilonf}) yields
\begin{multline}
\epsilon_f=x_1n+y_1(n_+-n_-)\\
+\frac{3^{5/3}\pi^{4/3}}{5}\left[x_2\left(n_+^{5/3}+n_-^{5/3}\right)+\left(y_2+\frac{y_3}{3}\right)\left(n_+^{5/3}-n_-^{5/3}\right)\right].\label{eq:fermionischeenergienaiveentwicklung}
\end{multline}
When neglecting all terms $\sim m_N^{-2}$ this expression reduces to that of (\ref{eq:fermionicenergy}), i.e. the expansion is consistent.

Using (\ref{eq:defbetai}) and (\ref{eq:defdeltai}), let us study the term $y_2+y_3/3$:
\begin{align}
	y_2+\frac{y_3}{3} &= y_2+\frac{\delta^2}{3 g_A\beta m_N^2}-\frac{y_2}{3}=\frac{\delta^2}{3 g_A\beta m_N^2}-\frac{g_A\beta}{6m_N^2}=\frac{2\delta^2-g_A^2\beta^2}{6g_A\beta m_N^2}\nonumber\\
	&= \frac{a}{12g_Am_N^2}\left(\cos\alpha_0\cot\alpha_0-2g_A^2\sin\alpha_0\right).\label{eq:divergierenderterm}
\end{align}
For $\alpha_0\to 0$ this term diverges to $+\infty$. Since $(n_+^{5/3}-n_-^{5/3})\leq 0$ for $0\leq n_+\leq n/2$ it follows that $\epsilon_f\to -\infty$ for $\alpha_0\to 0$. Obviously, this behavior is non-physical, namely the energy would no longer be constrained from below. Here, we are confronted with the problem we already mentioned at the beginning of this section, namely that the expansion of the fermionic energy in the way we did it is valid only for $\alpha_0$ not too small. A more detailed analysis of the expansion (see appendix \ref{ap:expansion}) yields the condition
\begin{equation}
	\frac{a}{2m_Ng_A\sin\alpha_0}\lesssim 1\label{eq:bedingunganaundalpha0}
\end{equation}
for the expansion to valid. Since the parameter $a$ will be known only after minimizing the energy with respect to it, it is not a priori possible to decide down to which $\alpha_0$ we may trust the expansion. We therefore now first go on using (\ref{eq:fermionischeenergienaiveentwicklung}) and find out whether some relevant information may be extracted.

\subsection{Expansion for $\alpha_0\sim\mathcal{O}(1)$}\label{sec:expansionforalpha0simo1}

Besides (\ref{eq:fermionischeenergienaiveentwicklung}), for the pionic contribution we use the same expression as before, i.e. (\ref{eq:pionicenergydensity}) together with (\ref{eq:masse1}). The total energy density is then again given by $\epsilon_\text{tot}=\epsilon_f+\epsilon_p$. After replacing the $x_i$ and $y_i$ with the expressions of (\ref{eq:defx1})-(\ref{eq:defy3}) the expression reads
\begin{multline}
\epsilon_\text{tot} = \left(m_N+\frac{\delta^2}{2m_N}\right)n+\frac{1}{2}g_A\beta\left(n_+-n_-\right)\left(1-\frac{\delta^2}{2m_N^2}\right)\\
+ \frac{3^{5/3}\pi^{4/3}}{10m_N}\left[n_+^{5/3}+n_-^{5/3}+\frac{2\delta^2-g_A^2\beta^2}{3g_A\beta m_N}\left(n_+^{5/3}-n_-^{5/3}\right)\right]\\
+ \frac{F^2}{2}a^2\sin^2\alpha_0-F^2M_\pi^2\cos\alpha_0.\label{eq:higherorderetotnochnichtersetzt}
\end{multline}
Finally, we use (\ref{eq:defbetai}) and (\ref{eq:defdeltai}) to replace $\beta$ and $\delta$ and substitute $n_-=n-n_+$ to get the total energy density as a function of $a$, $\alpha_0$, $n$ and $n_+$. As we did in section \ref{sec:thetotalenergydensity}, we now minimize $\epsilon_\text{tot}$ with respect to $a$. The equation $\partial\epsilon_\text{tot}/\partial a=0$ is a quadratic equation in $a$ possessing the two solutions
\begin{equation}
	a_\pm=\pm\frac{1}{30g_A^2(n-2n_+)\sin\alpha_0\cos^2\alpha_0}\left(\mp 20g_Am_N\left(n\cos^2\alpha_0+4F^2m_N\sin^2\alpha_0\right)+S\right)\label{eq:apmathigherorder},
\end{equation}
where $S$ is the square root
\begin{multline}
S=\Bigg\{400\left(g_Am_Nn\cos^2\alpha_0+4F^2g_Am_N^2\sin^2\alpha_0\right)^2\\
-240g_A^2(n-2n_+)\cos^2\alpha_0\sin\alpha_0\Big[(9\pi^4)^{1/3}\left(n_+^{5/3}-(n-n_+)^{5/3}\right)\cos\alpha_0\cot\alpha_0\\
+2g_A^2\sin\alpha_0\left\{(9\pi^4)^{1/3}\left((n-n_+)^{5/3}-n_+^{5/3}\right)-5m_N^2(n-2n_+)\right\}\Big]\Bigg\}^{1/2}.
\end{multline}
Since $a_-<0$ and $a_+>0$ for all values of $n$, $0\leq n_+\leq n/2$ and $0<\alpha_0\leq \pi/2$, we conclude that $a_+$ is the physical solution. $\partial^2\epsilon_\text{tot}/\partial a^2(a_+)>0$ confirms that $a_+$ minimizes $\epsilon_\text{tot}$. Hence, plugging $a_+$ into $\epsilon_\text{tot}$ we eliminate the parameter $a$ and are left with three parameters.

Next, we identify the non-spiral energy. Using de l'H\^opital we find
\begin{equation}
	\lim_{n_+\rightarrow \frac{n}{2}}a_+=0,
\end{equation}
so as before the spiral disappears for $n_+=n/2$. Setting $n_+=n/2$ and then $a=0$ in (\ref{eq:higherorderetotnochnichtersetzt}) we find
\begin{equation}
	\epsilon_\text{tot}\bigr|_{n_+=n/2,a=0}=m_Nn+\frac{3}{10m_N}\left(\frac{9}{4}\pi^4n^5\right)^{1/3}-F^2M_\pi^2\cos\alpha_0,\label{eq:higherordernonspiralmitalpha0}
\end{equation}
which is minimized for $\alpha_0=0$. Hence we can write
\begin{equation}
	\epsilon_\text{tot}^\text{ns}=m_Nn+\frac{3}{10m_N}\left(\frac{9}{4}\pi^4n^5\right)^{1/3}-F^2M_\pi^2\label{eq:higherorder_alphaorder1nsenergy},
\end{equation}
which is the same expression as (\ref{firstordernonspiralenergynonchiral}).

\begin{figure}[h]
\centering
\psfrag{al}{\small $\alpha_0$}
\psfrag{eps}{\small $\epsilon_\text{tot}\cdot 10^{-8}\,[\text{MeV}]$}
\subfloat[][$n_+=0$]{\includegraphics[width=0.4\textwidth]{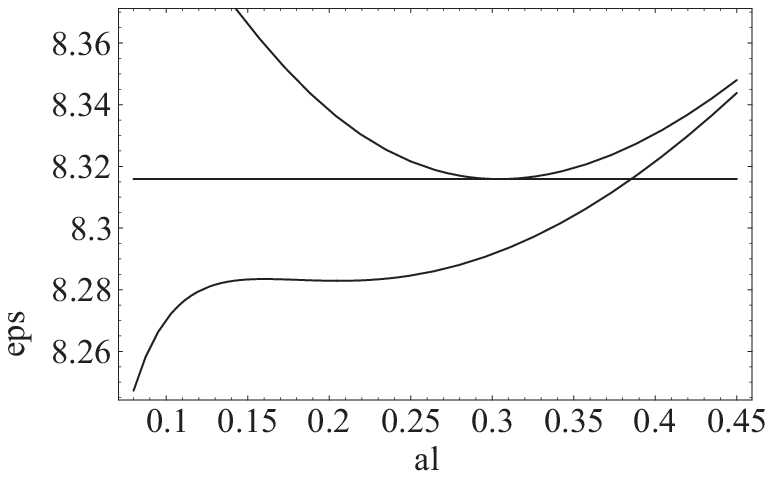}\label{fig:higherorderunkorratcrit}}
\hspace{8mm}
\subfloat[][$n_+=0.1n$]{\includegraphics[width=0.4\textwidth]{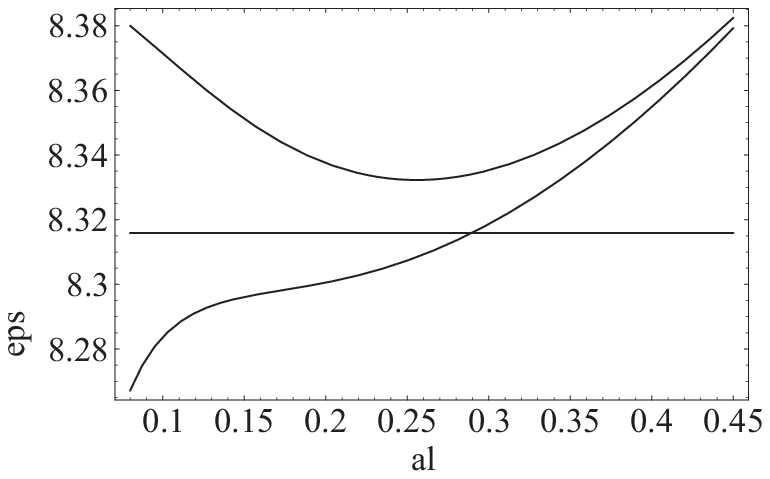}\label{fig:higherorderunkorratcrit_ndurch10}}
\caption{Lower and higher order energy densities at $n^{1/3}=102.27\,\text{MeV}$, at two different values of $n_+$, using the physical pion mass}\label{fig:lowerundhigherorderenergydensity}
\end{figure}

The two plots in fig. \ref{fig:higherorderunkorratcrit} and \ref{fig:higherorderunkorratcrit_ndurch10} show the energy densities both of lower and of higher order as functions of $\alpha_0$, as well as the non-spiral energy density (horizontal line). The particle density was chosen to be $n=102.27\,\text{MeV}$, i.e. the critical density we found in section \ref{sec:awayfromthechirallimit}.
The plot in fig. \ref{fig:higherorderunkorratcrit} shows the situation for $n_+=0$. The minimum of the lower order energy density (upper curve) touches $\epsilon_\text{tot}^\text{ns}$, confirming that the particle density is set to the the critical one of the lower order calculation. The higher order energy density (lower curve) however shows the problem we already predicted earlier, when having analyzed expression (\ref{eq:divergierenderterm}): The function does not show a minimum but it rather diverges towards $-\infty$ for small values of $\alpha_0$. This unphysical behavior is due to the fact that the expansion of the fermionic energy in the way we did it in (\ref{eq:energiezweiteordnung}) breaks down for too small angles. This is confirmed also quantitatively: Using (\ref{eq:apmathigherorder}) we find that in the case of $n^{1/3}=102.27\,\text{MeV}$ for $\alpha_0\lesssim 0.13$ the left hand side of (\ref{eq:bedingunganaundalpha0}) becomes larger than 1 and hence the expansion (\ref{eq:energiezweiteordnung}) starts to fail. This situation makes it impossible to identify a minimum of the higher order energy density and to decide whether it is lower than $\epsilon_\text{tot}^\text{ns}$ or not.

Things do not get better if we set $n_+\neq 0$. The plot in fig. \ref{fig:higherorderunkorratcrit_ndurch10} was obtained using the same parameters as in fig. \ref{fig:higherorderunkorratcrit}, but $n_+$ was set to 10\% of $n$. In accordance to what we found in section \ref{sec:awayfromthechirallimit} the lower order energy density rises; its minimum now lies above $\epsilon_\text{tot}^\text{ns}$. Also the higher order energy density increases and the fact that we cannot identify a minimum becomes even more distinctive. When going to lower values of $n$, the situation becomes even worse.

\begin{figure}[h]
\centering
\psfrag{al}{\small $\alpha_0$}
\psfrag{eps}{\small $\epsilon_\text{tot}\cdot 10^{-7}\,[\text{MeV}]$}
\includegraphics[scale=0.8]{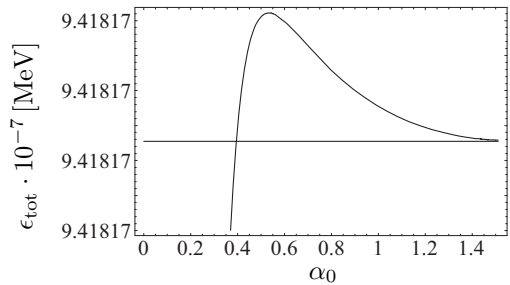}
\caption{$\epsilon_\text{tot}$ and $\epsilon_\text{tot}^\text{ns}$ at $n^{1/3}=47.36\,\text{MeV}$, $n_+/n=0.4995$ and $M_\pi=0$}\label{fig:higherorder_cl_1}
\end{figure}

Let us investigate the chiral limit. In section \ref{sec:ersteordnungchiralerlimes} we found that for $M_\pi=0$ $\alpha_0$ is fixed to $\pi/2$. If this is still the case here we might get a result as the expansion is useful for large values of $\alpha_0$. The plot in fig. \ref{fig:higherorder_cl_1} shows $\epsilon_\text{tot}$ at $n_+/n=0.4995$ and $\epsilon_\text{tot}^\text{ns}$, both at $n^{1/3}=47.36\,\text{MeV}$ and $M_\pi=0$.
\\
Indeed, this plot suggests that the minimum of the energy density is located at $\alpha_0=\pi/2$. The mentioned value for $n$ was found by the observation that, at $\alpha_0=\pi/2$, the minimum of $\epsilon_\text{tot}$ as a function of $n_+$ is located at $n_+=n/2$. Then, the root of the difference $\epsilon_\text{tot}^\text{ns}-\epsilon_\text{tot}$ at $n_+=n/2$ and $\alpha_0=\pi/2$ as a function of $n$ was found numerically\footnote{As it seems complicated to take the limits $\lim_{n_+\to n/2}\epsilon_\text{tot}$ and $\lim_{\alpha_0\to \pi/2}\epsilon_\text{tot}$ respectively we set $n_+=0.4995n$ and $\alpha_0=\pi/2-0.05$ in the numerical evaluation} to be $47.36^3\,\text{MeV}^3$. However, at this moment we cannot make any statements about the behavior of $\epsilon_\text{tot}$ for small values of $\alpha_0$ and we cannot exclude the possibility that the true minimum lies e.g. at $\alpha_0=0$. We will make up for that later in section \ref{sec:expansionforalpha0simmNhochminus1}, where we will find that there is no minimum at small values of $\alpha_0$. Accordingly, we have found the critical particle density in the chiral limit to be
\begin{equation}
	\mathring{n}_c=106.2\,\text{GeV}^3\quad\text{or}\quad \mathring{n}_c^{1/3}=47.36\,\text{MeV}.
\end{equation}

However, as we noticed above, in this expansion we are not able to find the critical densities for $M_\pi=40\,\text{MeV}$ or $M_\pi=135\,\text{MeV}$, since the minimum of $\epsilon_\text{tot}$ then seems to be located in a region where our expansion fails. We therefore have to try to do the expansion of the fermionic energy (\ref{eq:energieplus}) explicitly for small values of $\alpha_0$ and first try it for $\alpha_0\sim\mathcal{O}(m_N^{-1})$.

\subsection{Expansion for $\alpha_0\sim\mathcal{O}(m_N^{-1})$}\label{sec:expansionforalpha0simmNhochminus1}

Let us introduce a parameter $\eta$ of dimension of energy to write
\begin{equation}
	\alpha_0=\frac{\eta}{m_N}\label{eq:defvoneta}.
\end{equation}
Varying $\eta$ in an appropriate range of $\mathcal{O}(1)$ brings down $\alpha_0$ to $\mathcal{O}(m_N^{-1})$. (\ref{eq:defvoneta}) implies
\begin{align}
\beta&=a\frac{\eta}{m_N}+\mathcal{O}(m_N^{-3})\\
\delta&=\frac{a}{2}+\mathcal{O}(m_N^{-2})
\end{align}
and, when using (\ref{eq:skalarproduktpdelta}) and (\ref{eq:skalarproduktpbeta}), the internal square root of (\ref{eq:energieplus}) therefore may be expanded as follows:
\begin{align}
\sqrt{4(\vec{p}\cdot\vec{\delta})^2+g_A^2\{m_N^2\beta^2+(\vec{p}\cdot\vec{\beta})^2\}}&=\frac{a}{2}\sqrt{g_A^2\eta^2+p_3^2+\mathcal{O}(m_N^{-2})}\nonumber\\
&=\frac{a}{2}\sqrt{g_A^2\eta^2+p_3^2}\sqrt{1+\mathcal{O}(m_N^{-2})}\nonumber\\
&=\frac{a}{2}\sqrt{g_A^2\eta^2+p_3^2}+\mathcal{O}(m_N^{-2})\label{eq:entwinnerewurzelalpha0proptomnhochminus1},
\end{align}
where in the last step we have expanded the second square root under the assumption that the terms $\mathcal{O}(m_N^{-2})$ in the root are small. These are of the form
	\[\sim\frac{p_3^2\eta^2}{m_N^2(g_A^2\eta^2+p_3^2)},\quad \sim\frac{\eta^4}{m_N^2(g_A^2\eta^2+p_3^2)}.
\]
Since $p_3^2\ll m_N^2$ (see appendix \ref{ap:expansion}) the first term is indeed small as
	\[\frac{\eta^2}{g_A^2\eta^2+p_3^2}<1.
\]
The second term is larger than 1 only as soon as $\eta>g_Am_N$, if $p_3=0$. For $p_3^2>0$ $\eta$ may even be larger. According to (\ref{eq:defvoneta}) this would mean $\alpha_0>g_A=1.26$; we are however not interested in such high values for $\alpha_0$. Hence, in our regime the expansion of the internal square root should cause no problems.

Plugging (\ref{eq:entwinnerewurzelalpha0proptomnhochminus1}) into (\ref{eq:energieplus}) and replacing $\beta$ and $\delta$ now yields
\begin{align}
E_\pm&=\sqrt{m_N^2+p^2+\frac{a^2}{4}\pm	a\sqrt{g_A^2\eta^2+p_3^2}+\mathcal{O}(m_N^{-2})}\nonumber\\
&=m_N\sqrt{1+\frac{p^2}{m_N^2}+\frac{a^2}{4m_N^2}\pm\frac{a}{m_N^2}\sqrt{g_A^2\eta^2+p_3^2}+\mathcal{O}(m_N^{-4})}\nonumber\\
&=m_N\left(1+\frac{p^2}{2m_N^2}+\frac{a^2}{8m_N^2}\pm\frac{a}{2m_N^2}\sqrt{g_A^2\eta^2+p_3^2}+\mathcal{O}(m_N^{-4})\right)\nonumber\\
&=m_N+\frac{p^2}{2m_N}+\frac{a^2}{8m_N}\pm\frac{a}{2m_N}\sqrt{p_3^2+g_A^2\eta^2}+\mathcal{O}(m_N^{-3}).\label{eq:epmforalpha0simgaovermn}
\end{align}
The expansion of the square root in the second step is of course only allowed if
\begin{equation}
	\frac{a^2}{m_N^2}\ll 1\label{eq:higherorderskorrbedingungana}.
\end{equation}

As the dispersion relation (\ref{eq:epmforalpha0simgaovermn}) now contains $p_3$ under a square root, the surfaces of constant energy in momentum space and hence the Fermi volume are more complicated than before. Accordingly, the integration over the Fermi volume in the calculation of the particle number $N_\pm$ and the total fermionic energy $E_\text{tot}^\pm$ will be a bit trickier. The calculation is performed in appendix \ref{ap:deltaf}. There, it turns out that for the calculation of $n_-$ and $\epsilon_\text{tot}^-$ respectively we actually have to distinguish between two cases depending on the value of the spiral parameters $a$ and $\eta$, namely:
\begin{itemize}
\item{Case 1:} $a<2g_A\eta$
\item{Case 2:} $a>2g_A\eta$.
\end{itemize}
The particle number density $n_-^{(1)}$ and the energy density $\epsilon_\text{tot}^{(1)-}$ (the superscript (1) standing for case 1) are given by (\ref{eq:nmiti1}) and (\ref{eq:fermionischeenergiefall1}) respectively, while $n_-^{(2)}$ and $\epsilon_\text{tot}^{(2)-}$ are given by (\ref{eq:nmiti2}) and (\ref{eq:fermionischeenergiefall2}). For the calculation of $n_+$ and $\epsilon_\text{tot}^+$ no distinction is necessary and the results are found in (\ref{eq:nmiti1plus}) and (\ref{eq:fermionischeenergiefall1plus}).

\begin{figure}[h!]
\centering
\psfrag{aa}{\small $a\,[\text{MeV}]$}
\psfrag{eps}{\small $\epsilon_\text{tot}\cdot 10^{-7}\,[\text{MeV}]$}
\subfloat[][$n_-/n=0.5$]{\includegraphics[width=0.4\textwidth]{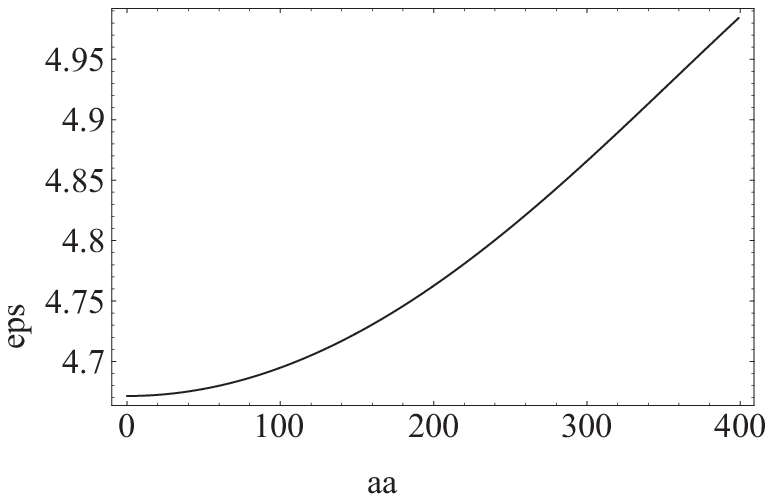}\label{fig:higherorder_korr_nospiral}}
\hspace{8mm}
\subfloat[][$n_-/n=0.65$]{\includegraphics[width=0.4\textwidth]{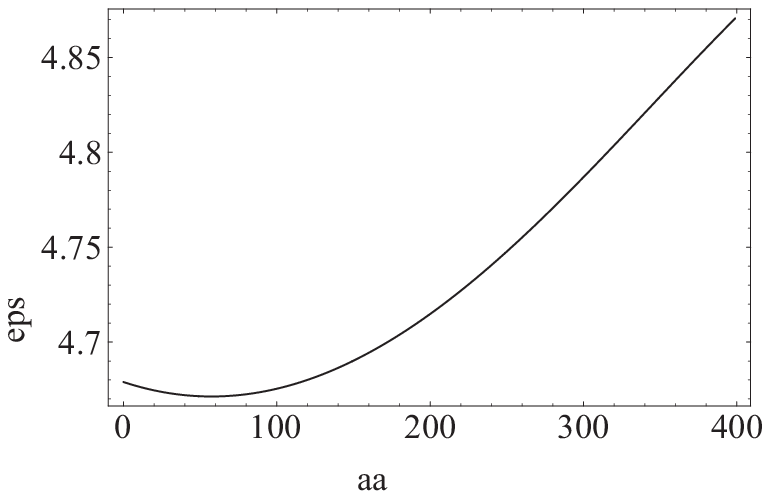}\label{fig:higherorder_korr_nm140400}}
\caption{$\epsilon_\text{tot}$ as a function of $a$ at $n^{1/3}=60\,\text{MeV}$, $\eta=15\,\text{MeV}$ and a physical pion mass.}\label{fig:higherorder_korr_minbeia0}
\end{figure}

Unfortunately, the expressions for $n_-^{(1)}$, $n_-^{(2)}$ and $n_+$, which are functions of the Fermi energy $E_F^\pm$, cannot be solved analytically for $E_F^\pm$ in order to express $\epsilon_\text{tot}^{(1)-}$, $\epsilon_\text{tot}^{(2)-}$ and $\epsilon_\text{tot}^+$ in terms of $n_-$ and $n_+$. Accordingly, we cannot minimize the total energy with respect to the parameter $a$ analytically in order to get rid of it. We rather have to do the replacements and the minimization numerically in order to find the total energy density $\epsilon_\text{tot}$ as a function of $\alpha_0$ at given values of $n$ and $n_+$ (or $n_-$). This is done in a Mathematica program according to the following scheme:
\begin{enumerate}
\item Split up a fixed particle number density $n$ into two parts $n_-$ and $n_+$, such that $n=n_-+n_+$.
\item Fix the value of $\alpha_0$ and hence of $\eta$.
\item Set $a$ to a particular value, e.g. $a=0$.
\item Depending on whether $a<2g_A\eta$ or $a>2g_A\eta$ numerically solve either (\ref{eq:nmiti1}) or (\ref{eq:nmiti2}) for $E_F^-$ and plug this into either (\ref{eq:fermionischeenergiefall1}) or (\ref{eq:fermionischeenergiefall2}) to find $\epsilon_\text{tot}^-$.
\item Numerically solve (\ref{eq:nmiti1plus}) for $E_F^+$ and plug this into (\ref{eq:fermionischeenergiefall1plus}) to find $\epsilon_\text{tot}^+$.
\item Calculate $\epsilon_\text{tot}=\epsilon_\text{tot}^-+\epsilon_\text{tot}^++\epsilon_p$, where $\epsilon_p$ is given by (\ref{eq:pionicenergydensity}).
\item Increase $a$ by a certain amount.
\item Repeat steps 4-7 up to some appropriate value of $a$.
\item From the set of all values of $\epsilon_\text{tot}$ obtained in such a way, choose the minimal one to be the total energy density for the given angle $\alpha_0$.
\item Increase $\alpha_0$ by a certain amount.
\item Repeat steps 3-10 up to a certain value of $\alpha_0$.
\end{enumerate}
As we target to a critical number density from where on a spiral configuration will be favored we still need to identify the non-spiral energy density. A non-spiral configuration is given for $a=0$. The plots in fig. \ref{fig:higherorder_korr_minbeia0} show $\epsilon_\text{tot}$ at fixed values of $n$ and $\alpha_0$ and at two different values for $n_-$. They were obtained by accomplishing the above steps 1-8 and show what we suppose: The non-spiral configuration with $a=0$ is given for $n_-=n_+=n/2$.
\\
Hence, let us calculate $\epsilon_\text{tot}$ for $a=0$ and $n_-=n_+=n/2$. Since $a=0<2g_A\eta$ we have to find $\epsilon_\text{tot}^-$ in the framework of the first case. For $a=0$ (\ref{eq:nmiti1}) reduces to
\begin{equation}
	n_-^{(1)}=\frac{m_N}{\pi^2}\int\limits_{m_N}^{E_F^-}dE\,\sqrt{2m_N(E-m_N)}=\frac{1}{3\pi^2}(2m_N(E_F^--m_N))^{3/2}\label{eq:higherorder_korr_nmbeia0}.
\end{equation}
This may be solved for $E_F^-$, and plugged into (\ref{eq:fermionischeenergiefall1}), leading to
\begin{equation}
	\epsilon_\text{tot}^-=m_Nn_-+\frac{3}{10m_N}\left(9\pi^4n_-^5\right)^{1/3}.\label{eq:higherorder_korr_epstotminus}
\end{equation}
For $a=0$ (\ref{eq:nmiti1plus}) reduces to exactly the same expression as (\ref{eq:higherorder_korr_nmbeia0}) and (\ref{eq:fermionischeenergiefall1plus}) to (\ref{eq:fermionischeenergiefall1}), except that we have to replace $E_F^-$ by $E_F^+$ and $n_-$ by $n_+$. $\epsilon_\text{tot}^+$ is therefore given also by (\ref{eq:higherorder_korr_epstotminus}) with $n_-\to n_+$. Finally, setting $n_-=n_+=n/2$ the total energy density reads
\begin{equation}
	\epsilon_\text{tot}=\epsilon_\text{tot}^-+\epsilon_\text{tot}^++\epsilon_p=m_Nn+\frac{3}{10m_N}\left(\frac{9}{4}\pi^4n^5\right)^{1/3}-F^2M_\pi^2\cos\alpha_0,
\end{equation}
which is minimized for $\alpha_0=0$, such that we get exactly the same non-spiral energy density as in (\ref{firstordernonspiralenergynonchiral}) or (\ref{eq:higherorder_alphaorder1nsenergy}) respectively.

Now that we have everything together we can ask for a critical particle number density. Based on the insight that at first order for physical pion masses $n_+$ will go to zero (see section \ref{sec:awayfromthechirallimit}) we now make the same assumption here and just take $n_-$ into account. Later we will verify this assumption and in fact learn that it is not exactly true anymore.

\begin{figure}[h!]
\centering
\psfrag{aa}{\small $a\,[\text{MeV}]$}
\psfrag{eps}{\small $\epsilon_\text{tot}\cdot 10^{-7}\,[\text{MeV}]$}
\subfloat[][$\eta=30\,\text{MeV}$ ($\alpha_0=0.032$)]{\includegraphics[width=0.4\textwidth]{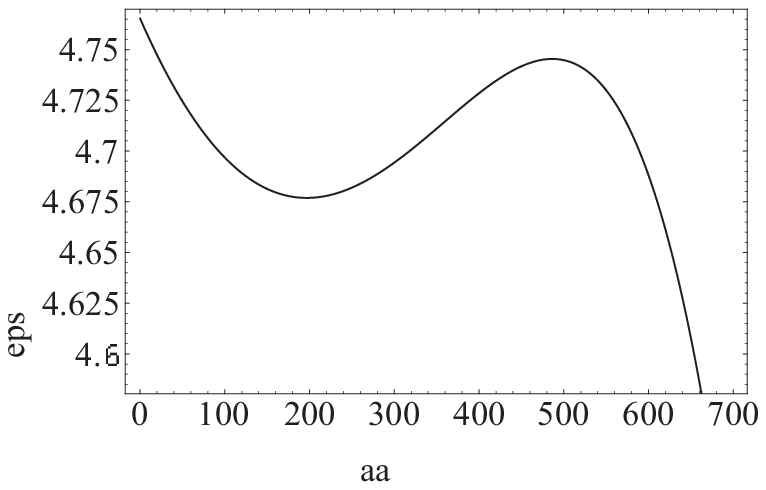}}
\hspace{8mm}
\subfloat[][$\eta=5\,\text{MeV}$ ($\alpha_0=0.005$)]{\includegraphics[width=0.4\textwidth]{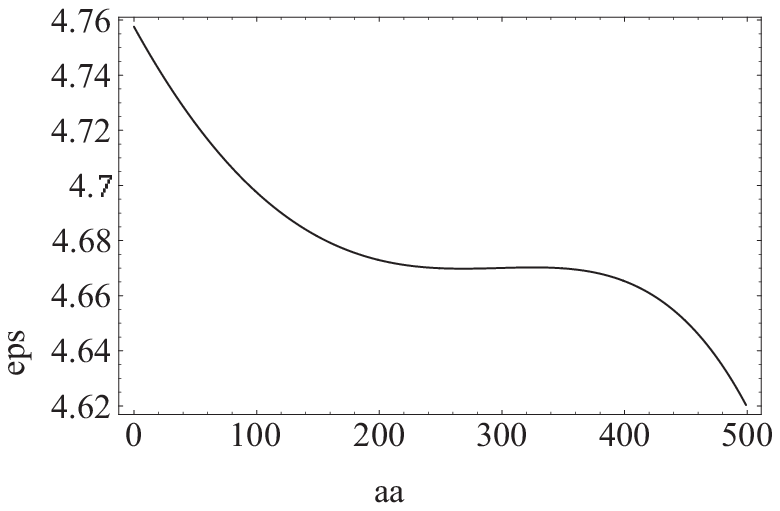}}
\caption{$\epsilon_\text{tot}$ as a function of $a$ at a density of $n^{1/3}=60\,\text{MeV}$}\label{fig:epstotasfunctionofa}
\end{figure}

The ``appropriate value'' of $a$ in above step 8 depends on the values of $n$ and $\eta$. It is clear that it must not be too large, otherwise the condition (\ref{eq:higherorderskorrbedingungana}) will not be satisfied and the expansion (\ref{eq:epmforalpha0simgaovermn}) is not valid anymore. Going to too small values of $\eta$ (for a given $n$) will manifest in the fact that $\epsilon_\text{tot}$ will not show a distinct minimum as a function of $a$. This can be seen in the plots of fig \ref{fig:epstotasfunctionofa}, which where also generated numerically through the above steps 1-8.
\\
It is obvious that in the first plot ($\alpha_0=0.032$) a clear minimum of $\epsilon_\text{tot}$ around $a=200\,\text{MeV}$ may be found. For values of $a$ larger that  $500\,\text{MeV}$ $\epsilon_\text{tot}$ drops again and in fact starts to diverge towards $-\infty$. For too large values of $a$ condition (\ref{eq:higherorderskorrbedingungana}) is not satisfied and the expansion not valid anymore. In the second plot ($\alpha_0=0.005$) $\epsilon_\text{tot}$ starts to diverge for smaller values of $a$. In that particular case the drop starts so early that we are not able to identify a local minimum. We should therefore hope that the minimum of $\epsilon_\text{tot}$ as a function of $\alpha_0$ around the critical density $n_c$ does not lie at too small values of $\alpha_0$, otherwise it can not be found in the framework of this expansion and we should do it for even smaller angles.

\begin{figure}[h]
\centering
\psfrag{ei}{\small 1}
\psfrag{zw}{\small 2}
\psfrag{dr}{\small 3}
\psfrag{vi}{\small 4}
\psfrag{alpha}{$\alpha_0$}
\psfrag{eps}{\small $\epsilon_\text{tot}\cdot 10^{-8}\,[\text{MeV}]$}
\includegraphics[scale=0.8]{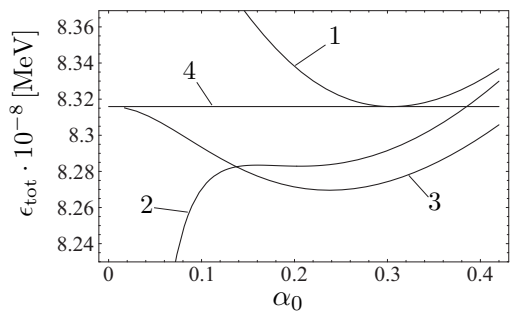}
\caption{1: First order energy density, 2: Higher order energy density of (\ref{eq:higherorderetotnochnichtersetzt}) and (\ref{eq:apmathigherorder}), 3: Higher order energy density in the $\alpha_0\sim\mathcal{O}(m_N^{-1})$ expansion, 4: Non-spiral energy density}\label{fig:higherordernaivundneu}
\end{figure}

\begin{figure}[h]
\centering
\psfrag{nn}{\small $n\,[\text{MeV}^3]$}
\psfrag{d}{\small $d\,[\text{MeV}]$}
\includegraphics[scale=0.8]{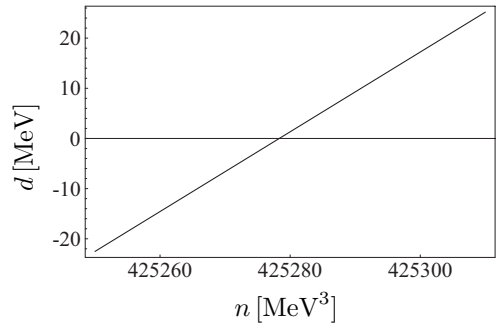}
\caption{Difference between $\epsilon_\text{tot}^\text{ns}$ and the minimum of $\epsilon_\text{tot}$ as a function of $n$}\label{fig:dathigherorder}
\end{figure}

Fig. \ref{fig:higherordernaivundneu} shows the same plot as in fig. \ref{fig:higherorderunkorratcrit}, i.e. the the lower and higher order energy densities as well as the non-spiral energy density at $n^{1/3}=102.27\,\text{MeV}$, supplemented with the new energy density obtained through above steps 1-11. Unlike the previous higher order energy density the latter one shows an obvious minimum which, as it lies below the non-spiral energy density, allows us to state that we are already above the critical number density $n_c$.
\newpage
In order to find $n_c$ we repeat the above 11 steps for a certain range of $n$, each time picking out the minimal value of $\epsilon_\text{tot}$ (as a function of $\alpha_0$) and calculating the difference $d$ between $\epsilon_\text{tot}^\text{ns}$ and the minimum (cf. the function $d_\text{max}(n)$ in section \ref{sec:awayfromthechirallimit}). The result is the plot in fig. \ref{fig:dathigherorder}.
\\ 
The root and thus the (alleged) critical number density is found to be located at
\begin{equation}
	n=425.3\,\text{GeV}^3\quad\text{or}\quad n^{1/3}=75.20\,\text{MeV}.
\end{equation}

\begin{figure}[h]
\centering
\psfrag{al}{\small $\alpha_0$}
\psfrag{eps}{\small $\epsilon_\text{tot}\cdot 10^{-8}\,[\text{MeV}]$}
\includegraphics[scale=0.8]{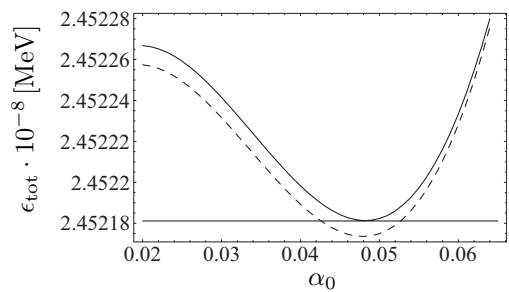}
\caption{The minimum of $\epsilon_\text{tot}$ drops as $n_+$ brings a contribution.}\label{fig:nplusnichtnulltiefer}
\end{figure}

\begin{figure}[h!]
\centering
\psfrag{al}{\small $\alpha_0$}
\psfrag{eps}{\small $\epsilon_\text{tot}\cdot 10^{-8}\,[\text{MeV}]$}
\subfloat[][$n=399.1\,\text{GeV}^3$, $n_+/n=0.07$]{\includegraphics[width=0.4\textwidth]{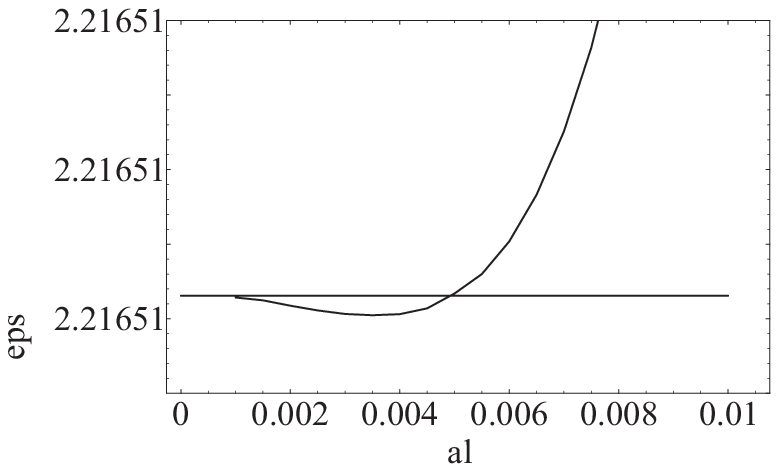}\label{fig:darueber}}
\hspace{8mm}
\subfloat[][$n=398.8\,\text{GeV}^3$, $n_+/n=0.3$]{\includegraphics[width=0.4\textwidth]{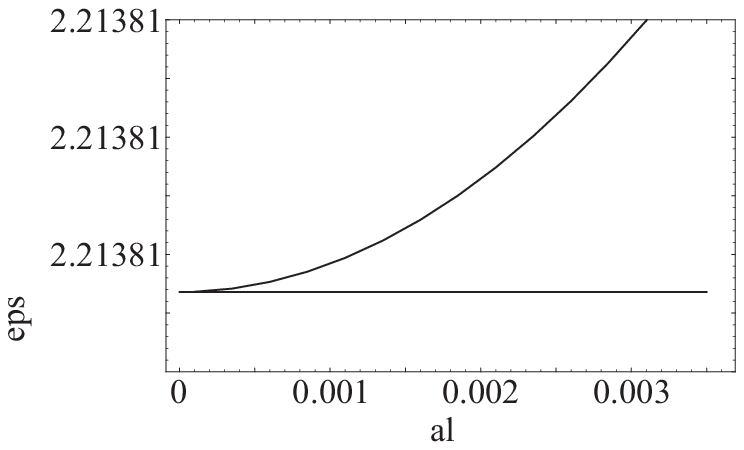}\label{fig:darunter}}
\caption{$\epsilon_\text{tot}$ and $\epsilon_\text{tot}^\text{ns}$ slightly above and below the critical particle density}
\end{figure}

We should now check whether the assumption that also here for physical pion masses only $E_-$ contributes to the fermionic energy density is true. To do so we apply the above 11 steps to the particle density $n^{1/3}=75.20\,\text{MeV}$ we just found but do not restrict ourselves to $n_+=0$. The plot in fig. \ref{fig:nplusnichtnulltiefer} shows $\epsilon_\text{tot}$ at $n^{1/3}=75.20\,\text{MeV}$ and $n_+=0$ (continuous graph) as well as $\epsilon_\text{tot}$ at the same particle density but with $n_+/n=0.001$ (dashed graph). The horizontal line corresponds to $\epsilon_\text{tot}^\text{ns}$.
\\
In contrast to our assumption the energy density and in particular its minimum does not increase but drops for $n_+\neq 0$. This differs from the properties observed earlier and shows that we actually have not yet found the critical particle number density. Since of course $\epsilon_\text{tot}^\text{ns}$ is not influenced by $n_+$ we have $n_c^{1/3}<75.20\,\text{MeV}$. The problem is now the following: The lower $n$ is chosen the more will move the minimum of $\epsilon_\text{tot}$ towards smaller values of $\alpha_0$. It then becomes numerically more and more difficult to identify a minimum. The lowest approximate value of $n$ where one is able to find a minimum of $\epsilon_\text{tot}$ which lies below $\epsilon_\text{tot}^\text{ns}$ lies in the region around $n=398.9\,\text{GeV}$. For densities lower than this value the minimum seems to have shifted to $\alpha_0=0$ and to have approached $\epsilon_\text{tot}^\text{ns}$. The plot in fig. \ref{fig:darueber} shows $\epsilon_\text{tot}$ at $n=399.1\,\text{GeV}^3$ and for $n_+/n=0.07$.
We can identify a minimum slightly below $\epsilon_\text{tot}^\text{ns}$ which shows that we are just above the critical particle density. On the other hand, the plot in fig. \ref{fig:darunter} was created using $n=398.8\,\text{GeV}^3$. At this density it is not possible to choose $n_+$ such that the minimum of $\epsilon_\text{tot}$ lies below the non-spiral energy density; we are thus below the critical particle density. From this we conclude that $n_c$ is located somewhere between $398.8\,\text{GeV}^3$ and $398.9\,\text{GeV}^3$. We take the larger value as an upper bound:
\begin{equation}
	n_c=398.9\,\text{GeV}^3\quad\text{or}\quad n_c^{1/3}=73.61\,\text{MeV}.
\end{equation}
Compared to the first order result (\ref{eq:lowerorderresult}) this is a 28\% decrease of the critical particle number density.\footnote{As we want to make a comparison on the energy scale we compare $n^{1/3}$ rather than $n$.}

In the same way as we found $n_c$, we also determine $\bar{n}_c$ (i.e. using $M_\pi=40\,\text{MeV}$):
\begin{equation}
	\bar{n}_c=172.1\,\text{GeV}^3\quad\text{or}\quad \bar{n}_c^{1/3}=55.62\,\text{MeV}.
\end{equation}

\begin{figure}[h]
\centering
\psfrag{al}{\small $\alpha_0$}
\psfrag{eps}{\small $\epsilon_\text{tot}\cdot 10^{-7}\,[\text{MeV}]$}
\includegraphics[scale=0.8]{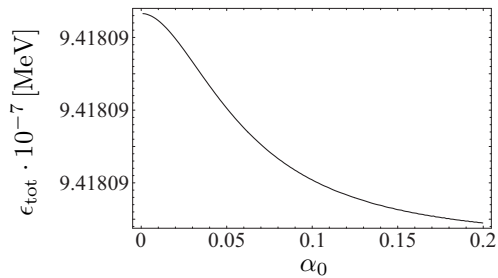}
\caption{$\epsilon_\text{tot}$ at $n^{1/3}=47.36\,\text{MeV}$, $n_+/n=0.4995$ and $M_\pi=0$}\label{fig:higherorder_cl}
\end{figure}

In section \ref{sec:expansionforalpha0simo1} we found the critical particle density in the chiral limit under the assumption that the minimum of $\epsilon_\text{tot}$ is at $\alpha_0=\pi/2$ and not in the region of small values of $\alpha_0$. With the expansion for $\alpha_0\sim\mathcal{O}(m_N^{-1})$ we can now close the gap and investigate the behavior of $\epsilon_\text{tot}$ for small values of $\alpha_0$. The plot in fig. \ref{fig:higherorder_cl} shows $\epsilon_\text{tot}$ as a function of $\alpha_0$ using the same parameters as in the discussion of the chiral limit in section \ref{sec:expansionforalpha0simo1}, i.e. $n^{1/3}=47.36\,\text{MeV}$, $n_+=0.4995n$ and $M_\pi=0$.
\\
From this plot we see that there is indeed no minimum of $\epsilon_\text{tot}$ at small values of $\alpha_0$. Hence, we subsequently justify the result for $\mathring{n}_c$ found in section \ref{sec:expansionforalpha0simo1}.

Apart from the chiral limit where $\alpha_0=\pi/2$, we found that, at the critical density, the minimum of $\epsilon_\text{tot}$ seems to lie at $\alpha_0=0$. This is a good sign. Whenever we figured out the non-spiral energy density $\epsilon_\text{tot}^\text{ns}$ we found that $\alpha_0$ will go to zero. It is therefore peculiar that in the lowest order calculations (see section \ref{sec:awayfromthechirallimit}) $\alpha_0$ seemed to adjust itself to a non-zero value at the critical density. The transition from the spiral the the non-spiral configuration would then show an unnatural discontinuity. Here, on the other hand, the energy density ``smoothly'' approaches the non-spiral energy density when shifting towards $n_c$ from above. As a non-spiral configuration (so far) always came with an equal filling of the states ($n_+=n_-=n/2$) we assume this to happen between $398.9\,\text{GeV}^3$ and $398.8\,\text{GeV}^3$ (for the physical pion mass).

\subsection{Expansion for $\alpha_0\sim\mathcal{O}(m_N^{-2})$}
\begin{figure}[h]
\begin{minipage}[h]{0.475\textwidth}
\centering
\psfrag{al}{$\alpha_0$}
\psfrag{aa}{$a_\text{min}\,[\text{MeV}]$}
\includegraphics[width=6cm]{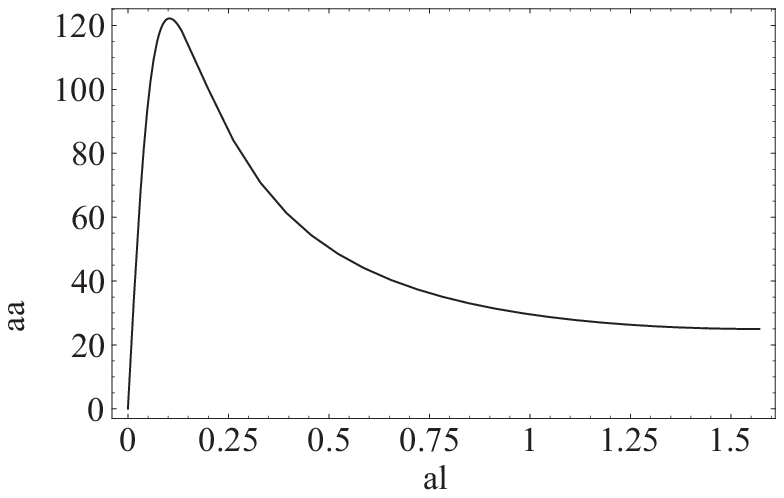}
\caption{$a_\text{min}$ at first order with $n^{1/3}=70\,\text{MeV}$ and $n_+=0$}\label{fig:a_firstorder}
\end{minipage}
\hfill
\begin{minipage}[h]{0.475\textwidth}
\vspace{5mm}
\centering
\psfrag{al}{$\alpha_0$}
\psfrag{aa}{$a_\text{min}\,[\text{MeV}]$}
\includegraphics[width=6cm]{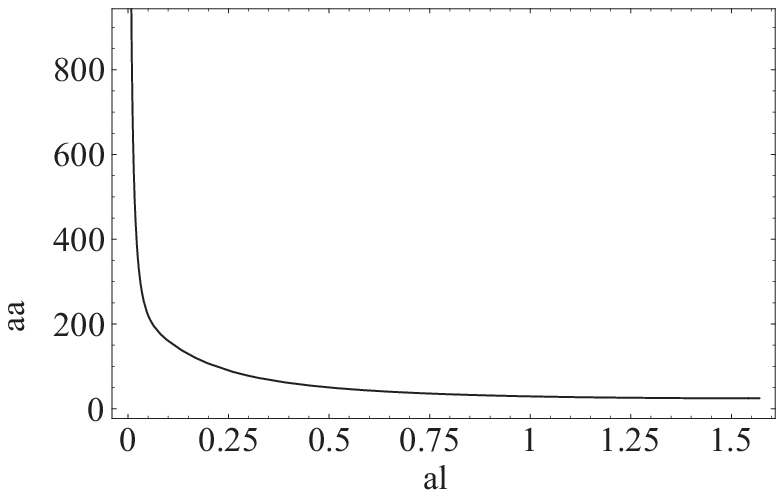}
\caption{$a_\text{min}$ at next order in the $\alpha_0\sim\mathcal{O}(1)$ expansion with $n^{1/3}=70\,\text{MeV}$ and $n_+=0$}\label{fig:a_higherorder}
\end{minipage}
\end{figure}

If this is true and the minimum of $\epsilon_\text{tot}$ indeed approaches $\alpha_0=0$ when $n\searrow n_c$, then one should verify whether the spiral parameter $a$ goes to zero for $\alpha_0\to 0$. The plot in fig. \ref{fig:darunter} gives a hint that this is in fact true since there the minimum touches $\epsilon_\text{tot}^\text{ns}$ \textit{and} this happens at $\alpha_0=0$. Additionally, the plot in fig. \ref{fig:a_firstorder} shows the parameter $a$ minimizing $\epsilon_\text{tot}$ of the first order calculation (see (\ref{eq:aminvonetot})) as a function of $\alpha_0$. It shows the described behavior (even though at first order the minimum of $\epsilon_\text{tot}$ does not approach $\alpha_0=0$ at the critical density).
\\
On the other hand, the plot in fig. \ref{fig:a_higherorder} shows the minimizing $a$ at next order in the $\alpha_0\sim\mathcal{O}(1)$ expansion, i.e. $a_+$ of (\ref{eq:apmathigherorder}). Here, $a\to\infty$ as $\alpha_0\to 0$, which would mean that the spiral would rotate faster and faster as we go to smaller values of $\alpha_0$. However, this would be bad as the energy expansion for too large values of $a$ fails, or, equivalently formulated, as a too quickly rotating spiral would imply too large spatial derivatives and therefore too large pion momenta for the ChPT to be valid.

\begin{figure}[h]
\centering
\psfrag{a}{\small $\xi$}
\psfrag{d}{\small $-\delta$}
\psfrag{p2}{\small $p_2$}
\psfrag{p3}{\small $p_3$}
\includegraphics[scale=0.8]{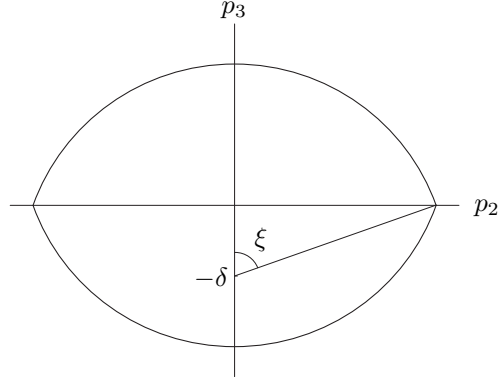}
\caption{Cut of a surface of constant energy for $E_+$ in the $p_2-p_3$ plane}\label{fig:fermiflaeche1}
\end{figure}

In order to verify our supposition that the spiral disappears for $\alpha_0\to 0$, let us therefore expand the fermionic energy at even smaller values of $\alpha_0$ than we did it previously, namely for $\alpha_0\sim\mathcal{O}(m_N^{-2})$. The expansion of (\ref{eq:energieplus}) with $\alpha_0=\kappa^2/m_N^2$ (analogon to (\ref{eq:defvoneta})) reads
\begin{equation}
	E_\pm=m_N+\frac{p^2}{2m_N}+\frac{a^2}{8m_N}\pm\frac{a|p_3|}{2m_N}+\mathcal{O}(m_N^{-3})\label{eq:higherorder_energie_alpha0mnhohminus2},
\end{equation}
which corresponds to (\ref{eq:epmforalpha0simgaovermn}) with $\eta=0$. As
	\[\delta=\frac{a}{2}\cos\frac{\kappa^2}{m_N^2}=\frac{a}{2}+\mathcal{O}(m_N^{-4}),
\]
we replace $a/2$ by $\delta$ in (\ref{eq:higherorder_energie_alpha0mnhohminus2}):
\begin{equation}
	E_\pm=m_N+\frac{p^2}{2m_N}+\frac{\delta^2}{2m_N}\pm\frac{\delta|p_3|}{m_N}\label{eq:higherorder_energie_alpha0mnhohminus2_2}.
\end{equation}
We now first treat $E_+$. We note that surfaces of constant energy in momentum space are symmetric with respect to $p_3$ and that we hence can integrate over $p_3\geq 0$ only and then multiply by two. Thus, we can drop the absolute value in (\ref{eq:higherorder_energie_alpha0mnhohminus2_2}) to get
\begin{equation}
	E_+=m_N+\frac{1}{2m_N}\left(p_1^2+p_2^2+(p_3+\delta)^2\right).
\end{equation}
From this we see that a surface of constant energy is a sphere centered at $p_3=-\delta$, however only for $p_3\geq 0$. For $p_3\leq 0$ we have the mirror image. Fig. \ref{fig:fermiflaeche1} shows a cut of a surface of constant energy in the $p_2-p_3$ plane.
\\
Shifting the integration variable $p_3$ to $p_3'=p_3+\delta$ allows us to integrate over the upper part of the sphere using ordinary spherical coordinates, where the polar angle $\theta$ runs from 0 to $\xi=\arccos \delta/p$. The absolute value $p$ of the momentum runs from $\delta$ to some momentum $p_{F+}$, which is the momentum reaching from $p_3=-\delta$ to the surface of the Fermi volume.\footnote{Since the Fermi volume is not a sphere we cannot call this momentum Fermi momentum. Nevertheless we denote it by $p_F$.} These considerations are only true if $p_{F+}\geq\delta$ (otherwise we will get no contribution from $E_+$). For large particle densities and if we restrict ourselves to small values of $a$ this is satisfied. For the particle number we then get (note the additional factor of 2)
\begin{align}
N_+&=4\left(\frac{L}{2\pi}\right)^3\int\limits_\delta^{p_{F+}}dp\,p^2\int\limits_0^{2\pi}d\varphi\int\limits_0^\xi d\theta\,\sin\theta=8\pi\left(\frac{L}{2\pi}\right)^3\int\limits_\delta^{p_{F+}}dp\,p^2\int\limits_{\delta/p}^1 dz\nonumber\\
&=\frac{L^3}{\pi^2}\int\limits_\delta^{p_{F+}}dp\,p^2\left(1-\frac{\delta}{p}\right)=\frac{L^3p_{F+}^3}{3\pi^2}\left(1-\frac{3\delta}{2p_{F+}}+\frac{\delta^3}{2p_{F+}^3}\right)\label{eq:alphamnquadratNplus},
\end{align}
where in the first step we performed the variable transformation $z=\cos\theta$. Analogously we find the total energy due to $E_+$ states:
\begin{align}
E_\text{tot}^+&=4\left(\frac{L}{2\pi}\right)^3\int\limits_\delta^{p_{F+}}dp\,p^2\int\limits_0^{2\pi}d\varphi\int\limits_0^\xi d\theta\,\sin\theta\left(m_N+\frac{1}{2m_N}p^2\right)\nonumber\\
&=m_NN_++\frac{L^3}{10\pi^2m_N}\left(p_{F+}^5-\frac{5}{4}\delta p_{F+}^4+\frac{\delta^5}{4}\right)\label{eq:alphamnquadratEplus}.
\end{align}

\begin{figure}[h]
\centering
\psfrag{z}{\small $\zeta$}
\psfrag{d}{\small $\delta$}
\psfrag{p2}{\small $p_2$}
\psfrag{p3}{\small $p_3$}
\includegraphics[scale=0.7]{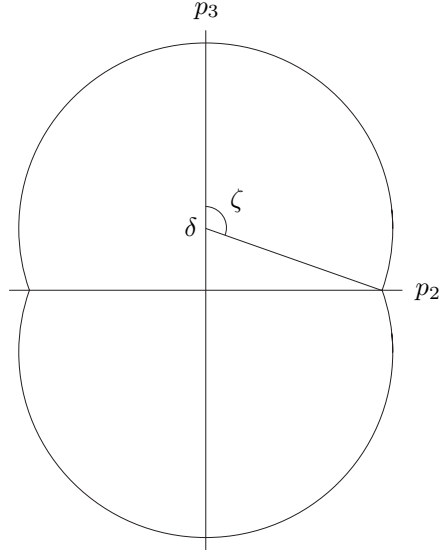}
\caption{Cut of a surface of constant energy for $E_-$ in the $p_2-p_3$ plane}\label{fig:fermiflaeche2}
\end{figure}

\begin{figure}[h]
\centering
\psfrag{x}{\small $n_+/n$}
\psfrag{d}{\small $\delta\,[\text{MeV}]$}
\psfrag{eps}{\small $\epsilon_\text{tot}\cdot 10^{-8}\,[\text{MeV}]$}
\includegraphics[scale=0.8]{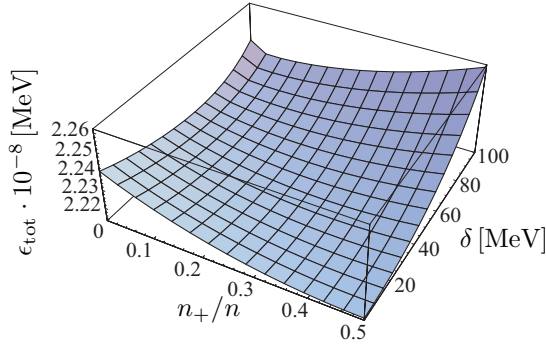}
\caption{The total energy density at very small values of $\alpha_0$ and $n^{1/3}=73.61\,\text{MeV}$}\label{fig:alphamnquadrat}
\end{figure}

Next we consider $E_-$. Due to symmetry we may take again only $p_3\geq 0$ into account. The dispersion relation may then be written as
\begin{equation}
	E_-=m_N+\frac{1}{2m_N}\left(p_1^2+p_2^2+(p_3-\delta)^2\right),
\end{equation}
which shows that a surface of constant energy is again a sphere but this time centered around $p_3=\delta$. For $p_3\leq 0$ we have once again the mirror image. Fig. \ref{fig:fermiflaeche2} shows again a cut of such a surface in the $p_2-p_3$ plane.
\\
After shifting the integration variable $p_3$ to $p_3'=p_3-\delta$ we again use spherical coordinates. Let us denote the absolute value of the momentum that reaches from $p_3=\delta$ to the Fermi surface by $p_{F-}$. Again under the assumption that $p_{F-}\geq\delta$ we now have to split the integration into two parts: A first part where $p\leq\delta$ and where the polar angle $\theta\in[0,\pi]$ and a second part where $p\geq\delta$ and where $\theta$ runs from 0 to $\zeta=\pi-\arccos\delta/p$:
\begin{align}
N_-&=4\left(\frac{L}{2\pi}\right)^3\int\limits_0^\delta dp\,p^2\int\limits_0^{2\pi}d\varphi\int\limits_0^\pi d\theta\,\sin\theta+\int\limits_\delta^{p_{F-}} dp\,p^2\int\limits_0^{2\pi}d\varphi\int\limits_0^\zeta d\theta\,\sin\theta\nonumber\\
&=16\pi\left(\frac{L}{2\pi}\right)^3\frac{\delta^3}{3}+8\pi\left(\frac{L}{2\pi}\right)^3\int\limits_\delta^{p_{F-}}dp\,p^2\int\limits_{-\delta/p}^1 dz\nonumber\\
&=\frac{2L^3\delta^3}{3\pi^2}+\frac{L^3}{\pi^2}\int\limits_\delta^{p_{F-}}p^2\left(1+\frac{\delta}{p}\right)=\frac{L^3p_{F-}^3}{3\pi^2}\left(1+\frac{3\delta}{2p_{F-}}-\frac{\delta^3}{2p_{F-}^3}\right).\label{eq:alphamnquadratNminus}
\end{align}
Similarly the total energy due to $E_-$ states is calculated, the result being
\begin{equation}
	E_\text{tot}^-=m_NN_-+\frac{L^3}{10\pi^2m_N}\left(p_{F-}^5+\frac{5}{4}\delta p_{F-}^4-\frac{\delta^5}{4}\right)\label{eq:alphamnquadratEminus}.
\end{equation}
Dividing (\ref{eq:alphamnquadratNplus}), (\ref{eq:alphamnquadratEplus}), (\ref{eq:alphamnquadratNminus}) and (\ref{eq:alphamnquadratEminus}) by $L^3$ we may write the particle number and energy densities in summary:
\begin{align}
n_\pm&=\frac{p_{F\pm}^3}{3\pi^2}\left(1\mp\frac{3\delta}{2p_{F\pm}}\pm\frac{\delta^3}{2p_{F\pm}^3}\right)\label{eq:alphamnquadratnplusminus}\\
\epsilon_\text{tot}^\pm &=m_Nn_\pm+\frac{1}{10\pi^2m_N}\left(p_{F\pm}^5\mp\frac{5}{4}\delta p_{F\pm}^4\pm\frac{\delta^5}{4}\right).\label{eq:alphamnquadratepsilonplusminus}
\end{align}
(\ref{eq:alphamnquadratnplusminus}) may be solved for $p_{F\pm}$; the expression reads
\begin{equation}
	p_{F\pm}=\frac{1}{2}\left[k(n_\pm,\mp\delta)\pm\delta\right]+\frac{\delta^2}{k(n_\pm,\mp\delta)}\label{eq:alphamnquadratpFplusminus},
\end{equation}
where
\begin{equation}
	k(n,\delta)=\left(12n\pi^2+\delta^3+2\pi\sqrt{36n^2\pi^2+6n\delta^3}\right)^{1/3}.
\end{equation}
Plugging (\ref{eq:alphamnquadratpFplusminus}) into (\ref{eq:alphamnquadratepsilonplusminus}) and substituting $n_-=n-n_+$ finally allows us to express the total fermionic energy density $\epsilon_f=\epsilon_\text{tot}^++\epsilon_\text{tot}^-$ in terms of of $n$ and $n_+$. The plot in fig. \ref{fig:alphamnquadrat} shows the total energy density\footnote{At this order $\epsilon_p=-F^2M_\pi^2+\mathcal{O}(m_N^{-4})$.} $\epsilon_\text{tot}=\epsilon_f+\epsilon_p$ at $n^{1/3}=73.61\,\text{MeV}$ as a function of $n_+$ and $\delta$.
\\
The plot shows a similar behavior for different values of $n$ and confirms what we wanted to show: For very small values of $\alpha_0$ the parameters $\delta$ and hence $a$ go to zero and the spiral disappears.

\section{Higher Orders in the Baryonic Sector}\label{sec:higherordersinthebaryonicsector}
Having gone away from the chiral limit and therefore including chiral breaking terms in the pion sector already in section \ref{sec:awayfromthechirallimit}, we should also consider these in the baryonic sector. This leads to the next order of the fermionic Lagrangian, which is given by (\ref{eq:secondorderfermioniclagrangian}). In our spiral configuration we have
\begin{align*}
\left[u_0,u_\mu\right]&=0\quad\text{(since $u_0=0$)}\\
\left[u_i,u_j\right]&=\left[c_i,c_j\right]=\left[c_{i,a}\tau^a,c_{j,b}\tau^b\right]=\left[c_{i,2}\tau^2,c_{j,2}\tau^2\right]=\beta_i\beta_j\left[\tau^2,\tau^2\right]=0,
\end{align*}
i.e. the term proportional to $c_4$ disappears. Furthermore, if we switch off all external fields except $s=\mathcal{M}=\text{diag}(m_u,m_d)$, $\chi$ of (\ref{eq:defchi}) becomes $\chi=2B\mathcal{M}$. The second order baryonic Lagrangian then reads
\begin{equation}
	\mathcal{L}_{\pi N}^{(2)}=2Bc_1\langle u^\dagger\mathcal{M}u^\dagger+u\mathcal{M}u\rangle\bar{\Psi}\Psi-\frac{c_2}{4m_N^2}\langle u_iu_j\rangle\left(\bar{\Psi}D^iD^j\Psi+\text{h.c.}\right)+\frac{c_3}{2}\langle u_iu^i\rangle\bar{\Psi}\Psi.
\end{equation}
Let us introduce the following notation:
\begin{equation}
	t^1=2Bc_1\langle u^\dagger\mathcal{M}u^\dagger+u\mathcal{M}u\rangle,\quad t^2_{ij}=-\frac{c_2}{4m_N^2}\langle u_iu_j\rangle,\quad t^3=\frac{c_3}{2}\langle u_iu^i\rangle.
\end{equation}
Using (\ref{eq:covariatderivativebaryons}), (\ref{eq:konstconvielb}) and the fact that the connection $\Gamma_\mu=d_\mu$ is antihermitian (see \ref{eq:ciunddi}) we find
\begin{align*}
\bar{\Psi}D^iD^j\Psi+\text{h.c.}&=\bar{\Psi}D^iD^j\Psi+\left(\bar{\Psi}D^iD^j\Psi\right)^\dagger=\bar{\Psi}D^iD^j\Psi+\left(\Psi^\dagger\gamma^0D^iD^j\Psi\right)^\dagger\\
&=\bar{\Psi}D^iD^j\Psi+\Psi^\dagger D^{j\dagger}D^{i\dagger}\gamma^0\Psi=\bar{\Psi}D^iD^j\Psi+\bar{\Psi}D^{j\dagger}D^{i\dagger}\Psi\\
&=\bar{\Psi}\left(D^iD^j+D^{j\dagger}D^{i\dagger}\right)\Psi=\bar{\Psi}\left[(\partial^i+d^i)(\partial^j+d^j)+(\partial^j-d^j)(\partial^i-d^i)\right]\Psi\\
&=\bar{\Psi}\left(2\partial^i\partial^j+d^id^j+d^jd^i\right)\Psi=\bar{\Psi}\left(2\partial^i\partial^j+\left\{d^i,d^j\right\}\right)\Psi.
\end{align*}
Remembering (\ref{eq:defdeltai}) we have $d^i=-i\delta^i\tau^3$ and therefore
\begin{equation}
	\left\{d^i,d^j\right\}=-\delta^i\delta^j\left\{\tau^3,\tau^3\right\}=-2\delta^i\delta^j,
\end{equation}
where of course the $2\times 2$ unit matrix is not written (exactly as with the term $2\partial^i\partial^j$). We can now write the Lagrangian as
\begin{equation}
	\mathcal{L}_{\pi N}^{(2)}=\left(t^1+t^3\right)\bar{\Psi}\Psi+2t^2_{ij}\bar{\Psi}\left(\partial^i\partial^j-\delta^i\delta^j\right)\Psi.
\end{equation}
We now go on as in section \ref{sec:singleparticlehamiltonian}, i.e. extract the single-particle Hamiltonian via Euler-Lagrange equations. Its individual parts read (using the full Lagrangian $\mathcal{L}_f=\mathcal{L}_{\pi N}^{(1)}+\mathcal{L}_{\pi N}^{(2)}$):
\begin{align*}
\partial^\mu\frac{\partial\mathcal{L}_f}{\partial\partial^\mu\bar{\Psi}}&=\underbrace{\partial^\mu\frac{\partial\mathcal{L}_{\pi N}^{(1)}}{\partial\partial^\mu\bar{\Psi}}}_{\hidewidth\substack{= 0,\text{ see}\\\text{sec. \ref{sec:singleparticlehamiltonian} }}\hidewidth}+\underbrace{\partial^\mu\frac{\partial\mathcal{L}_{\pi N}^{(2)}}{\partial\partial^\mu\bar{\Psi}}}_{=0}=0\\
\frac{\partial\mathcal{L}_f}{\partial\bar{\Psi}}&=\frac{\partial\mathcal{L}_{\pi N}^{(1)}}{\partial\bar{\Psi}}+\frac{\partial\mathcal{L}_{\pi N}^{(2)}}{\partial\bar{\Psi}}=\frac{\partial\mathcal{L}_{\pi N}^{(1)}}{\partial\bar{\Psi}}+\left(t^1+t^3\right)\Psi+2t^2_{ij}\left(\partial^i\partial^j-\delta^i\delta^j\right)\Psi.
\end{align*}
The term $\partial\mathcal{L}_{\pi N}^{(1)}/\partial\bar{\Psi}$ has already been calculated in section \ref{sec:singleparticlehamiltonian}, namely in (\ref{eq:eom}). In this term the sum $\gamma^\mu\partial_\mu\Psi$ appears. This sum was split up into the temporal and spatial part, the spatial part was brought to the other side of the equation of motion and the whole equation was multiplied by $\gamma^0$ in order to have the combination $i\partial_0\Psi$ isolated on one side of the equation. Since there is no additional time derivative originating from $\mathcal{L}_{\pi N}^{(2)}$, these steps can immediately be repeated:
\begin{align*}
\text{EOM: } 0 &= i\gamma^\mu\partial_\mu\Psi+i\gamma^id_i\Psi-m_N\Psi+\frac{g_A}{2}\gamma^i\gamma_5c_i\Psi+\left(t^1+t^3\right)\Psi+2t^2_{ij}\left(\partial^i\partial^j-\delta^i\delta^j\right)\Psi\\
i\gamma^0\partial_0\Psi &=\left[ -i\gamma^i\partial_i-i\gamma^id_i+m_N-\frac{g_A}{2}\gamma^i\gamma_5c_i-\left(t^1+t^3\right)-2t^2_{ij}\left(\partial^i\partial^j-\delta^i\delta^j\right)\right]\Psi\\
i\partial_0\Psi &= \gamma^0\left[\gamma^ip_i-i\gamma^id_i+m_N-\frac{g_A}{2}\gamma^i\gamma_5c_i-\left(t^1+t^3\right)-2t_{ij}^2\left(\partial^i\partial^j-\delta^i\delta^j\right)\right]\Psi\\
&= \hat{H}_1\Psi-\left(t^1+t^3\right)\gamma^0\Psi-2t_{ij}^2\gamma^0\left(\partial^i\partial^j-\delta^i\delta^j\right)\Psi.
\end{align*}
$\hat{H}_1$ is the first order Hamiltonian of (\ref{eq:fermionhamiltonian}), while the rest can now be identified as $\hat{H}_2$. Using $\partial^i\partial^j\Psi=\partial^i\left(-ip^j\Psi\right)=-ip^j\partial^i\Psi=-p^ip^j\Psi$, we get
\begin{equation}
	\hat{H}_2=-\left(t^1+t^3\right)\gamma^0+2t_{ij}^2\gamma^0\left(p^ip^j+\delta^i\delta^j\right).
\end{equation}
Comparing the $\gamma$ matrix structure of $\hat{H}_2$ with $\hat{H}_1$ we see that $\hat{H}_2$ contains only terms proportional to $\gamma^0$ (actually $\text{diag}(\gamma^0,\gamma^0)$). If $\hat{H}_1=a_1\gamma^0+g\Gamma$ is an operator (with $\Gamma$ being a linear combination of $\gamma$ matrices or products thereof, $g$ denoting the set of coefficients) leading to some eigenvalues $f(a_1,g)$, the eigenvalues of a new operator $\hat{H}=\hat{H}_1+a_2\gamma^0$ are easily found:
\begin{equation}
	\hat{H}=a_1\gamma^0+g\Gamma+a_2\gamma^0=(a_1+a_2)\gamma^0+g\Gamma\quad\longrightarrow\quad f(a_1+a_2,g).
\end{equation}
In $\hat{H}_1$ it is the mass term being proportional to $\gamma^0$. Hence, in order to find the new eigenvalues of the full Hamiltonian $\hat{H}=\hat{H}_1+\hat{H}_2$ we only have to replace
\begin{equation}
	m_N\longrightarrow m_N+2t_{ij}^2\left(p^ip^j+\delta^i\delta^j\right)-\left(t^1+t^3\right)\label{eq:masseersetzen}
\end{equation}
in the old eigenvalues of $\hat{H}_1$.

Let us calculate the remaining quantities: As $\mathcal{M}$ is diagonal it commutes with $u$ and $u^\dagger$ and we can write
	\[t^1=2Bc_1\langle\mathcal{M}U^\dagger+\mathcal{M}U\rangle.
\]
This trace we have already calculated in (\ref{eq:massentermberechnen}). Assuming again the isospin limit $m_u=m_d=m$ we therefore get
\begin{equation}
	t^1=8Bc_1m\cos\alpha_0=4M_\pi^2c_1\cos\alpha_0,
\end{equation}
where in the last step we used (\ref{eq:masse1}).

The trace $\langle u_i u_j\rangle$ we calculate using the constant vielbein of (\ref{eq:finaltransformedvielbein}), (\ref{eq:ciunddi}) and the definition (\ref{eq:defbetai}). For $t^2_{ij}$ we then get
\begin{equation}
	t^2_{ij}=-\frac{c_2}{4m_N^2}\langle c_ic_j\rangle=-\frac{c_2}{4m_N^2}\langle c_{i,2}\tau^2c_{j,2}\tau^2\rangle=-\frac{c_2\beta_i\beta_j}{4m_N^2}\langle\mathds{1}\rangle=-\frac{c_2\beta_i\beta_j}{2m_N^2}.
\end{equation}
In the replacement (\ref{eq:masseersetzen}) we now can write
\begin{align}
t_{ij}^2p^ip^j &= -\frac{c_2\beta_i\beta_j}{2m_N^2}p^ip^j=-\frac{c_2}{2m_N^2}\beta_i\beta_jp_ip_j=-\frac{c_2}{2m_N^2}\left(\vec{\beta}\cdot\vec{p}\right)^2\\
t_{ij}^2\delta^i\delta^j &= -\frac{c_2}{2m_N^2}\left(\vec{\beta}\cdot\vec{\delta}\right)^2.
\end{align}
Finally, we calculate the trace $\langle u_iu^i\rangle$ and find
\begin{equation}
	t^3=\frac{c_3}{2}\langle c_ic^i\rangle=\frac{c_3}{2}\beta_i\beta^i\langle\mathds{1}\rangle=c_3\beta_i\beta^i=-c_3\beta_i\beta_i=-c_3\beta^2.
\end{equation}
The replacement (\ref{eq:masseersetzen}) now reads
\begin{equation}
	m_N\longrightarrow m_N-\frac{c_2}{m_N^2}\left[\left(\vec{\beta}\cdot\vec{p}\right)^2-\left(\vec{\beta}\cdot\vec{\delta}\right)^2\right]+c_3\beta^2-4M_\pi^2c_1\cos\alpha_0.
\end{equation}

After having performed this replacement in the energy expression of (\ref{eq:energieplus}) we expand it in powers of $1/m_N$ up to order $1/m_N$ which yields\footnote{i.e. we again consider the lowest order. Even though in this expansion we found the peculiar behavior that the minimum of $\epsilon_\text{tot}$ does not approach $\alpha_0=0$ as $n\searrow n_c$, we still can compare the results with (\ref{eq:erstekritdichte}) and (\ref{eq:lowerorderresult}) in order to get an impression of the effects the higher order terms in the Lagrangian have.}
\begin{equation}
	E_\pm=m_N+\frac{p^2}{2m_N}+\frac{\delta^2}{2m_N}+c_3\beta^2-4c_1M_\pi^2\cos\alpha_0\pm\frac{1}{2}g_A\beta.
\end{equation}
Since at this order the energy only depends on $p^2$ we again talk about a Fermi sphere in momentum space and accordingly can make use of the formula (\ref{eq:particlenumberdensity}) and just slightly modify (\ref{eq:fermionischeenergiedichtenersterordnung}) to
\begin{equation}
	\epsilon_\text{tot}^\pm=\left(m_N+\frac{\delta^2}{2m_N}+c_3\beta^2-4c_1M_\pi^2\cos\alpha_0\pm\frac{1}{2}g_A\beta\right)n_\pm+\frac{(3\pi^2 n_\pm)^{5/3}}{10\pi^2 m_N}.
\end{equation}
Correspondingly, the total fermionic energy density now reads
\begin{align}
\epsilon_f &= \epsilon_\text{tot}^++\epsilon_\text{tot}^-\nonumber\\
&= \left(m_N+\frac{\delta^2}{2m_N}+c_3\beta^2-4c_1M_\pi^2\cos\alpha_0\right)n+\frac{g_A\beta}{2}\left(n_+-n_-\right)+\frac{(3\pi^2)^{5/3}}{10\pi^2m_N}\left(n_+^{5/3}+n_-^{5/3}\right)\label{eq:higherorderepsbaryonicsector}
\end{align}
Since the low energy constants $c_i$ are of order $m_N^{-1}$ (see section \ref{sec:nextorderbaryoniclagrangian}) the expression (\ref{eq:higherorderepsbaryonicsector}) shows us that, in terms of an expansion in powers of $m_N^{-1}$, we are actually not examining higher orders here, but rather completing the first order expansion.

In the pionic sector we stay at order $\mathcal{O}(p^2)$ and therefore again use (\ref{eq:pionicenergydensity}) for the pionic energy density. After having replaced $\beta$ and $\delta$ in terms of $a$ and $\alpha_0$ and setting $n_-=n-n_+$ the total energy density is
\begin{multline}
\epsilon_\text{tot}=\left(m_N+\frac{a^2\cos^2\alpha_0}{8m_N}+c_3a^2\sin^2\alpha_0-4c_1M_\pi^2\cos\alpha_0\right)n-\frac{g_A}{2}a\sin\alpha_0(n-2n_+)\\
+\frac{(3\pi^2)^{5/3}}{10\pi^2m_N}\left(n_+^{5/3}+(n-n_+)^{5/3}\right)+\frac{F^2}{2}a^2\sin^2\alpha_0-F^2M_\pi^2\cos\alpha_0.\label{eq:totalenergydensitymitahigherordersbaryonic}
\end{multline}
The value of $a$ that minimizes $\epsilon_\text{tot}$ is found to be
\begin{equation}
	a_\text{min}=\frac{2m_Ng_A\sin\alpha_0(n-2n_+)}{n\cos^2\alpha_0+4m_N(F^2+2c_3n)\sin^2\alpha_0}\label{eq:aminhigherorderbaryon},
\end{equation}
which, after plugging into (\ref{eq:totalenergydensitymitahigherordersbaryonic}) leads us to the total energy density as a function of $n$, $n_+$ and $\alpha_0$.

\subsection{Chiral Limit}\label{sec:higherorderbaryonchirallimit}
Even though we basically included higher order terms of the fermionic Lagrangian in order to take into account the effects of chiral symmetry breaking, $\mathcal{L}_{\pi N}^{(2)}$ still has terms that do not explicitly break chiral symmetry. Let us investigate their influence by considering the chiral limit, such that we can compare the results with those obtained in section \ref{sec:ersteordnungchiralerlimes}.

Varying $\epsilon_\text{tot}$ with $a$ replaced by $a_\text{min}$ we find in complete analogy to section \ref{sec:ersteordnungchiralerlimes} the condition $\sin 2\alpha_0=0$. To decide which one of the two solutions, $\alpha_0=0$ or $\alpha_0=\pi/2$, minimizes the energy density, we calculate the second derivative of $\epsilon_\text{tot}$ at $\alpha_0=\pi/2$ and get an expression very similar to that of (\ref{eq:ersteordnungzweiteablnachalpha0beipizweitel}):
\begin{equation}
	\frac{\partial^2\epsilon_\text{tot}}{\partial\alpha_0^2}\Bigr|_{\alpha_0=\frac{\pi}{2}}=\frac{g_A^2n(n-2n_+)^2}{16m_N(F^2+2c_3n)^2}.
\end{equation}
Again this is positive for $0\leq n_+\leq n/2$ and we conclude that, as before, in the chiral limit $\alpha_0=\pi/2$, i.e. the pion vector moves in the equatorial plane. The expression for the total energy density then simplifies to
\begin{equation}
	\epsilon_\text{tot}=m_N n-\frac{gA^2}{8(F^2+2c_3n)}(n-2n_+)^2+\frac{(3\pi^2)^{5/3}}{10\pi^2 m_N}\left(n_+^{5/3}+(n-n_+)^{5/3}\right).
\end{equation}
We vary this expression with respect to $n_+$,
\begin{equation}
	\frac{\partial\epsilon_\text{tot}}{\partial n_+}=\frac{(3\pi^2)^{5/3}}{6\pi^2m_N}\left(n_+^{2/3}-(n-n_+)^{2/3}\right)-\frac{g_A^2}{2(F^2+2c_3n)}(2n_+-n)\stackrel{!}{=}0,
\end{equation}
and find (cf. section \ref{sec:ersteordnungchiralerlimes}) that for small $n$ this equation has the only solution at $n_+=n/2$, which, because of (\ref{eq:aminhigherorderbaryon}), means a non-spiral configuration. Increasing $n$ above the critical particle density, the second derivative of $\epsilon_\text{tot}$ with respect to $n_+$ at $n_+=n/2$,
\begin{equation}
	\frac{\partial^2\epsilon_\text{tot}}{\partial n_+^2}\Bigr|_{n_+=\frac{n}{2}}=\frac{2(3\pi^2)^{5/3}}{9\pi^2m_N}\left(\frac{2}{n}\right)^{1/3}-\frac{g_A^2}{F^2+2c_3n},
\end{equation}
gets negative and the extremum at $n_+=n/2$ becomes a maximum i.e. the minimum lies somewhere at $n_+\neq n/2$ and the spiral configuration will be favored. The critical particle number density $n_c$ is the density where the second derivative becomes zero. For $c_3=-3.4m_N^{-1}$ (s. section \ref{sec:nextorderbaryoniclagrangian}) this is the case at
\begin{equation}
	\mathring{n}_c=79.2\,\text{GeV}^3\quad\text{or}\quad \mathring{n}_c^{1/3}=42.95\,\text{MeV}
\end{equation}
Using $c_3=-4.2m_N^{-1}$ the critical density even drops more:

\begin{equation}
	\mathring{n}_c=75.5\,\text{GeV}^3\quad\text{or}\quad \mathring{n}_c^{1/3}=42.26\,\text{MeV}
\end{equation}

\noindent Hence, due to the additional terms of the higher order baryonic Lagrangian $\mathcal{L}_{\pi N}^{(2)}$, compared with (\ref{eq:erstekritdichte}) the critical particle number density in the chiral limit drops by about $8-10\%$, depending on the exact value of $c_3$.

\subsection{Back to $M_\pi\neq 0$}\label{sec:higherorderbaryonawayfromchirallimit}
Let us now go back to (\ref{eq:totalenergydensitymitahigherordersbaryonic}). Since $\alpha_0\neq\pi/2$ for $M_\pi\neq 0$ we have to deal with both $n_+$ and $\alpha_0$. We treat this problem as we did in section \ref{sec:awayfromthechirallimit}. We notice that due to (\ref{eq:aminhigherorderbaryon}) a non-spiral configuration is given for $n_+=n/2$. This leads to
\begin{equation}
	\epsilon_\text{tot}(n_+=n/2,\alpha_0)=m_Nn+\frac{(3\pi^2n)^{5/3}}{2^{5/3}5\pi^2m_N}-M_\pi^2(F^2+4c_1n)\cos\alpha_0,
\end{equation}
which is minimized for $\alpha_0=0$ (if $c_1=-0.9m_N^{-1}$ ($-0.6m_N^{-1}$) the expression $F^2+4c_1n>0$ for $n^{1/3}<121.9\,\text{MeV}$ ($139.6\,\text{MeV}$)).

\begin{figure}[h]
\centering
\psfrag{dn}{\small $n^{1/3}\,[\text{MeV}]$}
\psfrag{f}{\small $d_\text{max}(n)\cdot 10^{-5}\,[\text{MeV}]$}
\includegraphics[width=6cm]{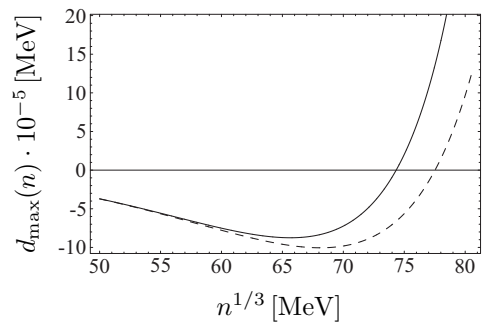}
\caption[]{$d_\text{max}(n)$ with the physical pion mass and for $c_1=-0.9m_N^{-1}$, $c_3=-4.2m_N^{-1}$ (continuous) and $c_1=-0.6m_N^{-1}$, $c_3=-3.4m_N^{-1}$ (dashed) respectively}\label{fig:higherorderdmax}
\end{figure}

As in (\ref{eq:differenzspiralnonspiral}) we now define the function $d(n_+,\alpha_0)$ as the difference between the energy densities of the non-spiral and the spiral configuration. Plotting then $d(n_+,\alpha_0)$ for different pion masses $M_\pi$ we find a behavior very similar to that found in the lower order calculation in section \ref{sec:awayfromthechirallimit}: With increasing pion mass the critical parameter $n_+^c$ (i.e. the value of $n_+$ for which $d(n_+^c,\alpha_0)=0$ for exactly one $\alpha_0$) goes to towards zero (cf. figure \ref{fig:contourplots}). For pion masses $M_\pi\gtrsim 20\,\text{MeV}$ we can restrict our analysis to the parameter subspace $n_+=0$. We then search the maximum values of $d(0,\alpha_0)$ for a given set of values for $n$, defining the function $d_\text{max}(n)$. The critical particle number density $n_c$ we are looking for satisfies $d_\text{max}(n_c)=0$, because then the energy density of the spiral configuration is equal to that of the non-spiral configuration. The plot in fig. \ref{fig:higherorderdmax} shows the function $d_\text{max}(n)$ for both $c_1=-0.9m_N^{-1}$, $c_3=-4.2m_N^{-1}$ (continuous graph) and $c_1=-0.6m_N^{-1}$, $c_3=-3.4m_N^{-1}$ (dashed graph) at the physical pion mass. The two sets of low energy constants $c_i$ are the ones given in section \ref{sec:nextorderbaryoniclagrangian}.
The critical densities for $c_1=-0.9m_N^{-1}$, $c_3=-4.2m_N^{-1}$ are numerically found to be at:
\begin{align}
n_c&=410.8\,\text{GeV}^3\quad\text{or}\quad n_c^{1/3}=74.34\,\text{MeV}\\
%n_{c,p}&=460.4\,\text{GeV}^3\quad\text{or}\quad n_{c,p}^{1/3}=77.22\,\text{MeV}\\
\bar{n}_c&=196.5\,\text{GeV}^3\quad\text{or}\quad \bar{n}_c^{1/3}=58.14\,\text{MeV},
%\bar{n}_{c,p}&=212.1\,\text{GeV}^3\quad\text{or}\quad \bar{n}_{c,p}^{1/3}=59.64\,\text{MeV},
\end{align}
while for $c_1=-0.6m_N^{-1}$, $c_3=-3.4m_N^{-1}$ we find
\begin{align}
n_c&=465.9\,\text{GeV}^3\quad\text{or}\quad n_c^{1/3}=77.52\,\text{MeV}\\
%n_{c,p}&=517.3\,\text{GeV}^3\quad\text{or}\quad n_{c,p}^{1/3}=80.28\,\text{MeV}\\
\bar{n}_c&=213.4\,\text{GeV}^3\quad\text{or}\quad \bar{n}_c^{1/3}=59.75\,\text{MeV}.
%\bar{n}_{c,p}&=228.3\,\text{GeV}^3\quad\text{or}\quad \bar{n}_{c,p}^{1/3}=61.11\,\text{MeV}
\end{align}
Compared to the critical particle number density we found at lower order in section \ref{sec:awayfromthechirallimit}, for $M_\pi=135\,\text{MeV}$ this means a $24-27\%$ decrease and for $M_\pi=40\,\text{MeV}$ a $15-17\%$ decrease of the numerical values, depending on the values of the low energy constants $c_1$ and $c_3$.

\chapter{Conclusion and Outlook}
In this thesis we have shown that in the framework of ChPT the pionic vacuum may change from a homogeneous to a spiral phase if the density of fermions is chosen to lie in the region of nuclear matter density. The numerical values for the critical fermionic density are collected here:

In the expansion of the fermionic energy up to $\mathcal{O}(m_N^{-1})$ but without the terms of $\mathcal{L}_{\pi N}^{(2)}$ (see sections \ref{sec:ersteordnungchiralerlimes} and \ref{sec:awayfromthechirallimit}) we found\footnote{As a reminder: The symbol $\circ$ denotes the chiral limit while a bar indicates a pion mass of $40\,\text{MeV}$.}
\begin{align}
\mathring{n}_c^{1/3}&=46.81\,\text{MeV} & \bar{n}_c^{1/3}&=70.16\,\text{MeV} & n_c^{1/3}&=102.27\,\text{MeV}.
%n_{c,p}^{1/3}&=102.30\,\text{MeV} & \mathring{n}_{c,p}^{1/3}&=46.6\,\text{MeV} & \bar{n}_{c,p}^{1/3}&=70.05\,\text{MeV}\nonumber
\end{align}
Including also the chiral breaking terms of $\mathcal{L}_{\pi N}^{(2)}$ bringing additional contributions of $\mathcal{O}(m_N^{-1})$ (see sections \ref{sec:higherorderbaryonchirallimit} and \ref{sec:higherorderbaryonawayfromchirallimit}), the values change to
%\begin{align}
%n_{c,0}^{1/3}&=74.34\,\text{MeV} & \mathring{n}_{c,0}^{1/3}&=42.26\,\text{MeV} & \bar{n}_{c,0}^{1/3}&=58.14\,\text{MeV}\\
%n_{c,p}^{1/3}&=77.22\,\text{MeV} & \mathring{n}_{c,p}^{1/3}&=42.85\,\text{MeV} & \bar{n}_{c,p}^{1/3}&=59.64\,\text{MeV}\nonumber
%\end{align}
%with $c_1=-0.9m_N^{-1}$, $c_3=-4.2m_N^{-1}$ or to
\begin{align}
\mathring{n}_c^{1/3}&=42.95\,\text{MeV} & \bar{n}_c^{1/3}&=59.75\,\text{MeV} & n_c^{1/3}&=77.52\,\text{MeV},
%n_{c,p}^{1/3}&=80.28\,\text{MeV} & \mathring{n}_{c,p}^{1/3}&=43.44\,\text{MeV} & \bar{n}_{c,p}^{1/3}&=61.11\,\text{MeV}\nonumber
\end{align}
when using $c_1=-0.6m_N^{-1}$, $c_3=-3.4m_N^{-1}$. Finally, going to higher orders in the expansion of the fermionic energy (see sections \ref{sec:expansionforalpha0simo1} and \ref{sec:expansionforalpha0simmNhochminus1}) led us to
\begin{align}
\mathring{n}_c^{1/3}&=47.36\,\text{MeV} & \bar{n}_c^{1/3}&=55.62\,\text{MeV} & n_c^{1/3}&=73.61\,\text{MeV}.
%n_{c,p}^{1/3}&=73.43\,\text{MeV} & \bar{n}_{c,p}^{1/3}&=55.48\,\text{MeV}\nonumber
\end{align}
The influence of the uncertainties in the low energy constants was found to be relatively small (see section \ref{sec:ersteordnungchiralerlimes}).

We have several possibilities of comparing the results originating from different order calculations. Let us first take a look at the result obtained using the physical pion mass. When excluding the terms of $\mathcal{L}_{\pi N}^{(2)}$ and going from the first order expansion of the fermionic energy to next order, the critical density drops from $102.27\,\text{MeV}$ to $73.61\,\text{MeV}$. This is a 28\% decrease, which is quite large. Staying at the lowest order of the fermionic energy expansion but including the terms of $\mathcal{L}_{\pi N}^{(2)}$ the critical density drops to $77.52\,\text{MeV}$, which is a 24\% decrease. However, as mentioned in section \ref{sec:higherordersinthebaryonicsector}, including the terms of the next order baryonic Lagrangian was in principle completing the expansion of $\mathcal{O}(m_N^{-1})$. It would therefore also be interesting to examine both steps at once, i.e. go to $\mathcal{O}(m_N^{-2})$ in the expansion of the energy eigenvalue and to include the terms of $\mathcal{L}_{\pi N}^{(2)}$.

Let us have a look at the results that were obtained using $M_\pi=40\,\text{MeV}$. The two ways of going to higher orders just mentioned imply corrections of 21\% and 15\% respectively. These are more moderate than in the case of the physical pion mass. Finally, in the chiral limit going to next order in the expansion of the fermionic energy yields an increase of the critical density of 1\% while including $\mathcal{L}_{\pi N}^{(2)}$ causes a decrease of 8\%.

Hence, the smaller the pion mass the smaller are the corrections. The corrections in the case of the physical pion mass should remind us to stay skeptical and that the numerical estimates we make are maybe not very reliable. However, they are nevertheless not too large and we may get confidence that the qualitative picture is correct.

The spiral configuration we investigated in this thesis was an ansatz that allowed analytic solutions of the Dirac equation. It is however by no means clear whether there is not another inhomogeneous pion configuration which could lower even more the total energy density of the many-particle system or which could happen at even smaller particle densities. Hence, the numerical values we found in this thesis should be understood as upper bounds of critical densities for a transition to an inhomogeneous pion configuration to happen.

What is the physical meaning of a spiral configuration we investigated? If we believe our calculations, then already in ordinary nuclear matter the pionic vacuum could differ from the vacuum outside the volume of dense matter. It is therefore not possible to apply an appropriate gauge transformation in order to transform the pion field everywhere to $U(x)=\mathds{1}$ and to establish the ground state of no pions. Hence, observed from outside, inside the dense matter pions could appear, which is known as \textit{pion condensation}. Different from free pions, which decay in a rather short time, our configuration is static and the pions should be considered as a collective phenomenon rather than free particles. A comparison for this might be the neutron; in the free case it decays after approximately 15 minutes, but as soon as it is enclosed in an atomic nucleus it turns out to be stable. Pion condensation in dense matter has been studied by several authors using different approaches, see e.g. \cite{Migdal:1978az, Baym:1973zk, sawyer, scalapino, campbell}.

Although the structure of the interior of neutron stars\footnote{For a short discussion on neutron stars and an estimation of their densities see appendix \ref{app:neutronstars}.} is not yet known in all details, there are models that include the phenomenon of pion condensation inside these superdense star remnants (see e.g. \cite{Fenyi:1992ke, Baym:1973zk, Khodel:2004nt, Suh:1999sw, Suh:2000ni, sawyer, campbell}).

The results in this thesis might hence give some alternative ideas in an approach towards an explanation of pion condensation in nuclear matter and neutron stars.\\

An interesting thing that could be done in further considerations is the analysis of higher order terms in the mesonic Lagrangian (see section \ref{sec:chirallagrangianorderp4}) and going to even higher orders in the baryonic Lagrangian. So far, the calculations at $\mathcal{O}(p^2)$ in the mesonic sector were only tree level calculations. This did not make it necessary to renormalize the low energy constants; we could take them as fixed and more or less precisely known values. Going to $\mathcal{O}(p^4)$ would however imply one-loop calculations with $\mathcal{L}_2$ and tree level calculations with $\mathcal{L}_4$. It is an interesting question how one could then calculate the energy of the system. Also if one goes to higher orders in the baryonic sector one will have to deal with loops. It is then by far not clear whether a Hamiltonian as we constructed it is still a well defined object. An alternative way of finding the energy of a nucleon in the background of pions could be to consider its propagator. This could be calculated using a well defined Lagrangian and it should be possible to make a conclusion for the nucleon energy via the pole of the propagator.

Another direction worth exploring is to switch on electromagnetic effects, i.e. to include virtual photons. Beside the quark mass matrix the quark charge matrix is an alternative way of breaking chiral symmetry explicitly and it would be interesting to find out the order of magnitude of the effects arising therefrom.

In the case of antiferromagnets the transition from a homogeneous to a spiral phase of the staggered magnetization implies a spontaneous breaking of translation symmetry on the lattice. The consequence is a further Goldstone boson (helimagnon). In the case of QCD we can expect a similar phenomenon. It might be interesting to study the properties of the resulting additional Goldstone bosons.

\chapter*{Acknowledgment}
First and foremost, I would like to thank Prof. Gilberto Colangelo for his steady support, for the time he always took when knocking at his door and that he still let me work independently. I sincerely thank also Prof. Uwe-Jens Wiese whose group found the spiral phase in doped antiferromagnets and who wondered whether this could also happen in QCD.

I would like to thank Stefan Lanz and Andreas Fuhrer to whom I could always ask my questions and who supported me especially at the beginning of my thesis. I am grateful to my office colleagues Christoph Weiermann, Martin Schmid and Vidushi Maillart for their support and their contribution to the confortable working atmosphere; it was a pleasure to share the office with them. Many thanks are addressed also to Florian K\"ampfer who introduced me into the very basics of antiferromagnets.

I am deeply indebted to Ottilia H\"anni, Ruth Bestgen and Esther Fiechter for guiding me through the unphysical and administrative aspects of my studies.

I am thankful to my parents for their support and care throughout all the years and for making my studies possible.

Anna, thank you very much for your constant understanding, support and patience.

\newpage
\begin{appendix}
\chapter{The Spiral Configuration}\label{app:spiral}
In this appendix we derive the static spiral configuration (\ref{eq:spiralconfiguration}) from the condition of having a constant connection and a constant mass term in the Lagrangian. For this purpose we use a different parametrization of the matrix $U$ than in (\ref{eq:parametrisierung}), namely
\begin{equation}
	U(x)=\frac{1}{F}\left[\sigma(x)\mathds{1}+i\vec{\tau}\cdot\vec{\pi}(x)\right],\quad \sigma(x)=\sqrt{F^2-\vec{\pi}^2(x)},
\end{equation}
where the three Hermitian fields $\pi^i$ describe pion fields. $\sigma(x)$ is chosen such that $UU^\dagger=\mathds{1}$. For the sake of simplicity let us absorb $F$ into the fields, such that we can write
\begin{equation}
	U(x)=\sigma(x)\mathds{1}+i\tau^i\pi^i(x),\quad \sigma(x)=\sqrt{1-\vec{\pi}^2(x)}.\label{eq:appmatrixU}
\end{equation}
We require the mass term $\langle\mathcal{M}\left(U+U^\dagger\right)\rangle$ to be constant:
\begin{equation}
	U+U^\dagger=2\sigma(x)\cdot\mathds{1}\stackrel{!}{=}\text{const.}\quad\Rightarrow\quad \sigma=\text{const.}\quad\Rightarrow\quad \vec{\pi}^2=\text{const.}\label{eq:piquadratconst}
\end{equation}
In order to calculate the connection $\Gamma_i$ we need the matrix $u$, such that $u^2=U$. We parametrize it as follows:
\begin{equation}
	u(x)=f\cdot\mathds{1}+ig^i\tau^i.
\end{equation}
Then\footnote{using $\tau^i\tau^j=i\epsilon^{ijk}\tau^k+\delta^{ij}\mathds{1}$ and, due to (anti-)symmetry, $g^ig^j\epsilon^{ijk}=0$.}
	\[u^\dagger u=f^2+g^ig^j\tau^i\tau^j=f^2+g^ig^i=\left(f^2+g^2\right)\mathds{1}\stackrel{!}{=}\mathds{1},
\]
from which we find
\begin{equation}
	f^2+g^2=1\quad\Rightarrow\quad f=\sqrt{1-g^2}\label{eq:fqdtplgqdr}.
\end{equation}
We calculate
	\[u^2=f^2+2ifg^i\tau^i-g^ig^j\tau^i\tau^j=f^2-g^ig^i+2ifg^i\tau^i=\left(f^2-g^2\right)\mathds{1}+2ifg^i\tau^i,
\]
such that the requirement $u^2=U$ yields
\begin{align}
f^2-g^2&=\sigma\label{eq:fqdrtminusgqudrt}\\
2fg^i&=\pi^i\label{eq:zwefge}.
\end{align}
Let us now calculate the connection $\Gamma_i$ (without external fields) according to (\ref{eq:defconnection}). The ingredients are easily found to be
\begin{align*}
u^\dagger\partial_i u&=f\partial_i f+if\tau^j\partial_i g^j-ig^j\tau^j\partial_i f+i\epsilon^{jkl}\tau^l g^j\partial_i g^k+g^j\partial_ig^j\\
u\partial_i u^\dagger &= f\partial_i f-if\tau^j\partial_i g^j+ig^j\tau^j\partial_i f+i\epsilon^{jkl}\tau^l g^j\partial_i g^k+g^j\partial_ig^j.
\end{align*}
Using (\ref{eq:fqdtplgqdr}) we have
	\[f\partial_i f+g^j\partial_i g^j=\frac{1}{2}\partial_i\underbrace{\left(f^2+g^jg^j\right)}_{1}=0,
\]
such that the connection may be written as
\begin{equation}
	\Gamma_i=i\epsilon^{jkl}g^j\partial_i g^k\tau^l\label{eq:connklfelder}.
\end{equation}
Plugging (\ref{eq:fqdtplgqdr}) into (\ref{eq:fqdrtminusgqudrt}) yields $1-2g^2=\sigma$ and since $\sigma=\text{const.}$ we conclude that $g^2=\text{const.}$ and therefore $f=\text{const.}$ For this reason, inserting (\ref{eq:zwefge}) into (\ref{eq:connklfelder}) leads to
\begin{equation}
	\Gamma_i=i\epsilon^{jkl}\frac{\pi^j}{2f}\partial_i\frac{\pi^j}{2f}\tau^l=\frac{i\epsilon^{jkl}}{4f^2}\pi^j\partial_i\pi^j\tau^l.
\end{equation}
If we now require the connection to be constant, we consequently must have
\begin{equation}
	\epsilon^{jkl}\pi^j\partial_i\pi^k=\text{const.},\quad\text{or}\quad \vec{\pi}\times\partial_i\vec{\pi}=\mathrel{\mathop:}\vec{d}_i=\text{const.}
\end{equation}
Due to (\ref{eq:piquadratconst}) we may write
\begin{align*}
\vec{\pi}&=\left(\pi_1,\,\pi_2,\,\sqrt{1-\pi_1^2-\pi_2^2}\right)\\
\partial_i\vec{\pi}&=\left(\partial_i\pi_1,\,\partial_i\pi_2,\,-\frac{\pi_1\partial_i\pi_1+\pi_2\partial_i\pi_2}{\sqrt{1-\pi_1^2-\pi_2^2}}\right)\\
\vec{d}_i=\vec{\pi}\times\partial_i\vec{\pi}&=\Bigg(-\frac{\pi_2}{\pi_3}(\pi_1\partial_i\pi_1+\pi_2\partial_i\pi_2)-\pi_3\partial_i\pi_2\,,\,\pi_3\partial_i\pi_1+\frac{\pi_1}{\pi_3}(\pi_1\partial_i\pi_1+\pi_2\partial_i\pi_2)\,,\\
&\hspace{4cm}\pi_1\partial_i\pi_2-\pi_2\partial_i\pi_1\Bigg).
\end{align*}
The relation $\pi_3=\sqrt{1-\pi_1^2-\pi_2^2}$ allows us to simplify $d_i^1$ and $d_i^2$:
\begin{align}
d_i^1&=-\frac{1}{\pi^3}\Bigg(\pi_2\pi_1\partial_i\pi_1+\underbrace{\pi_2^2\partial_i\pi_2+\pi_3^2\partial_i\pi_2}_{(1-\pi_1^2)\partial_i\pi_2}\Bigg)=\frac{\pi_1}{\pi_3}d_i^3-\frac{\partial_i\pi_2}{\pi_3}\label{eq:appdgl1}\\
d_i^2&=\frac{1}{\pi_3}\Bigg(\underbrace{\pi_1^2\partial_i\pi_1+\pi_3^2\partial_i\pi_1}_{(1-\pi_2^2)\partial_i\pi_1}+\pi_1\pi_2\partial_i\pi_2\Bigg)=\frac{\pi_2}{\pi_3}d_i^3+\frac{\partial_i\pi_1}{\pi_3}\label{eq:appdgl2},
\end{align}
where in the last step we inserted $d_i^3=\pi_1\partial_i\pi_2-\pi_2\partial_i\pi_1$. (\ref{eq:appdgl1}) and (\ref{eq:appdgl2}) are equivalent to
\begin{align}
\partial_i\pi_1&=d_i^2\pi_3-d_i^3\pi_2\label{eq:apppartipi1}\\
\partial_i\pi_2&=d_i^3\pi_1-d_i^1\pi_3\label{eq:apppartipi2},
\end{align}
which build, together with
\begin{equation}
	d_i^3=\pi_1\partial_i\pi_2-\pi_2\partial_i\pi_1,\label{eq:apppartipi3}
\end{equation}
a system of differential equations for the $\pi_i$. Inserting (\ref{eq:apppartipi1}) and (\ref{eq:apppartipi2}) into (\ref{eq:apppartipi3}) yields
	\[d_i^3=d_i^3\pi_1^2+d_i^3\pi_2^2-d_i^1\pi_1\pi_3-d_i^2\pi_2\pi_3\quad\Rightarrow\quad d_i^1\pi_1\pi_3+d_i^2\pi_2\pi_3+d_i^3\left(1-\pi_1^2-\pi_2^2\right)=0.
\]
The term in brackets is $\pi_3^2$, such that the equation reads
	\[\pi_3\left(d_i^1\pi_1+d_i^2\pi_2+d_i^3\pi_3\right)=0.
\]
This equation and thus (\ref{eq:apppartipi3}) is trivially satisfied due to the relation
\begin{equation}
	\vec{\pi}\cdot\vec{d}_i=0\label{eq:appsenkrecht},
\end{equation}
which clearly holds since $\vec{d}_i=\vec{\pi}\times\partial_i\vec{\pi}$. Thus, the two equations (\ref{eq:apppartipi1}) and (\ref{eq:apppartipi2}) suffice to solve the problem. The $\vec{d}_i$ are 3 vectors in the 3-dimensional isospin space. They cannot all be linearly independent, otherwise relation (\ref{eq:appsenkrecht}) could not be satisfied. Hence, at least one of the $\vec{d}_i$ must be linearly dependent from the others. However, if $\vec{\pi}$ is orthogonal to 2 independent constant vectors and $\vec{\pi}^2=\text{const.}$, then also $\vec{\pi}$ is a constant vector. This would be trivial. Consequently, all 3 vectors $\vec{d}_i$ must be linearly dependent, i.e. they all have the same direction. We can choose the coordinate system in the isospin space such that they point into the third direction; $\vec{d}_i=(0,0,a_i)$, where $a_i=|\vec{d}_i|$. Then, through relation (\ref{eq:appsenkrecht}), we immediately conclude that
\begin{equation}
	\pi_3=0,
\end{equation}
while (\ref{eq:apppartipi1}) and (\ref{eq:apppartipi2}) simplify to
\begin{align}
\partial_i\pi_1&=-a_i\pi_2\\
\partial_i\pi_2&=a_i\pi_1.
\end{align}
Since $a_i$ is a constant, the solution of this system of differential equations reads
\begin{align}
\pi_1(x)&=A\cos(a_ix_i+\varphi_0)\\
\pi_2(x)&=A\sin(a_ix_i+\varphi_0).
\end{align}
From $\sigma=\sqrt{1-\vec{\pi}^2(x)}$ it is clear that $\vec{\pi}^2(x)\leq 1$. On the other hand we have
	\[\vec{\pi}^2(x)=\pi_1^2(x)+\pi_2^2(x)+\pi_3^2(x)=A^2,
\]
showing that we must have $-1\leq A\leq 1$. This allows us to parametrize $A$ as
\begin{equation}
	A=\sin\alpha_0.
\end{equation}
According to (\ref{eq:appmatrixU}) the matrix $U(x)$ thus finally reads
\begin{equation}
U(x)=\cos\alpha_0\mathds{1}+i\sin\alpha_0\left(\cos\varphi(\vec{x})\tau^1+\sin\varphi(\vec{x})\tau^2\right),
\end{equation}
where
\begin{equation}
	\varphi(\vec{x})=\vec{a}\cdot\vec{x}+\varphi_0.
\end{equation}
Hence, through the requirement that the connection and the mass term are constant we are led to the spiral configuration (\ref{eq:spiralconfiguration}). The vielbein can then be made constant by an appropriate gauge transformation, as we did it in section \ref{sec:connectionandvielbein}.

\chapter{Analysis of the Energy Expansion}\label{ap:expansion}
Here we want to go through the expansion of the fermionic energy (\ref{eq:energieplus}) in order to find conditions for the expansion to be true. Let us first define the vector
\begin{equation}
	\gamma_i=\frac{1}{2}g_A\beta_i\label{eq:defvongamma}.
\end{equation}
With this, the energy eigenvalues of (\ref{eq:energieplus}) may be written as
\begin{equation}
	E_\pm=\sqrt{m_N^2+p^2+\delta^2+\gamma^2\pm 2\sqrt{m_N^2\gamma^2+(\vec{p}\cdot\vec{\delta})^2+(\vec{p}\cdot\vec{\gamma})^2}},
\end{equation}
or, when using (\ref{eq:skalarproduktpdelta}) and (\ref{eq:skalarproduktpbeta}),
\begin{equation}
	E_\pm=\sqrt{m_N^2+p^2+\delta^2+\gamma^2\pm 2\sqrt{m_N^2\gamma^2+p_3^2(\delta^2+\gamma^2)}}\label{eq:fermionischeenergieusingp3}.
\end{equation}
Now we go through the expansion (\ref{eq:energiezweiteordnung}) in more detail. First we process the internal square root:
\begin{align*}
\sqrt{m_N^2\gamma^2+p_3^2(\delta^2+\gamma^2)}&=m_N\gamma\sqrt{1+\vphantom{\frac{p_3^2}{m_N^2\gamma^2}(\delta^2+\gamma^2)}\smash{\underbrace{\frac{p_3^2}{m_N^2\gamma^2}(\delta^2+\gamma^2)}_{x}}}=m_N\gamma\left(1+\frac{x}{2}\right)+\mathcal{O}(x^2)\\[4ex]
&=m_N\gamma+\frac{p_3^2}{2m_N\gamma}(\delta^2+\gamma^2)+\mathcal{O}(m_N^{-3}),
\end{align*}
where we assumed that $x\ll 1$;
	\[\frac{p_3^2\delta^2}{m_N^2\gamma^2}+\frac{p_3^2}{m_N^2}\ll 1.
\]
For chiral perturbation theory to be valid we must have $p_3\ll m_N$. $p_3$ can at most be $p_F$ and at particle densities we talk about $p_F$ is indeed lower that $m_N$, according to the lower order result (\ref{eq:particlenumberdensity}). Consequently, we require
	\[\frac{p_3^2\delta^2}{m_N^2\gamma^2}=\frac{p_3^2}{m_N^2}\cdot\frac{\delta^2}{\gamma^2}\ll 1\quad\Rightarrow\quad \frac{\delta^2}{\gamma^2}\lesssim\mathcal{O}(1).
\]
Using (\ref{eq:defvongamma}), (\ref{eq:defbetai}) and (\ref{eq:defdeltai}) this condition reads
	\[\frac{\cos^2\alpha_0}{g_A^2\sin^2\alpha_0}\lesssim\mathcal{O}(1).
\]
For large values of $\alpha_0$ this is always satisfied and for small ones $\cos\alpha_0\approx 1$, such that we find
\begin{equation}
	g_A^2\sin^2\alpha_0\gtrsim 1\label{eq:higherordersbedingung1}
\end{equation}
for the above expansion of the internal square root to be allowed. If this is satisfied we can then handle the outer square root in the same manner:
\begin{align*}
E_\pm &=\sqrt{m_N^2+p^2+\delta^2+\gamma^2\pm 2m_N\gamma\pm\frac{p_3^2}{m_N\gamma}(\delta^2+\gamma^2)}\\
&=m_N\sqrt{1+\vphantom{\frac{p^2}{m_N^2}+\frac{\delta^2}{m_N^2}+\frac{\gamma^2}{m_N^2}\pm\frac{2\gamma}{m_N}\pm\frac{p_3^2}{\gamma m_N^3}(\delta^2+\gamma^2)}\smash{\underbrace{\frac{p^2}{m_N^2}+\frac{\delta^2}{m_N^2}+\frac{\gamma^2}{m_N^2}\pm\frac{2\gamma}{m_N}\pm\frac{p_3^2}{\gamma m_N^3}(\delta^2+\gamma^2)}_{x}}}.
\end{align*}
\\[2ex]
Here, we have to take care of the fact that $x$ contains a term $\sim m_N^{-1}$. We therefore have to expand up to $x^3$ producing terms $\sim m_N^{-2}$ and $\sim m_N^{-3}$ which yield terms $\sim m_N^{-1}$ and $\sim m_N^{-2}$ after multiplying the prefactor $m_N$ back. So we use
\begin{equation}
	\sqrt{1+x}=1+\frac{x}{2}-\frac{x^2}{8}+\frac{x^3}{16}+\mathcal{O}(x^4)\label{eq:wurzelentwickeln},
\end{equation}
which is again allowed for $x\ll 1$ only. For $a$ not too large the only term in $x$ that may cause problems is the last one, which blows up for small values of $\alpha_0$. So we must demand
	\[\frac{p_3^2\delta^2}{\gamma m_N^3}+\frac{p_3^2\gamma}{m_N^3}=\underbrace{\frac{p_3^2}{m_N^2}}_{\text{small}}\cdot\frac{\delta^2}{m_N\gamma}+\underbrace{\frac{p_3^2}{m_N^2}\cdot\frac{\gamma}{m_N}}_{\text{small}}\ll 1\quad\Rightarrow\quad \frac{\delta^2}{m_N\gamma}\lesssim\mathcal{O}(1),
\]
which is equivalent to
\begin{equation}
	\frac{a}{2m_Ng_A\sin\alpha_0}\lesssim 1.
\end{equation}
For large $a$ also the other terms in $x$ (except of the first one) may cause problems. Let us assume that we are allowed to do the expansion. Then, from $x^2$ we only pick the terms up to $\mathcal{O}(m_N^{-3})$, i.e.
	\[x^2=\pm\frac{4\gamma p^2}{m_N^3}\pm\frac{4\gamma\delta^2}{m_N^3}\pm\frac{4\gamma^3}{m_N^3}+\frac{4\gamma^2}{m_N^2}+\mathcal{O}(m_N^{-4}),
\]
while from $x^3$ we have only one term $\sim m_N^{-3}$:
	\[x^3=\pm\frac{8\gamma^3}{m_N^3}+\mathcal{O}(m_N^{-4}).
\]
Using (\ref{eq:wurzelentwickeln}) we are then led to
\begin{multline*}
E_\pm = m_N\Bigg(1+\frac{p^2}{2m_N^2}+\frac{\delta^2}{2m_N^2}+\frac{\gamma^2}{2m_N^2}\pm\frac{\gamma}{m_N}\pm\frac{p_3^2}{2\gamma m_N^3}(\delta^2+\gamma^2)\\
\mp\frac{\gamma p^2}{2m_N^3}\mp\frac{\gamma\delta^2}{2m_N^3}\mp\frac{\gamma^3}{2m_N^3}-\frac{\gamma^2}{2m_N^2}\pm\frac{\gamma^3}{2m_N^3}\Bigg)
\end{multline*}
	\[\hspace{-30mm}=m_N+\frac{p^2}{2m_N}+\frac{\delta^2}{2m_N}\pm\gamma \pm\frac{p_3^2}{2\gamma m_N^2}(\delta^2+\gamma^2)\mp\frac{\gamma p^2}{2m_N^2}\mp\frac{\gamma \delta^2}{2m_N^2},
\]
which is exactly the expression (\ref{eq:energiezweiteordnung}).

\chapter{Integration over the Fermi Volume in the $\alpha_0\sim\mathcal{O}(m_N^{-1})$ expansion}\label{ap:deltaf}
Here, we perform the integration over the Fermi volume for the dispersion relation
\begin{equation}
	E_\pm(p)=m_N+\frac{p^2}{2m_N}+\frac{a^2}{8m_N}\pm\frac{a}{2m_N}\sqrt{p_3^2+g_A^2\eta^2}+\mathcal{O}(m_N^{-3}),
\end{equation}
which results in the expansion of the fermionic energy for $\alpha_0\sim\mathcal{O}(m_N^{-1})$.

Let us first find $N_-$. The trick is to extend the momentum integration from the Fermi volume to arbitrary momenta by introducing $\Theta(E_F^--E_-(p))$, where $E_F^-$ is the Fermi energy of the $E_-$ states. Next, we insert a 1 by using the property
\begin{equation}
	\int\limits_{-\infty}^\infty dE\,\delta(E-E_-(p))=1.
\end{equation}
Hence,
\begin{align*}
N_-&=2\left(\frac{L}{2\pi}\right)^3\int\limits_{\hidewidth\substack{\text{Fermi}\\\text{volume}}\hidewidth}\! d^3p=2\left(\frac{L}{2\pi}\right)^3\int\limits_{\mathbb{R}^3}\! d^3p\;\Theta(E_F^--E_-(p))\\
&=2\left(\frac{L}{2\pi}\right)^3\int\limits_{-\infty}^\infty dE\int\limits_{\mathbb{R}^3}\! d^3p\;\delta(E-E_-(p))\Theta(E_F^--E_-(p)).
\end{align*}
For $E<0$ the delta function will always be zero as $E_-(p)>0$. If $E>E_F^-$ we should also have $E_-(p)>E_F^-$ for the delta function to give a contribution. But then $\Theta(E_F^--E_-(p))=0$. Altogether the $E$ integration runs from 0 to $E_F^-$ only:
\begin{equation}
	N=2\left(\frac{L}{2\pi}\right)^3\int\limits_0^{E_F^-}dE\underbrace{\int\limits_{\mathbb{R}^3}d^3p\;\delta(E-E_-(p))}_{=\mathrel{\mathop:}S(E)}=2\left(\frac{L}{2\pi}\right)^3\int\limits_0^{E_F^-}dE\;S(E).\label{eq:nmitsvone}
\end{equation}
Analogously we calculate the total fermionic energy:
\begin{align}
E_\text{tot}^-&=2\left(\frac{L}{2\pi}\right)^3\int\limits_{\hidewidth\substack{\text{Fermi}\\\text{volume}}\hidewidth}\! d^3p\;E_-(p)=2\left(\frac{L}{2\pi}\right)^3\int\limits_{-\infty}^\infty dE\int\limits_{\mathbb{R}^3}\! d^3p\;\delta(E-E_-(p))\Theta(E_F^--E_-(p))E_-(p)\nonumber\\
&=2\left(\frac{L}{2\pi}\right)^3\int\limits_0^{E_F^-}dE\int\limits_{\mathbb{R}^3}d^3p\;\delta(E-E_-(p))E_-(p)=2\left(\frac{L}{2\pi}\right)^3\int\limits_0^{E_F^-}dE\;E\int\limits_{\mathbb{R}^3}d^3p\;\delta(E-E_-(p))\nonumber\\
&=2\left(\frac{L}{2\pi}\right)^3\int\limits_0^{E_F^-}dE\;E\,S(E).\label{eq:emitsvone}
\end{align}

Let us evaluate the integral $S(E)$ which can be regarded as the surface of constant energy in momentum space. The integration over $p_1$ and $p_2$ may be easily performed by introducing polar coordinates $r\mathrel{\mathop:}=p_1^2+p_2^2$ and $\varphi$:
\begin{align*}
S(E)&=\int\limits_{\mathbb{R}^3}d^3p\;\delta(E-E_-(p))=\int\limits_{\mathbb{R}^3}d^3p\;\delta\left(E-m_N-\frac{a^2}{8m_N}-\frac{p^2}{2m_N}+\frac{a}{2m_N}\sqrt{p_3^2+g_A^2\eta^2}\right)\\
&=\int\limits_{\mathbb{R}^3}d^3p\;\delta\Bigg(\underbrace{E-m_N-\frac{a^2}{8m_N}}_{=\mathrel{\mathop:}A}-\frac{p_1^2+p_2^2}{2m_N}\underbrace{-\frac{p_3^2}{2m_N}+\frac{a}{2m_N}\sqrt{p_3^2+g_A^2\eta^2}}_{=\mathrel{\mathop:}-f(p_3)}\Bigg)\\
&=\int\limits_{-\infty}^\infty dp_3\int\limits_0^{2\pi}d\varphi\int\limits_0^\infty dr\;r\,\delta\Bigg(A-\frac{r^2}{2m_N}-f(p_3)\Bigg)=\pi\int\limits_{-\infty}^\infty dp_3\int\limits_0^\infty du\;\delta\Bigg(A-\frac{u}{2m_N}-f(p_3)\Bigg).
\end{align*}
In the last step we changed the integration variable to $u=r^2$. We next factor out $1/2m_N$ in the argument of the delta function and use the property
	\[\delta(ax)=\frac{1}{|a|}\delta(x)
\]
to get
	\[S(E)=2\pi m_N\int\limits_{-\infty}^\infty dp_3\int\limits_0^\infty du\;\delta\Bigg(2m_N(A-f(p_3))-u\Bigg).
\]
As the integration over $u$ runs from 0 to $\infty$ only, we have
	\[\int\limits_0^\infty du\;\delta\Bigg(2m_N(A-f(p_3))-u\Bigg)=\begin{cases}0, & \text{if }2m_N(A-f(p_3))<0\\ 1, & \text{if }2m_N(A-f(p_3))>0 \end{cases}=\Theta(A-f(p_3))
\]
and thus
	\[S(E)=2\pi m_N\int\limits_{-\infty}^\infty dp_3\;\Theta(A-f(p_3)).
\]
In order to evaluate $S(E)$ we therefore basically have to find the roots of the function
\begin{equation}
	g(p_3)\mathrel{\mathop:}=A-f(p_3)=E-m_N-\frac{a^2}{8m_N}-\frac{p_3^2}{2m_N}+\frac{a}{2m_N}\sqrt{p_3^2+g_A^2\eta^2}\label{eq:defvongvonp3}.
\end{equation}
Since $g(p_3)$ is even in $p_3$ the integration may run from 0 to $\infty$ only, bringing a factor 2;
\begin{equation}
	S(E)=4\pi m_N\int\limits_0^\infty dp_3\;\Theta(g(p_3))\label{eq:svone}.
\end{equation}
Let us analyze the behavior of $g(p_3)$. Setting $g'(p_3)=0$ we find its extrema to be located at
\begin{equation}
	p_3=0\quad,\quad p_3=\frac{1}{2}\sqrt{a^2-4g_A^2\eta^2}\label{eq:lagederextremevong}
\end{equation}
(considering only positive values of $p_3$). We now obviously have to distinguish two cases:
\begin{enumerate}
\item $a<2g_A\eta$: The square root gets imaginary and we only have one extremum at $p_3=0$. Since $\lim_{p_3\to\infty}g(p_3)=-\infty$ it is a maximum. The function $g(p_3)$ hence looks like in fig. \ref{fig:higherorders_fall1} and has, depending on the value of $E$ which shifts the graph vertically, one root $b_2$ or none.
\begin{figure}[h!]
\centering
\psfrag{p3}{\small$p_3$}
\psfrag{g}{\small$g(p_3)$}
\psfrag{b2}{\small$b_2$}
\psfrag{b1}{\small$b_1$}
\subfloat[][$a<2g_A\eta$]{\includegraphics[width=0.3\textwidth]{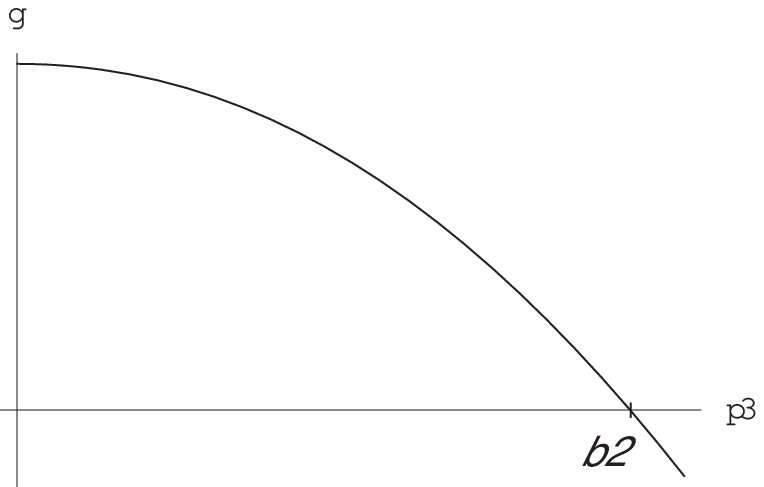}\label{fig:higherorders_fall1}}
\hspace{8mm}
\subfloat[][$a>2g_A\eta$]{\includegraphics[width=0.3\textwidth]{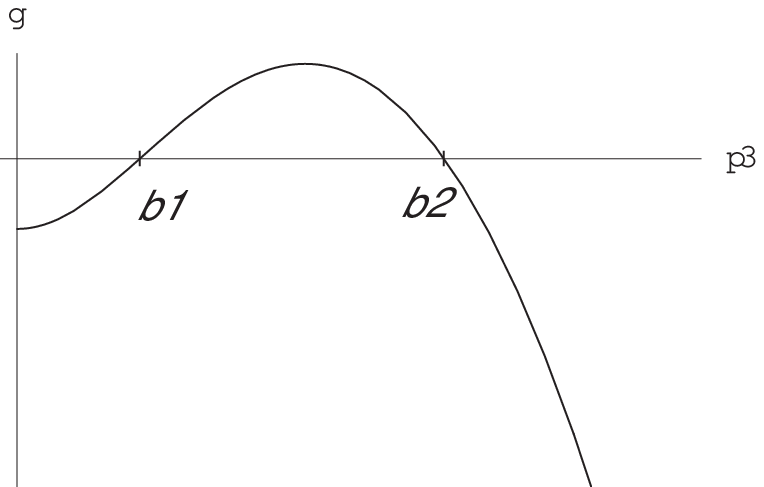}\label{fig:higherorders_fall2}}
\caption{Different behavior of $g(p_3)$ depending on the value of $a$}
\end{figure}\\
\item $a>2g_A\eta$: For $p_3\geq 0$ $g(p_3)$ now has two extrema; the one at $p_3=0$ is a minimum and the other one a maximum. The function thus has the form indicated in fig. \ref{fig:higherorders_fall2} and has, depending on the vertical shift due to $E$ two ($b_1$ and $b_2$), one ($b_2$) or no root.
\end{enumerate}
We now first deal with case 1. In order to evaluate $S(E)$ of (\ref{eq:svone}) we have to know from which value $E_l$ of $E$ on $g(p_3)$ has a root. Below this value $g(p_3)$ is never positive and hence $S(E)=0$. Consequently $E_l$ is the lower integration limit in the integrals of (\ref{eq:nmitsvone}) and (\ref{eq:emitsvone}) respectively. By setting $g(0)=0$ $E_l$ is found to be
\begin{equation}
	E_l=m_N+\frac{a^2}{8m_N}-\frac{ag_A\eta}{2m_N}\label{eq:untereintgrenze}.
\end{equation}
For $E\geq E_l$ the root $b_2$ of $g(p_3)$ lies at
\begin{equation}
	b_2=b_2(E)=\sqrt{\frac{a^2}{4}+2m_N(E-m_N)+a\sqrt{2m_N(E-m_N)+g_A^2\eta^2}}\label{eq:nullstelleb2},
\end{equation}
and we thus find
\begin{equation}
	S^{(1)}(E)=\Theta(E-E_l)4\pi m_Nb_2(E),
\end{equation}
the superscript denoting case 1. With (\ref{eq:nmitsvone}) we are now able to calculate $N$:
\begin{align*}
N^{(1)}_-=8\pi m_N\left(\frac{L}{2\pi}\right)^3\int\limits_{E_l}^{E_F^-}dE\;b_2(E),
\end{align*}
or
\begin{equation}
	n^{(1)}_-=\frac{m_N}{\pi^2}\int\limits_{E_l}^{E_F^-}dE\;b_2(E)\label{eq:nmiti1}.
\end{equation}
The integral may be calculated analytically by changing the integration variable to $x=2m_N(E-m_N)+g_A^2\eta^2$, leading to an integral of the type $\int dx\;\sqrt{x+c_1+c_2\sqrt{x}}$. Staying on the same path as in the previous calculations we would now try to solve this equation for $E_F^-$ and plug it into the expression for $E_\text{tot}^-$ (which still needs to be evaluated) in order to express the latter in terms of $n_-$. However, it does not seem to be possible to do so, since the result of the integration contains $E_F^-$ in a nontrivial way. We therefore will have to do this replacement numerically, i.e. for given values of $n_-$ and $a$ we numerically solve (\ref{eq:nmiti1}) for $E_F^-$ (which is possible) and can thus calculate $E_\text{tot}^-$. Let us find the corresponding expression for $E_\text{tot}^-$, picking up (\ref{eq:emitsvone}):
	\[E_\text{tot}^{(1)-}=8\pi m_N\left(\frac{L}{2\pi}\right)^3\int\limits_{E_l}^{E_F^-}dE\;E\,b_2(E).
\]
This integral may also be calculated analytically even though resulting in an even more complicated expression. The total fermionic energy density due to $n_-$ finally reads
\begin{equation}
	\epsilon_\text{tot}^{(1)-}=\frac{m_N}{\pi^2}\int\limits_{E_l}^{E_F^-}dE\;E\,b_2(E)\label{eq:fermionischeenergiefall1}.
\end{equation}

Next, we have to study case 2, i.e. $a>2g_A\eta$. For this purpose we have to determine for which ranges of $E$ the function $g(p_3)$ possesses no, two and one root respectively. The value $E_l'$ below which $g(p_3)$ has no root is found by solving $g(p_{3,\text{max}})=0$ for $E$, where $p_{3,\text{max}}$ is the location of the maximum, given by the second equation of (\ref{eq:lagederextremevong}). The result is
\begin{equation}
	E_l'=m_N-\frac{g_A^2\eta^2}{2m_N}.
\end{equation}
The next crucial value of $E$ is again where $g(0)=0$. This value is of course in turn given by $E_l$ of (\ref{eq:untereintgrenze}). Hence, for $E_l'\leq E\leq E_l$ the function $g(p_3)$ possesses two roots $b_1$ and $b_2$, where
\begin{equation}
	b_1=b_1(E)=\sqrt{\frac{a^2}{4}+2m_N(E-m_N)-a\sqrt{2m_N(E-m_N)+g_A^2\eta^2}},
\end{equation}
and $b_2$ is given by (\ref{eq:nullstelleb2}). For $E>E_l$ $b_2$ is again the only root to deal with. Altogether we obtain
\begin{equation}
	S^{(2)}(E)=4\pi m_N\begin{cases}0, & E<E_l'\\ b_2(E)-b_1(E), & E_l'\leq E\leq E_l\\ b_2(E), & E>E_l\end{cases}.
\end{equation}
Consequently we have to split the integration in (\ref{eq:nmitsvone}) and (\ref{eq:emitsvone}) into two parts:
\begin{align}
N^{(2)}_-&=8\pi m_N\left(\frac{L}{2\pi}\right)^3\left\{\int\limits_{E_l'}^{E_l}dE\;[b_2(E)-b_1(E)]+\int\limits_{E_l}^{E_F^-}dE\;b_2(E)\right\}\nonumber\\
&=\frac{L^3m_N}{\pi^2}\left\{\int\limits_{E_l'}^{E_F^-}dE\;b_2(E)-\int\limits_{E_l'}^{E_l}dE\;b_1(E)\right\}\nonumber\\
n^{(2)}_-&=\frac{m_N}{\pi^2}\left\{\int\limits_{E_l'}^{E_F^-}dE\;b_2(E)-\int\limits_{E_l'}^{E_l}dE\;b_1(E)\right\}\label{eq:nmiti2}
\end{align}
\begin{align}
E_\text{tot}^{(2)-}&=8\pi m_N\left(\frac{L}{2\pi}\right)^3\left\{\int\limits_{E_l'}^{E_l}dE\;E\,[b_2(E)-b_1(E)]+\int\limits_{E_l}^{E_F^-}dE\;E\,b_2(E)\right\}\nonumber\\
&= \frac{L^3m_N}{\pi^2}\left\{\int\limits_{E_l'}^{E_F^-}dE\;E\,b_2(E)-\int\limits_{E_l'}^{E_l}dE\;E\,b_1(E)\right\}\nonumber\\
\epsilon_\text{tot}^{(2)-}&=\frac{m_N}{\pi^2}\left\{\int\limits_{E_l'}^{E_F^-}dE\;E\,b_2(E)-\int\limits_{E_l'}^{E_l}dE\;E\,b_1(E)\right\}.\label{eq:fermionischeenergiefall2}
\end{align}

Now we have to repeat this calculation with $E_+=E_+(p)$. The only thing that changes is the function $g(p_3)$ of (\ref{eq:defvongvonp3}), which in the case of $E_+$ now reads (let us denote it with a superscript +):
\begin{equation}
	g^+(p_3)=E-m_N-\frac{a^2}{8m_N}-\frac{p_3^2}{2m_N}-\frac{a}{2m_N}\sqrt{p_3^2+g_A^2\eta^2}.
\end{equation}
This function always has a maximum at $p_3=0$ only, i.e. we simply need to repeat the above first case. $E_l$ of (\ref{eq:untereintgrenze}) will be replaced by
\begin{equation}
	E_l^+=m_N+\frac{a^2}{8m_N}+\frac{ag_A\eta}{2m_N},
\end{equation}
while the root $b_2$ in (\ref{eq:nullstelleb2}) becomes
\begin{equation}
	b_2^+(E)=\sqrt{\frac{a^2}{4}+2m_N(E-m_N)-a\sqrt{2m_N(E-m_N)+g_A^2\eta^2}}.
\end{equation}
Equations (\ref{eq:nmiti1}) and (\ref{eq:fermionischeenergiefall1}) may again be used with the replacements $E_F^-\to E_F^+$, $E_l\to E_l^+$ and $b_2(E)\to b_2^+(E)$:
\begin{align}
n_+&=\frac{m_N}{\pi^2}\int\limits_{E_l^+}^{E_F^+}dE\;b_2^+(E)\label{eq:nmiti1plus}\\
\epsilon_\text{tot}^+&=\frac{m_N}{\pi^2}\int\limits_{E_l^+}^{E_F^+}dE\;E\,b_2^+(E)\label{eq:fermionischeenergiefall1plus}.
\end{align}

\chapter{Neutron Stars}\label{app:neutronstars}
During its lifetime a star produces its energy mainly through the fusion reactions
\begin{align*}
\sideset{^1}{}{\mathop{\text{H}}}+ \sideset{^1}{}{\mathop{\text{H}}} &\to \sideset{^2}{}{\mathop{\text{H}}}+e^++\nu+1.44\,\text{MeV}\\
\sideset{^2}{}{\mathop{\text{H}}}+ \sideset{^1}{}{\mathop{\text{H}}} &\to \sideset{^3}{}{\mathop{\text{He}}}+\gamma+5.49\,\text{MeV}\\
\sideset{^3}{}{\mathop{\text{He}}}+ \sideset{^3}{}{\mathop{\text{He}}} &\to \sideset{^4}{}{\mathop{\text{He}}}+ \sideset{^1}{}{\mathop{\text{H}}}+ \sideset{^1}{}{\mathop{\text{H}}}+12.85\,\text{MeV}.
\end{align*}
As soon as all the hydrogen is burned, gravitation compresses the helium until other fusion reaction start. With increasing diversity of possible nuclear reactions the synthesis of heavier nuclei is pushed further until most of the matter consists of Fe, Si and close-by elements. This is then the point where the fusion reactions stop and the thermal pressure cannot compensate the gravitational pressure anymore. If the star has sufficient mass, the thereupon incipient contraction crushes the atomic structure resulting in a mixture of electrons and nuclei, where the electrons can be treated as an ideal Fermi gas (i.e. neglecting interactions between them) in a first approximation. It is then the Fermi pressure of the electrons that prevents the object (called white dwarf) from a further contraction. Let us calculate this pressure by taking the electrons as degenerate Fermi gas at $T=0$.

Since we consider free electrons the dispersion relation of one electron is given by (let us introduce factors of $c$ and $\hbar$ in this section)
\begin{equation}
	E(p)=\sqrt{m_e^2c^4+p^2c^2},
\end{equation}
where $m_e$ is the electron mass and $p=|\vec{p}|$ the electron momentum. (\ref{eq:particlenumberdensity}) gives the connection between the Fermi momentum $p_F$ and the electron number density $n_e$:
\begin{equation}
	p_F=\hbar\left(3\pi^2n_e\right)^{1/3}.\label{eq:neutronstarfermiimpuls}
\end{equation}
At $T=0$ all states up to $p_F$ are filled and the total energy of the electrons is given by (cf. (\ref{eq:totaleenergiepm}))
\begin{align}
E_e&=2\frac{V}{(2\pi\hbar)^3}\int\limits_{\hidewidth\substack{\text{Fermi}\\\text{sphere}}\hidewidth}\!d^3p\, E(p)=\frac{8\pi V}{(2\pi\hbar)^3}\int\limits_0^{p_F}dp\,p^2\sqrt{m_e^2c^4+p^2c^2}\nonumber\\
&=\frac{m_e^4c^5}{\pi^2\hbar^3}V\int\limits_0^{x_F}dx\,x^2\sqrt{1+x^2}=\frac{m_e^4c^5}{\pi^2\hbar^3}Vf(x_F)\label{eq:neutronstartotelenergie},
\end{align}
where we have introduced the dimensionless momenta
\begin{equation}
	x=\frac{p}{m_ec}\;,\quad x_F=\frac{p_F}{m_ec}
\end{equation}
and the function
\begin{equation}
	f(x_F)=\int\limits_0^{x_F}dx\,x^2\sqrt{1+x^2}=\begin{cases}\frac{1}{3}x_F^3\left(1+\frac{3}{10}x_F^2+...\right), & x_F\ll 1\\ \frac{1}{4}x_F^4\left(1+\frac{1}{x_F^2}+...\right), & x_F\gg 1\end{cases}\label{eq:neutronstarfvonxf}.
\end{equation}
The pressure $P_e$ then follows from the thermodynamical relation $dE=TdS-PdV$ with $T=0$, leading to $P_e=-\partial E_e/\partial V$. When taking the derivative of (\ref{eq:neutronstartotelenergie}) with respect to $V$ we have to keep in mind, that $x_F$ also depends on $V$; due to (\ref{eq:neutronstarfermiimpuls}) we have $x_F=CV^{-1/3}$ with some constant $C$. So we get
	\[\frac{\partial}{\partial V}f(x_F)=\frac{\partial f}{\partial x_F}\frac{\partial x_F}{\partial V}=-x_F^2\sqrt{1+x_F^2}\frac{C}{3V^{4/3}}=-\frac{x_F^3}{3V}\sqrt{1+x_F^2},
\]
leading to
\begin{equation}
	P_e=-\frac{\partial E_e}{\partial V}=\frac{m_e^4c^5}{\pi^2\hbar^3}\left(\frac{x_F^3}{3}\sqrt{1+x_F^2}-f(x_F)\right)\label{eq:neutronstarPe}.
\end{equation}
Here we see why only the Fermi pressure of the electrons is important. If the particles are non-relativistic ($x_F\ll 1$) the Fermi pressure goes like $1/m$, i.e. only the light electrons have an important contribution. As the electrons get relativistic the star becomes unstable, see below.

We want to express $P_e$ as a function of the (for simplicity constant) matter density
\begin{equation}
	\rho=\sigma n_em_N\label{eq:neutronstarichte},
\end{equation}
where $\sigma$ is the mean number of nucleons (of mass $m_N$) per electron and where the contribution of electrons to the matter density is neglected. Solving (\ref{eq:neutronstarichte}) for $n_e$ and plugging that into $x_F$ yields
\begin{equation}
	x_F=\frac{\hbar}{m_ec}\left(\frac{3\pi^2\rho}{\sigma m_N}\right)^{1/3}\label{eq:neutronstarxF}.
\end{equation}
Solving this for $\rho$ and setting $x_F=1$ we get some characteristic density $\rho_c$ (which will turn out to be the typical density of a white dwarf star):
\begin{equation}
	\rho_c=\frac{\sigma m_N}{3\pi^2\hbar^3}(m_ec)^3\label{eq:neutronstarrhoc}.
\end{equation}
Since $x_F\propto \rho^{1/3}$ the case $x_F\ll 1$ in (\ref{eq:neutronstarfvonxf}) corresponds to $\rho\ll\rho_c$, while $x_F\gg 1$ means $\rho\gg\rho_c$. Expanding (\ref{eq:neutronstarPe}) together with (\ref{eq:neutronstarfvonxf}) in powers of $x_F$ and using (\ref{eq:neutronstarxF}) we are thus led to
\begin{equation}
	P_e=\frac{m_e^4c^5}{\pi^2\hbar^3}\begin{cases}\frac{1}{15}x_F^5, & x_F\ll 1\\ \frac{1}{12}x_F^4, & x_F\gg 1\end{cases}=\begin{cases} K_1\rho^{5/3}, & \rho\ll\rho_c\\ K_2\rho^{4/3}, & \rho\gg\rho_c\end{cases},
\end{equation}
with $K_1$ and $K_2$ containing the prefactors. In both cases we hence have a so called polytropic equation of state, i.e. a relation $P=K^\gamma$. A necessary condition for the stability of a star is however $\gamma\geq 4/3$, which follows from an analysis of the total energy of the star when varying its radius. Through the so called Lane-Emden function the case $\rho\gg\rho_c$ (which means that the electrons become relativistic) may be connected to the mass of the star as follows:
\begin{equation}
	M_C=\frac{5.87}{\sigma^2}M_\text{sun},
\end{equation}
where $M_C$ is known as the Chandrasekhar mass limit. Consequently, as $M\to M_C$ the star becomes unstable. For a white dwarf consisting of iron, $\sigma=56/26\approx 2.15$ and hence $M_C=1.27M_\text{sun}$. Already before reaching this mass limit the reaction
\begin{equation}
	p+e^-\to n+\nu_e\label{eq:neutronstarinversebeta}
\end{equation}
starts to take place (as soon as the Fermi energy of the electrons exceeds the energy that is released in the $\beta$ decay, which is about $1.5m_ec^2$), which implies a breakdown of the Fermi pressure and a further collapse of the star. A typical white dwarf therefore has the approximately the mass of our sun and $\rho_c$ of (\ref{eq:neutronstarrhoc}) is indeed the characteristic density of a white dwarf. Here we talk about approximately two tons per cubic centimeter, while its radius corresponds roughly to that of the earth.

For a star of mass $M\geq M_C$ the Fermi pressure of the electrons is not able to prevent the matter from further contraction. The denser the material the more protons and electrons are converted into neutrons via the reaction (\ref{eq:neutronstarinversebeta}). The neutrinos escape and what is left is an object that predominantly consists of neutrons; a so called neutron star. It is then the Fermi pressure of the neutrons that keep the object stable. Hence, the above formulas may also be used for a neutron star if we replace $m_e$ by $m_N$ and set $\sigma=1$. The characteristic density of a neutron star is therefore
\begin{equation}
	\rho_c=\frac{m_N^4c^3}{3\pi^2\hbar^3},
\end{equation}
which is about $6\cdot 10^{18}\,\text{kg/m$^3$}$. Accordingly, the particle number density of the neutrons is $n=\rho_c/m_N\approx 4\cdot 10^{45}\,\text{m}^{-3}=4\,\text{fm}^{-3}$ and is thus of the same order as in an atomic nucleus (cf. (\ref{eq:nuclearmatterdensityinfm})). The calculation of the mass and the radius of the neutron star with the methods used in the case of white dwarfs is however only possible for densities well below $\rho_c$. In this region the radius of such a dense object can be estimated to be approximately $10\,\text{km}$. In order to find out the mass limit one has to go to higher densites, where relativistic corrections and other processes become important. Different model calculations led to mass limits for neutron stars between $1.5M_\text{sun}$ and $3M_\text{sun}$. If the mass of a star reaches this mass limit the neutrons become relativistic and the object unstable. Then, the gravitational collapse (most likely) cannot be stopped by anything and the star will form a black hole.

\end{appendix}

\end{document}